\documentclass[traditabstract]{aa}
\usepackage{graphicx}
\usepackage{txfonts}
\usepackage{longtable,lscape}

\newcommand{\eps}[1]{\mbox{log~$\epsilon$(#1)}}
\newcommand\iso[2]{$\phantom{}^{\rm #1}$#2}

\def\etal{\mbox{\rm et al.\ }}

\def\rpro{\mbox{$r$-process}}
\def\spro{\mbox{$s$-process}}
\def\sri{\mbox{$s$-rich}}
\def\spo{\mbox{$s$-poor}}
\def\ncap{\mbox{$n$-capture}}
\def\teff{\mbox{T$_{\rm eff}$}}
\def\logg{\mbox{log~{\it g}}}
\def\vmicro{\mbox{$\xi_{\rm t}$}}

\def\deg{\mbox{$\rm {^{\circ}}$}}

\begin{document}

\title{The two metallicity groups of the globular cluster M22: a chemical
  perspective\thanks {Based on data collected at: Anglo-Australian Telescope 
    with the University College London Echelle Spectrograph, Apache Point
    Observatory with the ARC Echelle Spectrograph, Lick Shane 3.0m Telescope
    with the Hamilton Echelle Spectrograph, McDonald Smith 2.7m Telescope with
    the Robert G.\ Tull Coud\'e Spectrograph, and European Southern
    Observatory with the FLAMES/UVES spectrograph.}
    \thanks {Tables 3,
    4, 5, and 6 are only available in electronic form at the CDS via
    anonymous ftp to cdsarc.u-strasbg.fr}}

\author{
A.\, F.\, Marino\inst{1,2},
C.\, Sneden\inst{3},
R.\, P.\, Kraft\inst{4},
G.\, Wallerstein\inst{5},
J.\, E.\, Norris\inst{6},
G.\, Da Costa\inst{6},
A.\, P.\, Milone\inst{7,8},
I.\, I.\, Ivans\inst{9},
G.\, Gonzalez\inst{5,10},
J.\, P.\, Fulbright\inst{4,11}
M.\, Hilker\inst{12},
G.\, Piotto\inst{2},
M.\, Zoccali\inst{13},
P.\, B.\, Stetson\inst{14}}

\offprints{A.\ F.\ Marino}

\institute{  Max-Planck-Institut f\"{u}r Astrophysik, Karl-Schwarzschild-Str. 1,
             85741 Garching bei M\"{u}nchen, Germany\\
             \email{amarino@MPA-Garching.MPG.DE}
             \and
             Dipartimento  di   Astronomia,  Universit\`a  di Padova,
             Vicolo dell'Osservatorio 3, Padova, I-35122, Italy\\
             \email{anna.marino@unipd.it, giampaolo.piotto@unipd.it}
             \and
             Department of Astronomy and McDonald Observatory,
             The University of Texas, Austin, TX 78712, USA\\
             \email{chris@verdi.as.utexas.edu}
             \and
             UCO/Lick Observatory, Department of Astronomy and
             Astrophysics, University of California, Santa Cruz,
             CA 95064, USA\\ 
             \email{kraft@ucolick.org}
             \and
             Department of Astronomy, University of Washington,
             Seattle, WA 98195, USA\\ 
             \email{wall@astro.washington.edu}
             \and
             Research School of Astronomy and Astrophysics,
             Australian National University, Cotter Road, Weston,
             ACT 2611, Australia\\
             \email{jen@mso.anu.edu.au, gdc@mso.anu.edu.au}
             \and
            Instituto de Astrof\'isica de Canarias, La
            Laguna, Tenerife, Spain\\
             \email{milone@iac.es}   
             \and
            Departamento de Astrof\'isica, Universidad de La Laguna, Tenerife, Spain
             \and
             Department of Physics and Astronomy, The University of Utah,
             Salt Lake City, UT 84112, USA\\
             \email{iii@physics.utah.edu}
             \and
             Physics Department, Grove City College,
             100 Campus Drive,
             Grove City, Pennsylvania 16127, USA\\
             \email{gonzalezg@gcc.edu}
             \and
             Department of Physics and Astronomy, Johns Hopkins
             University, Baltimore, MD 21218, USA\\
             \email{jfulb@skysrv.pha.jhu.edu}
             \and
             European Southern Observatory, Karl-Schwarzschild-Str. 2,
             85748 Garching bei M\"{u}nchen, Germany\\
             \email{mhilker@eso.org} 
             \and
             Pontificia Universidad Cat\'olica de Chile,
             Departmento de Astronom\'ia y Astrofis\'ica, Casilla 306,
             Santiago 22, Chile\\
             \email{mzoccali@astro.puc.cl} 
             \and
             Herzberg Institute of Astrophysics,
             National Research Council Canada, 5071 West Saanich Road,
             Victoria, BC V9E 2E7\\
             \email{Peter.Stetson@nrc-cnrc.gc.ca}
             }

\date{Received Xxxxx xx, xxxx; accepted Xxxx xx, xxxx}
%__________________________________________________________________
%
\abstract{
We present a detailed chemical composition analysis of 35 red giant
stars in the globular cluster M22.
High resolution spectra for this study were obtained at five
observatories, and analyzed in a uniform manner.
We have determined abundances of representative light proton-capture,
$\alpha$, Fe-peak and neutron-capture element groups.
Our aim is to better understand the {\it peculiar} chemical enrichment
history of this cluster, in which two stellar groups are characterized
by a different content in iron, neutron capture elements Y, Zr and Ba,
and $\alpha$ element Ca.
The principal results of this study are:
\textit{(i)} substantial star-to-star metallicity scatter
($-$2.0~$\lesssim$~[Fe/H]~$\lesssim$~$-$1.6);
\textit{(ii)} enhancement of \spro/\rpro\ neutron-capture abundance
ratios in a fraction of giants, positively correlated with metallicity;
\textit{(iii)} sharp separation between the \spro-rich and \spro-poor
groups by [La/Eu] ratio;
\textit{(iv)} possible increase of [Cu/Fe] ratios with increasing [Fe/H],
suggesting that this element also has a significant \spro\ component; and
\textit{(v)} presence of Na-O and C-N anticorrelations in both the
stellar groups.
}


\titlerunning{The two metallicity groups in M22}
\authorrunning{Marino et al.}

\maketitle

%%%%%%%%%%%%%%%%%%%%%%%%%%%%%%%%%%%%%%%
\section{Introduction}
\label{introduction}
%%%%%%%%%%%%%%%%%%%%%%%%%%%%%%%%%%%%%%%
In recent years, a large amount of observational evidence, both from
high resolution spectroscopy and from photometry, has established that
globular clusters (GC) can host more than one stellar population.
Photometric evidence of multiple stellar populations has been recently
observed in many GCs in the form of multiple main sequences
(Bedin \etal\ 2004, Piotto \etal 2007, Anderson \etal 2009, and Milone
\etal 2010), split 
sub-giant branches (Milone \etal 2008, and Piotto 2009), and multiple
red-giant branches (e.g.\ Marino \etal 2008, Lee \etal 2009, Lind
\etal\ 2011a).

Spectroscopic clues, often easy to detect simply by inspection of
high-resolution spectra,  arise in derived elemental abundance variations.
Most GCs are mono-metallic (Carretta \etal\ 2009a) , i.e. they have no detectable dispersion in
[Fe/H].\footnote{
We adopt standard stellar abundance notation (Helfer, Wallerstein,
\& Greenstein 1959): for elements A and B,
[A/B] $\equiv$ log$_{\rm 10}$(N$_{\rm A}$/N$_{\rm B}$)$_{\star}$ $-$
log$_{\rm 10}$(N$_{\rm A}$/N$_{\rm B}$)$_{\odot}$.
We also define
\eps{A} $\equiv$ log$_{\rm 10}$(N$_{\rm A}$/N$_{\rm H}$) + 12.0, and
and use [Fe/H] synonymously with stellar overall metallicity.}
But all clusters studied to date with substantial 
stellar samples exhibit star-to-star variations in the light elements 
C, N, O, Na, Mg, and Al.
These variations do not occur randomly, but have obvious correlations
and anticorrelations, which clearly point to the actions of high-temperature
proton fusion cycles that have processed C and O into N, (unseen) Ne to
Na, and Mg to Al.

Proton-capture reactions to effect Ne$\rightarrow$Na conversion require
fusion-region temperatures T~$\gtrsim$ $40\times10^{\rm 6}$~K,
and Mg$\rightarrow$Al conversion can only occur with even higher
temperatures.
Such conditions cannot be achieved in the H-fusion layers of the 
presently observed low-mass ($\simeq$0.8M$_\odot$) GC stars.
This suggests that the stars observed now were formed from cluster
material that had been polluted in proton-capture products by a prior 
generation of cluster stars.
Today, it is widely accepted that the observed variations in light
elements provide strong evidence for the presence of different
generations of stars in GCs, with the younger stars born in a medium
enriched in the material ejected by earlier generation stars.
However the debate on the possible polluters is still open
(Ventura \etal\ 2001, D'Antona \& Ventura 2007, Decressin \etal\ 2007).

The mono-metallicity of most GCs has excluded supernovae from being 
responsible for the chemical enrichment of the intra-cluster medium
from which the second generation formed, because supernovae ejecta 
would be enriched in Fe-peak elements.
We know of some exceptions, most notably $\omega$~Centauri.
This most massive cluster is an anomaly among GCs because its Fe 
variations are huge, spanning more than 1~dex.
However, there have been long-held suspicions that M22 has metallicity
variations.
A large number of photometric and spectroscopic studies have attempted
to verify or disprove this idea.
For example, Hesser \etal (1977) and Peterson \& Cudworth (1994) showed
a significant spread along the RGB in M22 in observed ($B-V$) and
Str\"omgren colours.
Norris \& Freeman (1983) used low-resolution spectra of about 100 red 
giants (RGB) to demonstrate the existence of star-to-star variations in the 
strengths of the \ion{Ca}{II} H\&K lines, which they interpreted as
a $\sim$0.30~dex spread in Ca abundances.

However, M22 lies nearly in the Milky Way plane, toward the Bulge 
([($l$,~$b$) $\sim$ (10\deg,~7\deg)]).
As such, it suffers significant dust extinction, with probable 
reddening variations across the face of the cluster.
This has limited conclusions that could be drawn from the photometric
and spectroscopic variations present in M22.
Early spectroscopic studies claiming a spread in metallicity, with
$-$1.4~$<$ [Fe/H]~$<$ $-$1.9, included those of Pilachowski \etal\ (1982),
based on 6 stars, and Lehnert \etal\ (1991), based on 4 stars.
On the other hand, neither Cohen (1981), nor Gratton (1982) found a
significant M22 metallicity variation within the three stars
that each analyzed.
These early studies are not necessarily in contradiction because
they all are conclusions from small sample sizes.

Recently, intrinsic Fe variations in M22 have been confirmed definitively
by Marino \etal\ (2009, hereafter M09) and Da Costa \etal\ (2009, hereafter
DC09). 
In particular M09, from the analysis of high resolution UVES spectra of
17 stars, found that M22 shows a complex chemical pattern that resembles
the extreme case of $\omega$~Cen (see also Da Costa \& Marino 2010).
Stars in M22 show intrinsic variations in Fe, albeit significantly
smaller than the ones observed in $\omega$~Cen, i.e.\ while the
difference between the mean Fe abundances of different groups of stars
is 0.14 dex (M09), stars in $\omega$~Cen span a range of $\sim$1.5 dex.
Additionally, M09 argued for the presence of two different stellar
populations in this cluster characterized by significant differences in
neutron-capture (\ncap) elements Y, Ba, and Zr.
Light proton-capture elements vary in M22 as they do in most clusters
as described above, but their abundance variations are uncorrelated to
those of the \ncap\ elements.
The two M22 populations also appear to have different [Ca/Fe] ratios,
but no linkage to the proton-capture elements is evident.

The multiple populations of M22 are now clearly manifest in a photometric
split in the sub-giant branch (SGB) region revealed in Hubble Space
Telescope images (M09; Piotto 2009).
The split SGB points towards the presence of two stellar
generations, which are likely related to chemical 
composition differences.
However, M09 argued that evolutionary models cannot entirely 
reproduce the size of the split by considering only the metallicity 
spread (but see also DC09).
Probably the origin of the split is more complex, and could involve also
variations in the total CNO abundance, as proposed by Cassisi \etal\ (2008)
and Ventura \etal\ (2009) for NGC~1851, in contrast to the usual
assumption of constant C+N+O inside stars in a given GC.

Since the 1980's we have known that large CO band strength variations 
(Frogel, Persson \& Cohen 1983) are present among M22 stars, and these 
are accompanied by often strong CN enhancements (Cohen 1981,
Norris \& Freeman 1983). 
More recently, Kayser \etal\ (2008) confirmed the presence of
both CN-weak and CN-strong stars in M22, with the majority of
stars being CN-weak group.
Their pager found no evidence for a CN-CH anticorrelation.  
Indeed, Norris \& Freeman had noted that in M22, contrary to $normal$ 
GCs, CN strengths are positively correlated with those of CH.
They also found that CH and CN are correlated with Ca, with the
correlation being tighter between CH and Ca.
This suggests that a common source is responsible for the C and Ca
enhancements, and a different mechanism is responsible for the
N enhancements.
This further argues that a simple conversion of C to N by CN cycling
cannot be the unique cause of the spread in CN index.
Brown, Wallerstein, \& Oke (1990) studied CNO abundances in seven 
M22 RGB stars, and found evidence in two of the stars for a large
overabundance of N, corresponding to a higher CNO total abundance.
This behavior cannot be explained with enrichment from material that
simply has undergone the complete CNO cycle.

The various chemical anomalies of M22 stars indicate that this GC has
undergone a complex, and still unclear, chemical enrichment history.
Similarly to $\omega$~Cen, Fe and Ca variations suggest that core-collapse
supernovae have had a role in the pollution of the intra-cluster medium
from which the present generation of slightly Fe-enriched stars formed.
At the same time, the presence of two different groups in some \ncap\
elements suggests that at the same epoch a number of low mass AGB stars
($\sim$3 M$_{\odot}$, Ventura \etal\ 2009) experienced the third
dredge-up and polluted the intra-cluster medium with {\spro\ and triple
$\alpha$ products.

In this study we present a new high-resolution spectroscopic analysis
of a larger sample of RGB stars in M22 than has been conducted to date.
Altogether our sample now consists of 35 stars.
Our aim is to study the chemical properties of different stellar
populations in this cluster in order to re-construct its chemical
enrichment history.
The layout of this paper is as follows: \S\ref{data} is a brief
overview of the data set; \S\ref{modelatm} and \S\ref{abunds} contain
descriptions of model atmosphere parameter and abundance derivations;
\S\ref{results} presents the abundance results; \S\ref{rgb} demonstrates
the links between our abundances and spectroscopic/photometric
indices of M22 giants; and \S\ref{conclusions} summarizes and
discusses our findings.

%%%%%%%%%%%%%%%%%%%%%%%%%%%%%%%%%%%%%%%%%%%%%%%%%%%%%%%%%%%%%%%%%%%%%%%%%%
\section{DATA SOURCES\label{data}}
%%%%%%%%%%%%%%%%%%%%%%%%%%%%%%%%%%%%%%%%%%%%%%%%%%%%%%%%%%%%%%%%%%%%%%%%%%

High-resolution spectra of M22 giants were obtained at five observatories.
In Table~\ref{tab-datasources} we summarize the properties of these
instruments, giving their resolving powers, approximate spectral range
coverage, typical signal-to-noise values per pixel for the reduced spectra,
the number of observed stars and references to information on the instruments.
Here are some notes about the observations and references to more
detailed information.

\begin{itemize}

\item Anglo-Australian Telescope, University College London
Echelle Spectrograph (hereafter called AAT):
observations and reductions were described by Norris, Da Costa,
\& Tingay (1995).
The reductions were accomplished with a combination of tasks from
the IRAF\footnote{
IRAF is distributed by NOAO, which is operated by AURA,
under cooperative agreement with the US National Science Foundation.}
and FIGARO\footnote{ \sf {http://www.aao.gov.au/figaro/}}
software packages.
That paper also tabulates their equivalent widths, but we do not
use those measurements in this paper; for uniformity in our analyses we
re-measured equivalent widths from the original spectra.
Comparison of our equivalent widths with those of Norris \etal
shows good agreement, with a mean difference of 
$-$1.8~m\AA\ ($\sigma$ = 4.5~m\AA) and no trend with line strength.

\item Apache Point Observatory, 3.5m telescope, ARC Echelle Spectrograph
(hereafter, APO):
Observation and reduction procedures were identical to those described by
Laws \& Gonzalez (2003).
In particular, the data were reduced using standard routines in IRAF
for flat fielding, wavelength calibration and background subtraction.

\item Lick Shane 3.0m Telescope, Hamilton Echelle (hereafter, LICK):
We followed the procedures outlined in Ivans \etal\ (1999).
The reductions were carried out using the VISTA software package
(Goodrich \& Veilleux 1988).

\item McDonald Smith 2.7m Telescope, Robert G. Tull Coud\'e Spectrograph
(hereafter, MCD):
Observations and reductions were identical
to those employed by Ivans \etal\ (1999) for their high-resolution
study of stars in the GC M4.

\item ESO Very Large Telescope, Ultraviolet and Visual Echelle Spectrograph
(programmes: 68.D-0332 and 71.D-0217; hereafter UVES):
See M09 for a discussion of the observations and reductions.

\end{itemize}

Equivalent widths (EWs) were measured by fitting Gaussian profiles to
isolated stellar absorption lines.
For each line, we selected a region of 10~\AA\ centered on the line
itself to estimate continuum placement.
This value is a good compromise between having enough points to build
reasonable statistics, and avoiding regions where the spectrum is not
sufficiently flat.
Then we built the histogram of the distribution of the flux where the peak
is a rough estimation of the continuum.
We refined this determination by fitting a parabolic curve to the peak and
using the vertex as our continuum estimation.
Finally, the continuum determination was revised by eye and corrected by
hand if a clear discrepancy with the spectrum was found.
We rejected the EW of a transition if any of the following problems were
detected: non-Gaussian line profile; central observed wavelength mismatch
with the expected line list wavelength; line breadth either substantially
broader or narrower with respect to the mean FWHM.
We verified that the Gaussian shape was a good approximation for our
spectral lines, so no Lorentzian correction was applied.

In Table~\ref{tab-datastars} we provide a list of basic data
for each observed star, together with the instrument source.
The different spectroscopic data sets were analyzed separately,
but the results are combined in the abundance discussion.
The table lists magnitudes and colours as observed (no corrections
for dust extinction), and the adopted extinction $A_V$ for each star.
The broadband $V$, $B$, and $I$ magnitudes are from Stetson's
photometric database\footnote{available at
\sf {http://www3.cadc-ccda.hia-iha.nrc-cnrc.gc.ca/community/STETSON/}}
while the $K$ magnitudes are from the 2MASS Point Source
Catalog (Skrutskie et al. 2006).
The Str\"omgren colours $b-y$ and $m_1$ are taken from Richter,
Hilker, \& Richtler (1999).

We computed extinction values $A_V$ in the following manner.
M22 has large mean reddening, $E(B-V)$~=~0.34 (Harris 1996).
Additionally, there is evidence for differential reddening
at a level of nearly a tenth of a magnitude in $E(B-V)$ across
the face of the cluster as evidenced by a spread in
the colours and magnitudes of its evolutionary sequence.
We corrected the cluster mean $E(B-V)$ for differential reddening
with a method that will be described by Milone \etal (2011).
This technique is similar to the ones adopted by Sarajedini \etal
(2007) and Milone \etal (2009).
Briefly, we first defined the fiducial main sequence for the cluster.
Then for each star we estimated how much other observed stars in its
cluster spatial vicinity systematically lie to the red or the blue of
the fiducial sequence.
This systematic colour offset is indicative of the local differential
reddening.
We applied the M22 mean $E(B-V)$ plus the differential correction
to the observed $(B-V)$ values of red giant(s) that occur near the
same spatial location in the cluster.
Finally, we adopted the Cardelli \etal\ (1989) recommended ratio
of total to selective extinction, $R_V$~=~3.1, to convert
$E(B-V)$ to $A_V$ values.
We also used the Cardelli \etal\ extinction curve to perform
extinction corrections to the other colours.

%%%%%%%%%%%%%%%%%%%%%%%%%%%%%%%%%%%%%%%%%%%%%%%%%%%%%%%%%%%%%%%%%%%%%%%%%%%%%%%%%%%%%%%

\begin{table*}
\begin{center}
\caption{Sources of the Spectroscopic Data.}\label{tab-datasources}
\begin{tabular}{ccccccc}
\hline \hline
Instrument\tablefootmark{a}&$R$\tablefootmark{b}&$\lambda$ range&$S/N$      &$S/N$       &\# stars   &Reference \\
                          &                    &\AA            &$@$5000\AA &$@$7000\AA  &           &  \\
\hline
AAT-UCLES  & 38,000 & 5050$-$7300 &  50 & 100 &10& Diego et al.\ (1990) \\
           &        &             &     &     &  & Norris et al.\ (1995) \\
APO-ARCES  & 37,500 & 4000$-$9000 &  50 &  75 & 5& Wang et al.\ (2003) \\
MCD-2DC    & 60,000 & 4200$-$9000 &  70 & 120 & 6& Tull et al.\ (1995) \\
LICK-HAM   & 50,000 & 4800$-$8800 &  50 &  90 & 9& Vogt et al.\ (1987) \\
           &        &             &     &     &  & Valenti et al.\ (1995) \\
VLT-UVES   & 45,000 & 3300$-$4500 & 100 & 120 &16& Dekker et al.\ (2000) \\
           &        & 4800$-$6800 &     &     &  & Marino et al.\ (2009) \\
\hline
\end{tabular}
\end{center}
\tablefoottext{a}{Short-hand notation adopted in this paper.}\\
\tablefoottext{b}{$R \equiv \lambda/\Delta\lambda$}

\end{table*}

%%%%%%%%%%%%%%%%%%%%%%%%%%%%%%%%%%%%%%%%%%%%%%%%%%%%%%%%%%%%%%%%%%%%%%%%%%%%%%%%%%%%%%%%

%%%%%%%%%%%%%%%%%%%%%%%%%%%%%%%%%%%%%%%%%%%%%%%%%%%%%%%%%%%%%%%%%%%%%%%%%%
\section{MODEL ATMOSPHERE PARAMETERS\label{modelatm}}
%%%%%%%%%%%%%%%%%%%%%%%%%%%%%%%%%%%%%%%%%%%%%%%%%%%%%%%%%%%%%%%%%%%%%%%%%%

Our spectroscopic analysis of M22 giant stars was 
conducted with the local thermodynamic equilibrium (LTE) line analysis 
code MOOG (Sneden 1973).
In order to keep the present analysis consistent with that of M09,
we used interpolated model atmospheres from the grid of Kurucz (1992).
Those models include convective overshooting, which is not part of
subsequent grids in this series (e.g., Castelli \& Kurucz 2004).
In \S\ref{abunderror} we will comment on the (small) effect of
using these overshoot models in our analysis.

Atmospheric parameters for these models were estimated from Fe spectral
lines.
Effective temperatures \teff\ were derived by removing trends in
\ion{Fe}{I} abundances with excitation potential, and the microturbulent
velocities \vmicro\ were set by removing trends with EW.
Gravities were determined by satisfying the ionization equilibrium
between \ion{Fe}{I} and \ion{Fe}{II} abundances.
This process was done iteratively until a final interpolated model was
obtained.
In this manner we specified model atmosphere parameters entirely based on
the spectra; thus they are independent of photometric information.
This is an important advantage when analyzing M22, with its star-to-star
reddening variations.
More details about the line list, atmospheric parameters determination,
and abundance measurements can be found in Marino \etal (2008) and M09.

The adopted atmospheric parameters for the program stars, together with
the telescope source of the spectra, are listed in Table~\ref{tab-model}.
By comparing the atmospheric parameters for the same stars observed with
different sources, no evidence for systematic offsets have been found.

To investigate internal uncertainties related to the atmospheric parameters
we applied the same procedure used in Marino \etal\ (2008) and M09 to 
which we refer the reader for a more detailed description.
The procedure was applied separately for spectra taken with
different instruments because of the different quality of the data.
Briefly, first we calculated, for the stars, the errors
associated with the slopes of the best least squares fit in the
relations between \ion{Fe}{I} abundance vs. E.P. The average of the
errors corresponds to the typical error on the slope. Then, we
selected, for each set of data observed at different telescopes, a
star at intermediate temperature. For these stars, we fixed the
other parameters and varied the temperature until the slope of the
line that best fits the relation between abundances and E.P. became
equal to the respective mean error. 
These differences in temperature represent an estimate of the 
error in temperature itself. 
A similar procedure was applied for \vmicro, but using the relation
between abundance and EWs. 
For gravities, determined by imposing the ionization equilibrium 
for iron, we considered the averaged uncertainties 
$\overline {\sigma_{\rm {star}}[\rm {FeII/H}]}$ and $\overline
{\sigma_{\rm {star}}[\rm {FeI/H}]}$ 
(where $\sigma_{\rm {star}}[\rm {Fe/H}]$ 
is the dispersion of \ion{Fe}{I} and \ion{Fe}{II} abundances 
derived by the various spectral lines in each spectrum as given by 
MOOG, divided by $\sqrt { {N_{\rm lines}-1}}$), and
varied the gravity of the representative stars such that the relation:

\begin{equation}
{\rm [FeI/H]} - \overline{\sigma_{\rm star}{\rm [FeI/H]}} = {\rm [FeII/H]} + \overline{\sigma_{\rm star}\rm [FeII/H]}
\end{equation}

\noindent was satisfied. 

The resulting atmospheric parameter uncertainties for different data 
sets are listed in Table~\ref{tab-errors} together with the abundance 
uncertainties (see Sect.~4.2).

In order to understand the uncertainties associated with our
spectroscopically-derived temperatures and gravities, we first tested 
the assumption that RGB stars with the same photometric
properties ought to have the same \teff\ and \logg\ values.
In Fig.~\ref{referee} we plot our adopted \teff\ as a function of 
the $(B-V)$ colour, and \logg\ as a function of $V$.
The mean trends were computed with linear least square regressions.
The rms of the differences between individual points and the
mean curve is quoted in each panels of Fig.~\ref{referee}, 
and are not dissimilar from our estimated \teff\ and \logg\ uncertainties.
These values are an estimate of the differences that we expect in 
temperatures and gravities among stars with similar magnitude and colours.  

In Fig.~\ref{t} we compare our spectroscopically-derived
model atmosphere parameters with these quantities derived by other methods.
In the upper-left panel we show the comparison of our effective temperatures,
determined from \ion{Fe}{I} excitation equilibrium (\teff),
with the ones obtained from photometry.
Photometric temperatures $\rm {T_{\it (V-K)}}$ have been derived from
the colours $(V-K)$ corrected for differential reddening (as explained
in \S\ref{data}), by using the calibrations of Alonso \etal 
(1999, 2001), assuming a mean reddening of $E(B-V)$=0.34 (Harris 1996).
The scatter of the points around the line of perfect agreement is
$\sim$50~K, that is what we expect from observational errors.
Since we used colours corrected for differential reddening, 
we have minimized this effect in the derivation of $\rm {T_{\it (V-K)}}$.

The upper-right panel shows our spectroscopic gravities
determined from \ion{Fe}{I}/\ion{Fe}{II} ionization equilibrium
(\logg$_{\rm adopted}$) as a function of gravities \logg$_{\rm photometry}$ that
were obtained with standard relations by using $\rm {T_{\it (V-K)}}$,
bolometric corrections from Alonso \etal\ (1999), and a distance
modulus of $(m-M)_{V}=13.60$ (Harris 1996).
The dispersion around the line of perfect agreement is $\sim$0.20~dex,
and is in both axis, i.e.\ partly in $(V-K)$ due to uncertain reddening
(which is much more in $(V-K)$ than $(B-V)$) and the uncertainty
introduced by the NLTE effects in \ion{Fe}{I}/\ion{Fe}{II} as discussed
by Kraft \& Ivans (2003).
Indeed, while there are good reasons to base [Fe/H] on the lines
of \ion{Fe}{II} (Kraft \& Ivans 2003) our resolution limits the number
of unblended \ion{Fe}{II} lines so that it is necessary to base [Fe/H]
on lines of \ion{Fe}{I}. 
However, the scatter of points in the gravity plane in the upper 
panel of Fig.~\ref{t} from the line of perfect agreement exhibits 
no obvious \teff\-dependence. 
This is emphasized in the gravity difference plot in lower-right panel 
of the figure.
Finally, in the lower-left panel we show our adopted spectroscopic 
\logg\ values compared with estimates based on spectroscopic gravity 
calibrations of GCs provided by Ku{\v c}inskas  \etal (2006).

These tests demonstrate that our estimates of the atmospheric 
parameters are quite reliable and that NLTE effects may not be 
important in determining these quantities.
Of course, some offsets between the photometric and spectroscopic
parameters may be present, as it seems for temperature with 
spectroscopic values $\sim$20-30 K higher than T$_{V-K}$, 
but these offsets are comparable with our uncertainties.

\begin{figure*}
\centering
\includegraphics[width=12.3cm]{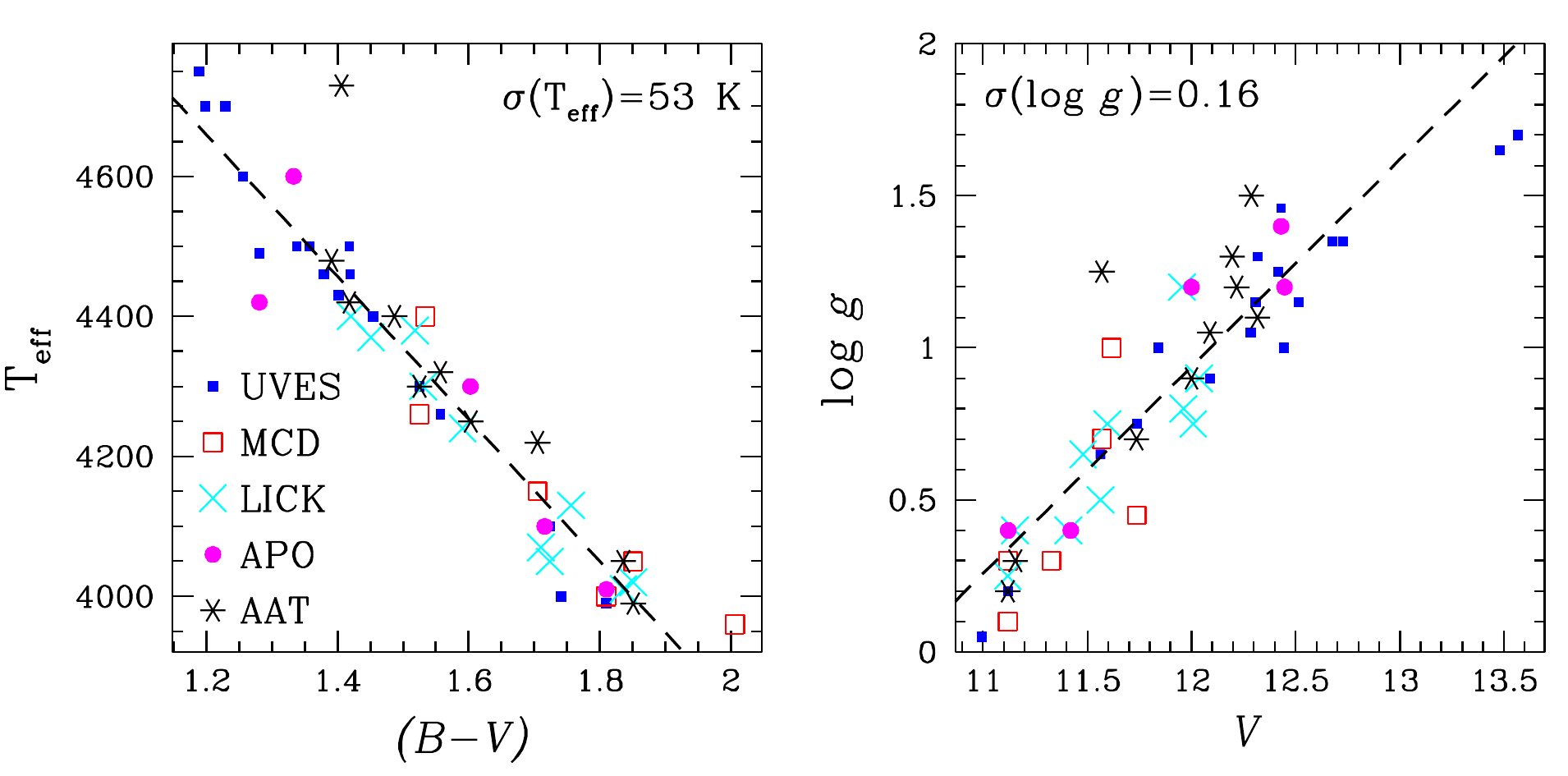}
\caption
{
Adopted \teff\ and \logg\ as a function of $(B-V)$ colour and $V$ mag
respectively.
Different data-sets were represented by different symbols (as quoted
in the left panel). 
}
\label{referee}
\end{figure*}

\begin{figure}
\centering
\includegraphics[width=9.1cm]{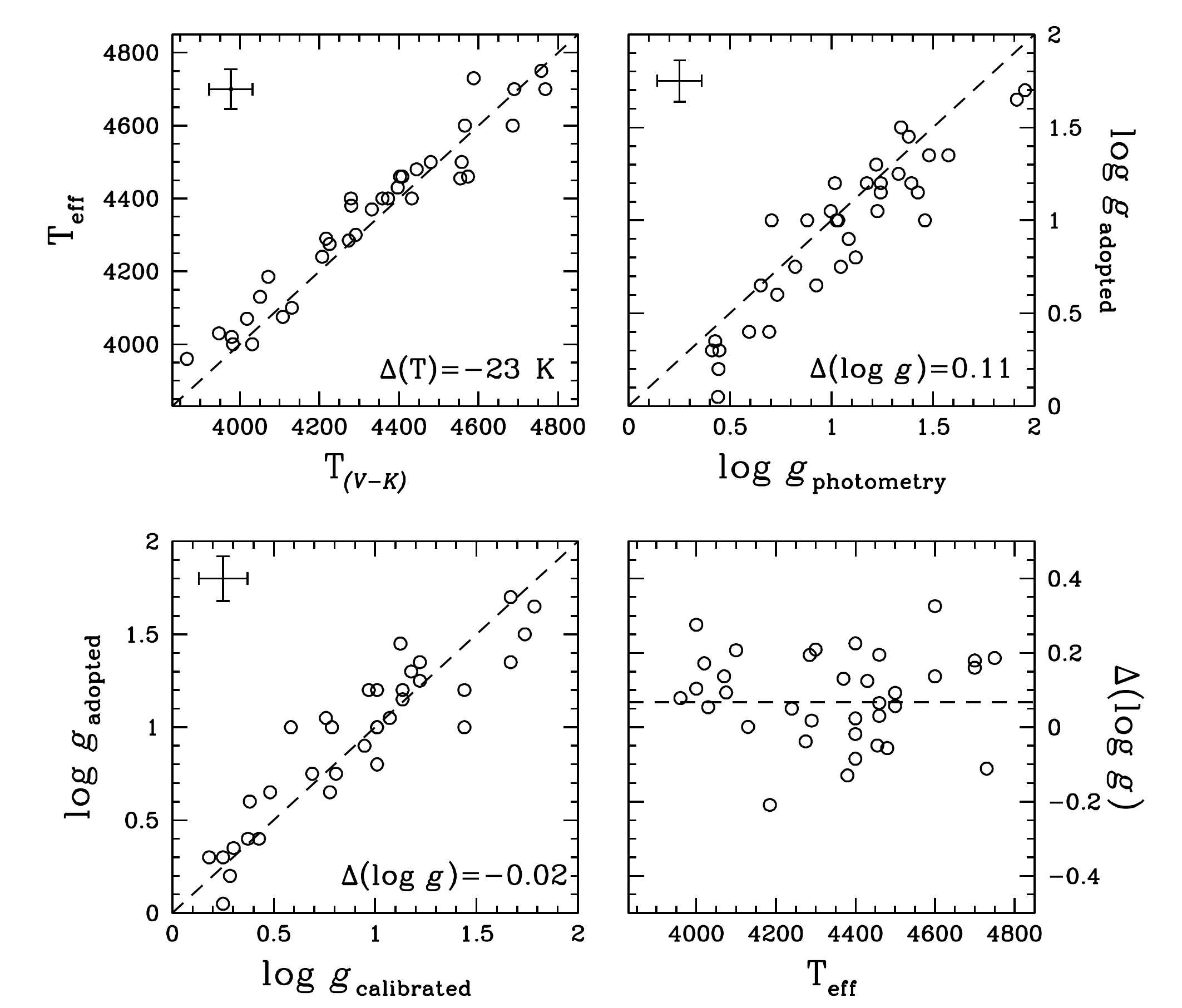}
\caption{
{\it Upper-left panel}: Adopted temperatures derived from the excitation 
potential equilibrium (\teff$_{spec}$) as a function of photometric
 temperatures obtained from colours $(V-K)$.
{\it Upper-right panel}: Adopted gravities from ionization equilibrium
 (\logg$\rm {_{adopted}}$) as a function of gravities obtained from the
 estimated mass (\logg$\rm {_{photometry}}$), radius, and adopting a distance
 modulus of $(m-M)_{V}$=13.60 (Harris 1996). 
{\it Lower-left panel}: Adopted gravities versus the calibrated \teff-logg\
 relation from Ku{\v c}inskas \etal\ (2006).
The dashed line in these three panels indicates perfect agreement.
The mean difference between the adopted values and the comparison ones
has been quoted in each panel (as $\Delta$(comparison-adopted)). 
{\it Lower-right panel}: Distances $\Delta$(\logg) of each
point in the upper panel, in the $\rm {\logg_{photometry}}$-$\rm
{\logg _{adopted}}$ plane, from the line of perfect agreement, as a
function of the adopted \teff.}
\label{t}
\end{figure}

%%%%%%%%%%%%%%%%%%%%%%%%%%%%%%%%%%%%%%%%%%%%%%%%%%%%%%%%%%%%%%%%%%%%%%%%%%
\section{ABUNDANCE DERIVATIONS\label{abunds}}
%%%%%%%%%%%%%%%%%%%%%%%%%%%%%%%%%%%%%%%%%%%%%%%%%%%%%%%%%%%%%%%%%%%%%%%%%%

Using the model atmospheres and analysis code described in
\S\ref{modelatm}, we determined abundances for Fe, $\alpha$ elements
(Mg, Si, Ca, Ti), $p$-capture elements (C, N, O, Na, and Al), Fe-peak
elements Cu and Zn, and several \ncap\ elements (Y, Zr, Ba, La, Nd, Eu).
To accomplish this we employed EWs for Fe, the $\alpha$ elements, Y, Ba, and Nd.
For C, N, O, Na, Zr, La, Eu, Zn and Cu, because of significant atomic
blends, isotopic/hyperfine structure issues, or weak molecular band
(CN) contamination, we derived abundances by comparing synthetic and
observed spectra.
We list the results for light and heavy elements in Table~\ref{tab-ablight}
and Table~\ref{tab-abheavy} respectively.
In this section we comment on the transitions that we used, and
on uncertainties in the resulting abundances.

%%%%%%%%%%%%%%%%%%%%%%%%%%%%%%%%%%%%%%%%%%%%%%%%%%%%%%%%%%%%%%%%%%%%%%%%%%
\subsection{Spectral Features\label{transitions}}
%%%%%%%%%%%%%%%%%%%%%%%%%%%%%%%%%%%%%%%%%%%%%%%%%%%%%%%%%%%%%%%%%%%%%%%%%%

\textit{Proton-capture elements:} We determined Na abundances from 
spectral synthesis of the \ion{Na}{I} doublets at 5680~\AA\ and 
6150~\AA, and O abundances from the synthesis of the forbidden 
[\ion{O}{I}] line at 6300~\AA.
Aluminum was determined from EWs of the doublet at 6667~\AA.\
For the UVES data, the O, Na, and Al abundances are those reported in M09.
We applied NLTE corrections from Lind \etal\ (2011b) to the Na
abundances. 
These corrections are not available for gravities $\lesssim$1.00. 
However our \ion{Na}{I} line strengths are relatively insensitive to 
gravity.
We determined NLTE corrections for the lowest gravity stars using the
Lind corrections for \logg\~=~1.0, but future NLTE corrections for
Na abundances will be welcome.
In the following discussion both NLTE and LTE Na abundances will be presented. 

For 14 (out of 35) stars we were able to determine C and N abundances.
Carbon was measured from spectral synthesis of the CH G-band
($A^2\Delta−X^2\Pi$) heads near 4314 and 4323~\AA.
Nitrogen was derived from synthesis of the 2-0 band of the CN red system
($A^2\Pi-X^2\Sigma$) near 8005~\AA\ (available for the MCD spectra),
and from the CN blue system ($B^2\Sigma-X^2\Sigma$) bandhead at
$\sim$4215~\AA\ (for UVES data).
The synthesis linelist for the blue CN band is described in
Hill \etal\ (2002).
The linelists for the CH band and the CN red system were provided
by B. Plez (CH band, private communication), and V. Smith (CN band,
private communication).
As an example of the molecular band calculations, in
Fig.~\ref{CN8000} we show synthetic/observed spectral matches of the CN
red system for two stars IV-102 and III-3 that have nearly identical
atmospheric parameters \teff, \logg, and \vmicro.
Superimposed on the observed spectra are synthetic models at constant
C, O, while varying N around the best fit value of $\Delta$N=$\pm$0.20 dex.
The star IV-102, similarly to some other stars in our sample, has
very weak features of CN bands; hence we could estimate only an upper
limit to its N abundance.

Of course, in the C abundance computations we used the
previously-determined O contents of each star,
and for N, both observed C and O abundances needed to be employed.
Unfortunately we could measure C and N only for MCD and UVES data.
Spectra obtained with several of the other instruments do not have
sufficient S/N in the spectral region around the CH bands to determine
meaningful values or limits for C.
For UVES only a sub-sample of six stars have available data in the
spectral range covering molecular bands of CN and CH.

Isotopic ratios \iso{12}{C}/\iso{13}{C} were derived for
the six MCD stars, since relatively strong and isolated features of
\iso{12}{CN} and \iso{13}{CN} are available for the 2-0
band of the CN red system.
First, several iterations of synthesis were done to obtain a satisfactory
match to the strengths of the \iso{12}{CN} features, after
which syntheses were calculated with different values of the
\iso{12}{C}/\iso{13}{C} ratio.
The isotopic ratio was derived from several \iso{13}{CN} features,
with the highest weight given to the blended triplet of lines at
8004.7~\AA.
Other features in the same "window" provide a check on the ratio.

\textit{$\alpha$ elements:}
We determined abundances from EWs of the same Si, Mg, Ca, and Ti
(I and II) lines used in M09.
We measured Ca and Ti from usually about 10 transitions, while for
Mg and Si we had few lines, typically about four for Si, and one or
two for Mg.

\textit{Heavy Fe-peak elements: } We determined abundances for Cu
from synthesis of the \ion{Cu}{I} lines at 5105, 5218, and 5782~\AA.
Both hyperfine and isotopic splitting were included in the analysis,
with well-studied spectral line component structure from the Kurucz
(2009)\footnote{Available at: \sf {http://kurucz.harvard.edu/}} compendium.
Solar-system isotopic fractions were assumed in the computations:
f(\iso{63}{Cu})~=~0.69 and f(\iso{65}{Cu})~=~0.31.
For Zn we analyzed the \ion{Zn}{I} lines at 4722 and 4810~\AA.
These lines have no significant hyperfine or isotopic substructures,
and were treated as single absorbers in our syntheses.
However, the S/N of our spectra that extend down to 4722~\AA\ is poor,
thus yielding larger abundance uncertainties.

\textit{Neutron-capture elements:}
For UVES data, Y and Ba are from M09, to which we add new measurements
for Zr, La, Nd, and Eu.
Zirconium abundances for UVES data were calculated by M09 from EWs.
Here we determine Zr from spectral synthesis of just the 5112~\AA\
\ion{Zr}{II} line.
Hence, to homogeneously analyze our data, the UVES Zr abundances were
re-determined with syntheses.

We determined Y, Ba, and Nd contents from EWs of isolated spectral lines.
For Ba abundances we employed the 5853, 6141, and 6496~\AA\
\ion{Ba}{II} lines.
Since these lines have (very narrow) hyperfine and isotopic substructures,
and suffer blending by other atomic species to greater or lesser degrees,
we used a blended-line EW analysis option in our synthesis code.

Lanthanum abundances were derived from spectral synthesis of the
\ion{La}{II} lines at 6262, 6390, and 6774~\AA.
Hyperfine splitting for the 6262 and 6390~\AA\ lines was taken into
account with the laboratory data from Lawler \etal\ (2001a).
Hyperfine data are not available for the 6774~\AA\ line, but
it is weak enough that no substantial abundance error results from
treating the line as a single absorber.
As examples of La syntheses, in Fig.~\ref{LaSynth} we show the 6262
and 6390~\AA\ lines in the LICK spectrum of star III-14 (left panels)
and in the UVES spectrum of star III-52 (right panels).
These stars were chosen for display because they have nearly the same
\teff\ and \logg\ values, but have contrasting La line strengths.

For Eu we computed spectral syntheses of the \ion{Eu}{II} line at 6645~\AA,
considering the hyperfine and isotopic splitting structure given
in Lawler \etal\ (2001b).
We did not obtain Eu from AAT data since those spectra do not cover
the 6645~\AA\ spectral region.
Due to the poor S/N ($<$30) and line crowding we could not obtain
reliable abundances from stronger \ion{Eu}{II} lines in the blue-violet region.

\begin{figure}
\centering
\includegraphics[width=9.3cm]{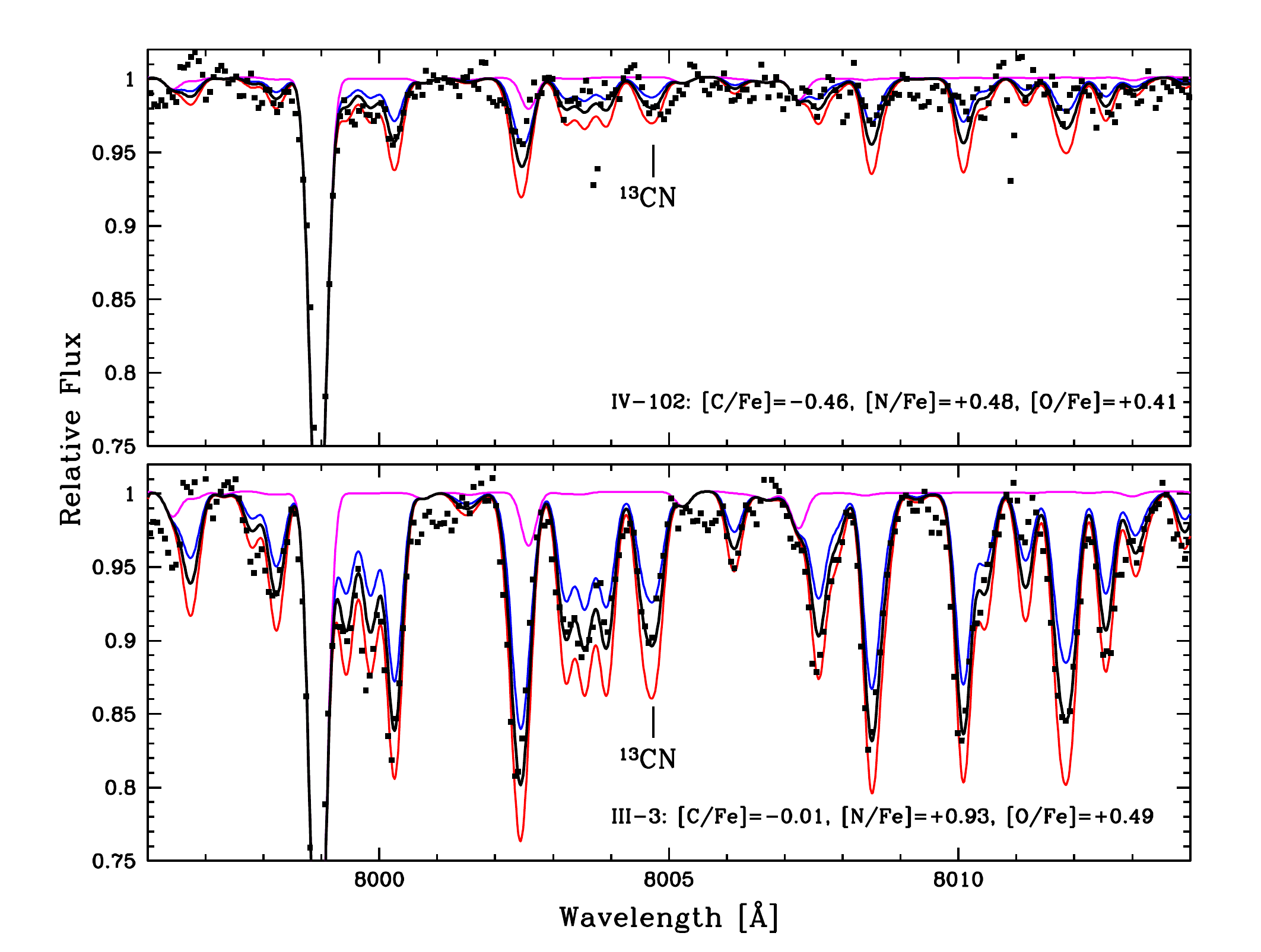}
\caption{Spectral region around the CN features at 8000~\AA\ from McD
data for the two stars IV-102 (upper panel) and III-3 (lower panel).
The observed spectrum is shown as points.
Synthetic spectra with N value of best fit (black line), with no N
(magenta line), and with $\Delta$(N)=$\pm$0.2 dex deviations from the
best fit (blue and red lines) are superimposed to the spectrum of the stars.}
\label{CN8000}
\end{figure}

\begin{figure}
\centering
\includegraphics[width=9.3cm]{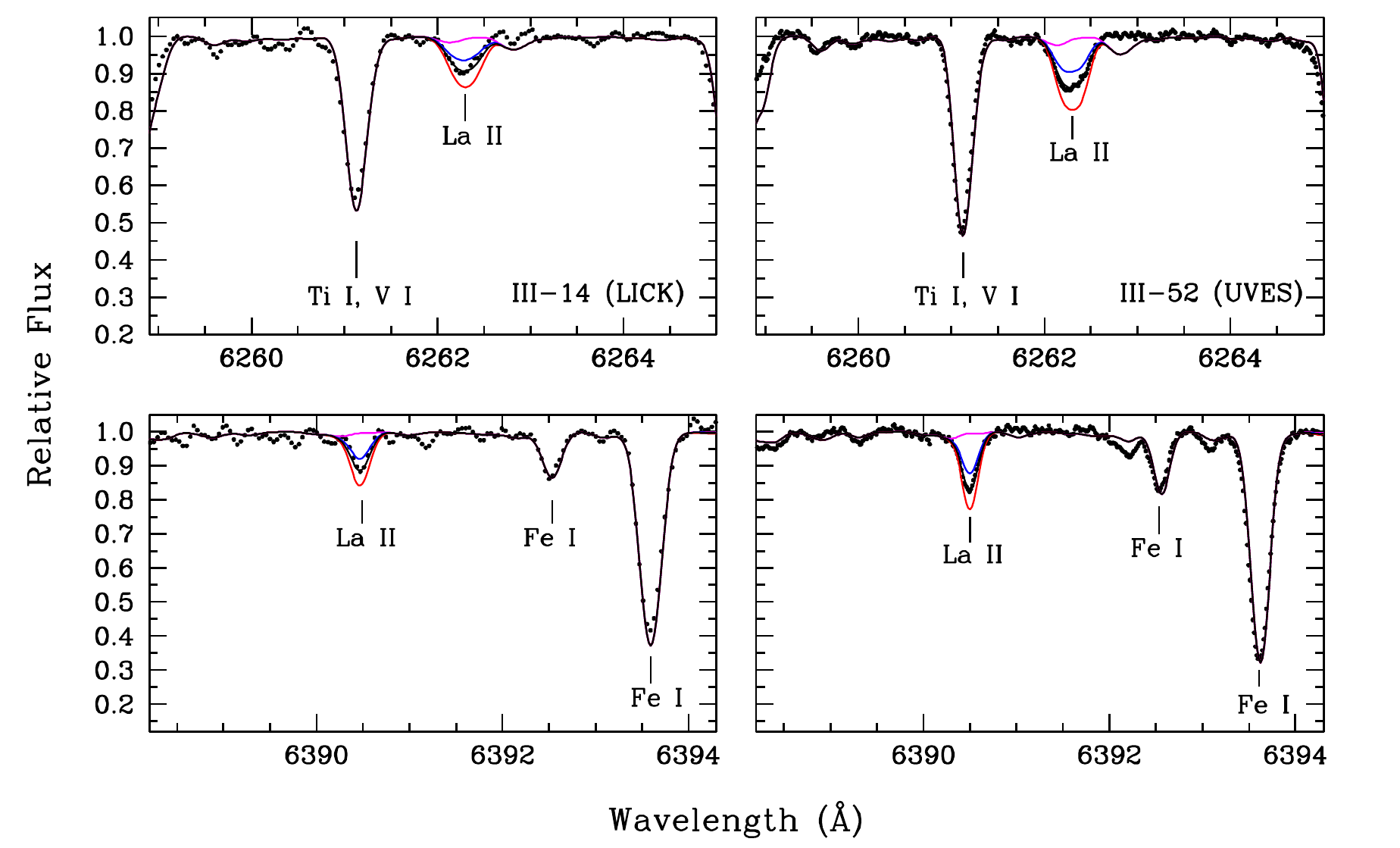}
\caption{Observed and synthetic spectra around the La lines at 6262~\AA\
  and 6390~\AA\ for the \spo\ star III-14 observed at LICK (left panels),
  and the \sri\ star III-52 observed with UVES (right panels).
  In each panel the points represent the observed spectrum.
  The magenta line is the spectrum computed with no contribution
  from \ion{La}{II}; the black line is the best-fitting synthesis (with
  the La abundance given in Table~\ref{tab-abheavy}); and the red and
  blue lines are the syntheses computed with La abundances altered by
  $\pm$0.2 dex from the best value.}
\label{LaSynth}
\end{figure}

%%%%%%%%%%%%%%%%%%%%%%%%%%%%%%%%%%%%%%%%%%%%%%%%%%%%%%%%%%%%%%%%%%%%%%%%%%
\subsection{Abundance Uncertainties\label{abunderror}}
%%%%%%%%%%%%%%%%%%%%%%%%%%%%%%%%%%%%%%%%%%%%%%%%%%%%%%%%%%%%%%%%%%%%%%%%%%

Stars with repeated observations, for which we reported model atmospheres
and abundances derived from different sources, suggest good
agreement (within observational errors) of their results.
For these stars, we will employ the averaged abundance results
in subsequent discussions.

To verify how model atmosphere uncertainties influence the derived
chemical compositions, we repeated the abundance derivations for
one representative star at intermediate temperature for each set of data.
For this exercise we changed only one atmospheric parameter each time.
Results of these calculations are listed in Table~\ref{tab-errors}.
Assuming that the atmospheric parameter uncertainties are uncorrelated,
we estimate total  sensitivities for absolute abundances to be
typically $\sim$0.15-0.20.
The abundance ratios [X/Fe] have sensitivities of $\sim$0.05-0.10 for
LICK and MCD spectra, and slightly higher for APO and AAT data due to
their lower resolution (see Table~\ref{tab-errors}).

An additional source of abundance internal errors is
the uncertainty in the EW measurements.
In the case of Fe this contribution is small since a large
number of transitions (typically N$_{\rm \rm{\ion {Fe}{1}}}$~$\simeq$ 40) 
are available.
The uncertainty can be estimated as 
$\sigma_{EW}$/$\sqrt{N_{\rm {\ion{Fe}{I}}}-1}$, which on average we 
estimated as $\sim$0.02~dex for all data sources.
However, for those species with only few (or even one) transitions, such
as \ion{Mg}{I} and \ion{Si}{I}, the error introduced by EWs measurements
became important, $\sim$0.10~dex.
Particular caution should also be noted for Ba abundances that have
been measured from only strong and blended lines.

In Table~\ref{tab-errors} we do not list the sensitivities of either
C or N on the atmospheric parameters.
The dominant source of uncertainty for the abundances derived from
molecular bands is the continuum placement.
For C the values we obtained for the two CH bandheads generally agree with
each other within $\sim$0.15 dex, and we adopted the average of the
two measurements as our final C abundance.
In the case of N, the continuum placement errors are not critical
in the the CN red system available for MCD data.
It could became important for the CN band at $\sim$4215~\AA\ that we
used for the UVES spectra, however the relatively high S/N ($\sim$50-60)
around the CN band of these data served to reduce such problems.
Random uncertainties for [C/Fe] and [N/Fe] are estimated to be 0.10
and 0.20 dex, respectively.

Recall from \S\ref{modelatm} that we have used model atmospheres 
from the Kurucz (1992) grid, which include convective overshooting.
To test the sensitivity of our abundances to this effect, we repeated 
the analysis using models without overshooting (Castelli \& Kurucz 2004)
for few stars representative of the entire range in temperature.
The main systematic results of this exercise are: (a) a decrease 
of $\lesssim$0.10 in \logg\ values; (b) a decrease in [Fe/H] of 
$\sim$0.05; (c) an increase in [N/Fe] of $\sim$0.10 dex; and (d)
negligible effect on other [X/Fe] abundance ratios.
Overshooting in the model atmospheres does not materially affect our results.

As a final estimate of errors associated with each abundance 
measurement, we assume the rms of the abundances of stars with the
same chemical properties.
This error includes both the errors introduced by atmospheric
uncertainties, and errors due to EW measurements. 
Further details on how the stars with the same chemical properties
were selected are given in Sect.~\ref{results}.

\begin{figure}
\centering
\includegraphics[width=8.8cm]{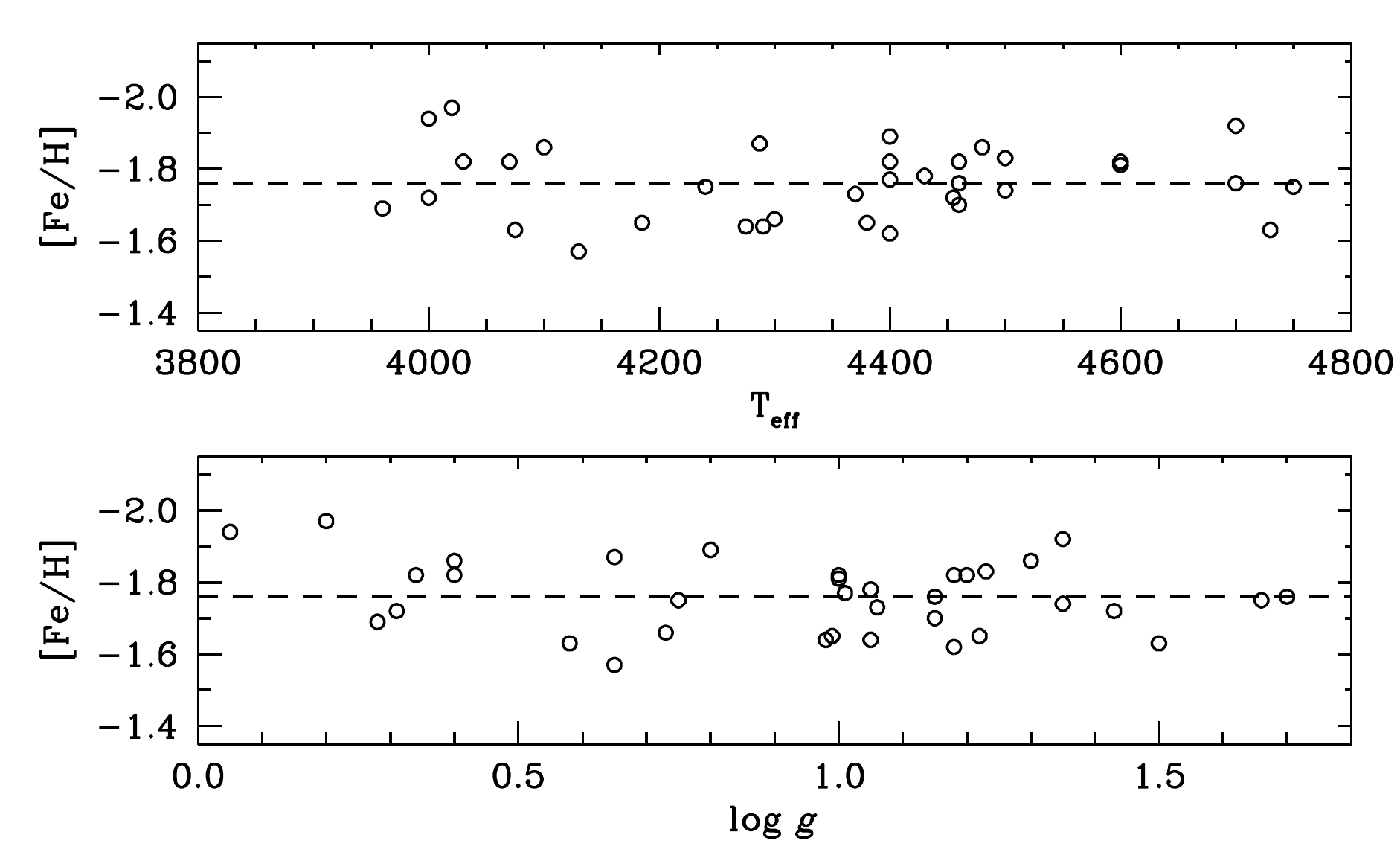}
\caption{Correlations of derived [Fe/H] metallicities as functions
  of effective temperature (top panel) and gravity (bottom panel).
  The dashed line in each panel indicates the mean from the complete
  35-star sample, $<$[Fe/H]$>$~= $-$1.76 .}
\label{fetefflogg}
\end{figure}

%%%%%%%%%%%%%%%%%%%%%%%%%%%%%%%%%%%%%%%%%%%%%%%%%%%%%%%%%%%%%%%%%%%%%%%%%%
\section{ABUNDANCE RESULTS\label{results}}
%%%%%%%%%%%%%%%%%%%%%%%%%%%%%%%%%%%%%%%%%%%%%%%%%%%%%%%%%%%%%%%%%%%%%%%%%%

In this section, we consider our abundance results of \ncap, light,
and $\alpha$ elements in M22, expanding on the discussion of M09.

For our entire 35-star sample, we obtain a mean metallicity of
[Fe/H]~= $-$1.76~$\pm$~0.02 dex ($\sigma$~=~0.10).
However, this simple mean obscures the fact that the total
metallicity spread is more than a factor of two:
$-$1.57~$\geq$ [Fe/H]~$\geq$ $-$1.97, a range that cannot be explained
by observational/analytical uncertainties.
This point was demonstrated previously in M09 and in DC09.  
There are four stars in common between this work and DC09,
who used intermediate resolution spectra at the \ion{Ca}{II} triplet to 
derive [Fe/H] values for 41 M22 red giants.  
For these stars the mean difference in [Fe/H], in the sense of this 
paper minus DC09, is small: $-$0.01$\pm$0.04 dex (sigma 0.09 dex).
In Fig.~\ref{fetefflogg} we plot individual [Fe/H] values as
functions of \teff\ and \logg.
It is clear that there are no metallicity trends with either parameter;
the scatter is the same at all M22 giant branch positions.
Below we consider the metallicity spread in concert with other M22 
abundance anomalies.

A summary of our results for the 18 non-Fe species is displayed in
Fig.~\ref{provafig1}, where we show relative abundances [X/Fe]
as a function of [Fe/H].
The numerical ranges of both quantities are the same in all panels
of this figure so that one can compare the variations of different
elements with changing metallicity in M22.
The figure organization differentiates between the ``lighter''
elements (Z~$<$~26, shown in the two left-hand columns of panels),
and the ``heavier'' elements (shown in the two right-hand columns).
We have represented stars belonging to two different 
metallicity groups in M22 with different symbols:  blue crosses for
more metal-poor stars, and red filled circles for less 
metal-poor stars.
We will justify and expand this distinction in \S\ref{ncapel}.
The error bar in each panel (and in the next figures of this paper) 
is an estimate of the uncertainty associated with individual abundance
measures, calculated as the rms of the abundances of stars in the 
same metallicity group (see Sect.~\ref {abunderror}).
For this estimate we used values from the more metal-poor group only.
Of course, this is an overestimate of the error if intrinsic 
abundance variations are present in each group. 
The light proton-capture elements (C, N, O, Na, and Al) 
exhibit intrinsic abundance variations within each
metallicity group (see \S\ref{proton}).
For these elements, we calculated the rms for stars in metal-poor group
that have [Na/Fe]$<$+0.2 dex.

Among the elements investigated in this paper, we found a 
small abundance trend with temparature for Cu, Zn, and Y relative to Fe.
However, they affect in a similar way both M22 metallicity groups,
and may possibly explain the some of the internal scatter of these 
abundances in the two groups, but they do not influence our basic results.
In the next few subsections we consider the abundance trends
among and between elements of different nucleosynthetic groups.

\begin{figure*}
\centering
\includegraphics[width=14.6cm]{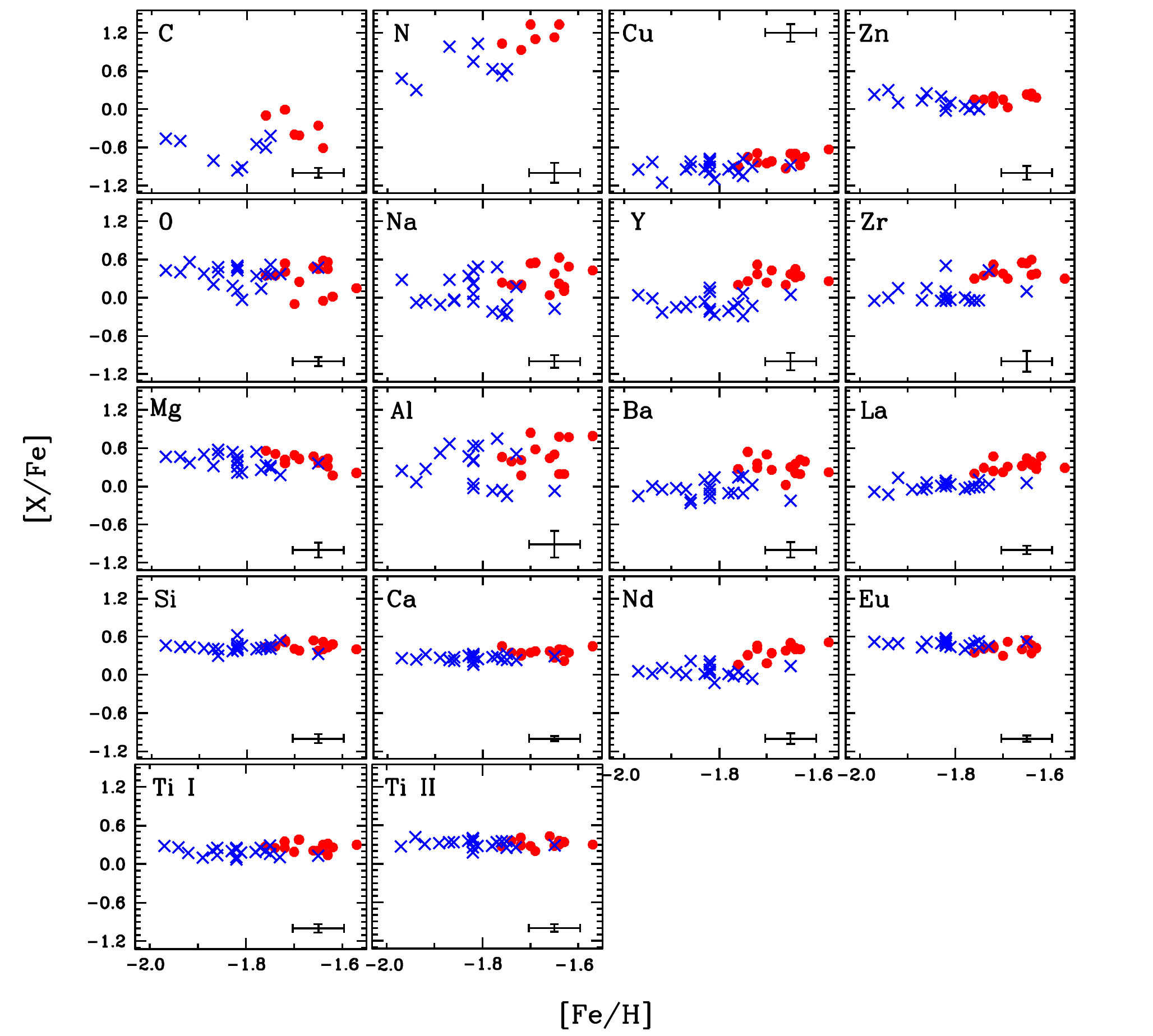}
\caption{Summary of the abundance results.
  For all 18 non-Fe species, their [X/Fe] relative abundances are plotted
  versus their [Fe/H] metallicities.
  The horizontal and vertical ranges are identical in all panels.
  Filled circles are used for stars with \spro\ enhancements
  and $\times$ symbols are for stars without such enhancements;
  see \S\ref{ncapel} for definitions of these two stellar groups.
  Error bars in each panel represent estimated errors for single
    measurements.}
\label{provafig1}
\end{figure*}

%%%%%%%%%%%%%%%%%%%%%%%%%%%%%%%%%%%%%%%%%%%%%%%%%%%%%%%%%%%%%%%%%%%%%%%%%%
\subsection{The Neutron-Capture Elements\label{ncapel}}
%%%%%%%%%%%%%%%%%%%%%%%%%%%%%%%%%%%%%%%%%%%%%%%%%%%%%%%%%%%%%%%%%%%%%%%%%%

As described in the previous section, we follow M09 in using 
different symbols in Fig.~\ref{provafig1} and subsequent figures 
to segregate stars into two metallicity groups.
But Fig.~\ref{provafig1} clearly suggests that relative \ncap\ abundances
also vary with metallicity.
Abundances of just Y, Zr, and Ba were reported in our earlier work.
The solar-system abundances of these three elements are due 
overwhelmingly to the \spro: Y 72\%; Zr 81\%, and Ba 85\% 
(e.g., Table~10 of Simmerer \etal\ 2004).
Therefore M09 called stars ``\sri'' if [Y/Fe]~$>$~0 and
``\spo'' if [Y/Fe]~$<$~0.
This suggested link between \ncap-rich stars in M22 with the \spro\
is sensible but not definitive, because Y, Zr, Ba also can be
synthesized in the \rpro\ (e.g, see the review by Sneden \etal\
2008, which has references to individual \rpro-rich stars).

Here we can make a cleaner test by comparing the abundances of
elements with sharply contrasting solar-system \spro/\rpro\ origins:
La (75\% \spro) and Eu (only 3\% \spro).
In Fig.~\ref{laabund2} we show abundance ratios [Eu/Fe], [La/Fe],
and [La/Eu] as functions of [Fe/H].
Panels (a) and (b) are enlargements of their respective panels shown
in Fig.~\ref{provafig1}.
From these data one sees that relative abundances of Eu have no dependence 
on metallicity, while those of La exhibit a positive correlation.
Therefore, in agreement with Da Costa \& Marino (2010),  there is no 
doubt that the variations in [La/Fe] are due to variations in amounts 
of \spro\ material.

We have employed spectrum syntheses to derive the La and Eu abundances,
because the spectral features of both \ion{La}{II} and \ion{Eu}{II} have
significant hyperfine substructure, and the \ion{Eu}{II} lines also have
isotopic splitting.
However, the differences between the \sri\ and \spo\ groups as defined
by M09 can be easily seen in the spectra without any detailed analyses.
In Fig.~\ref{laeuspec} we show the La and Eu transitions in
stars with similar atmospheric parameters but very different
derived [Fe/H] and [La/Eu] ratios.
The \sri\ star III-3 (\teff/\logg/\vmicro/[Fe/H] =
4000/0.30/2.20/$-$1.72, Table~\ref{tab-model}) clearly has much stronger
La lines than does the \spo\ star IV-102 (4020/0.20/2.20/$-$1.97),
while its Eu lines are perhaps even weaker than those of IV-102.
Inspection of other contrasting pairs of stars yields the same conclusion.
The La/Eu ratios are very different in the \spo\ and \sri\ stars.

In panel (c) of Fig.~\ref{laabund2} we plot the [La/Eu] ratios of
our sample; here the separation between \sri\ and \spo\ stars is even
more clear than in the [La/Fe] ratios shown in panel (b).
For the entire M22 sample, $<$[La/Eu]$>$~=~$-$0.30
(Table~\ref{tab-mean}), and the gap between the smallest [La/Eu]
value of the \sri\ stars and the largest [La/Eu] value of the \spo\
stars is nearly 0.2~dex.
Therefore for the remainder of this paper we redefine \sri\ stars
as those with [La/Eu]~$>$~$-$0.3; filled red circles will be used
to identify them in the figures.
Similarly, the \spo\ stars hereafter are those with [La/Eu]~$<$~$-$0.30;
they will be plotted with blue crosses in the figures.
This empirically-set dividing line between the two groups is indicated
in panel (c) of Fig.~\ref{laabund2}.
All \sri\ stars also have [La/Fe]~$\geq$~$+$0.2, as we indicate
with the line in panel (b).

With this defined division by [La/Eu] ratio, 14 stars (40\% of the
total sample) are \sri\ and 21 stars (60\%) are \spo.
In Table~\ref{tab-mean} we give the mean abundances for the whole sample
and for the \sri\ and \spo\ subsets.
For two elements A and B, we can define the difference in their
abundance ratio between the \sri\  and \spo\ stars as
$\Delta^{rich}_{poor}$[A/B]~$\equiv$ [A/B]$_{\small
  {\sri}}$~$-$~[A/B]$_{\small {\spo}}$.
For La, $\Delta^{rich}_{poor}$[La/Fe]~= $+$0.32$\pm$0.02 (Table~\ref{tab-mean}).
Nearly identical results are obtained for the other four \spro-dominated
elements in our survey:
$\Delta^{rich}_{poor}$[Y/Fe]~= $+$0.41$\pm$0.04,
$\Delta^{rich}_{poor}$[Zr/Fe]~= $+$0.34$\pm$0.05,
$\Delta^{rich}_{poor}$[Ba/Fe]~= $+$0.36$\pm$0.05, and
$\Delta^{rich}_{poor}$[Nd/Fe]~= $+$0.32$\pm$0.04.
This consistency is illustrated in Fig.~\ref{otherncap}, whose four
panels show correlations of [La/Eu] versus [Y/Eu], [Zr/Eu], [Ba/Eu],
and [Nd/Eu].
Note the relatively large star-to-star scatter in Ba abundances compared
to other elements.
This is due to difficulties associated with deriving reliable abundances
for this species which, as described in \S\ref{transitions}, is
represented by three lines that have hyperfine and isotopic splitting
and are general saturated in M22 giant star spectra.

For each star we have formed average \spro-element abundance
ratios [\spro/Fe] and [\spro/Eu], where \spro\ here represents
the five elements Y, Zr, Ba, La, and Nd in most cases.
In some cases we were not able to derive abundances for one or
more of the \spro-elements; their means were formed from the available
abundances.
In Fig.~\ref{avgspro} we display these results.
For the \sri\ stars we derive
$<$[\spro/Fe]$>$~= $+$0.35~$\pm$~0.02 ($\sigma$~=~0.06) and
$<$[\spro/Eu]$>$~= $-$0.07~$\pm$~0.02 ($\sigma$~=~0.06),
while for the \spo\ stars we derive
$<$[\spro/Fe]$>$~= $-$0.01~$\pm$~0.01 ($\sigma$~=~0.06) and
$<$[\spro/Eu]$>$~= $-$0.49~$\pm$~0.01 ($\sigma$~=~0.05).
These means and $\sigma$ values are also shown in Fig.~\ref{avgspro}.
All the \ncap\ abundance data displayed in the bottom panel of
Fig.~\ref{avgspro} strongly support the notion of a bimodal separation
between the \sri\ and \spo\ groups, as suggested by M09, with
a typical [\spro/Eu] difference of $\sim$0.4 between the groups.

Taking the [La/Y] abundance ratio as a [heavy-$s$/light-$s$]
indicator, we find that the \sri\ stars show also slightly lower [La/Y]
abundance: $\Delta^{rich}_{poor}$[La/Y]~=~$-$0.12$\pm$0.05 
(Tab.~\ref{tab-mean}). 
This is suggestive of additional light-$s$ synthesis products 
from the progenitors to the \sri\ stars.

The mean [Fe/H] values for the \spo\ and the \sri\ groups
are $<$[Fe/H]$>$$_{\small {\rm \spo}}$~=  $-$1.82$\pm$0.02
and $<$[Fe/H]$>$$_{\small {\rm \sri}}$~=~$-$1.67$\pm$0.01 respectively (see
Table~\ref{tab-mean}). 
The two groups have a different [Fe/H] with a mean difference of 
0.15$\pm$0.02 dex, a 5$\sigma$ effect.
The standard deviations of the [Fe/H] values are 0.07 and 0.05 for
the \spo\ and \sri\, which might indicate that there is no [Fe/H] spread
within each group, if the relative errors are of the order of 0.05 dex.
Some small overlap in metallicity between the two \textit{s} (and Fe)-groups,
at roughly $-1.8$~$\lesssim$~[Fe/H]~$\lesssim$~$-1.7$ cannot be excluded, as apparent 
in the top panel of Fig.~\ref{avgspro}, but the difference in [Fe/H] 
between \sri\ and \spo\ groups is not much larger than the estimated 
observational error associated to [Fe/H] of $\sim$0.07.

As demonstrated in M09 (see their Fig.~12), due to the limits imposed
by our observational errors, the relatively small difference in [Fe/H] between
the two stellar groups in M22 is much more clearly recognizable by separating
stars on the basis of their \textit{s}-richness.
We note however that a similar difference in [Fe/H] as the one found in
M09 and confirmed here, is also confirmed by DC09 who
found that two groups of stars with mean metallicities of [Fe/H]$-$1.63 and
[Fe/H]=$-$1.89.

Our focus in the rest of the present paper is to study the abundance
behavior of different elements in stars belonging to the \spo\ or \sri\
groups, as defined by their [La/Eu] ratios.\footnote{
As is often the case with shorthand labels, the \sri\ and \spo\ designations
should not be taken too literally. In M22, the \spo\ population alternatively
could have been labeled ``$r$-rich'', because they have
[\spro/Fe]~$\simeq$~0.0 and [\rpro/Fe]~$\simeq$~$+$0.4. The \sri\ population
could have been labeled ``$r+s$-rich'' because they have overabundances of all
\ncap\ elements relative to Fe. What is secure is the addition of much more
\spro\ than \rpro\ material in the higher metallicity stars of M22.}

\begin{figure}
\centering
\includegraphics[width=9.8cm]{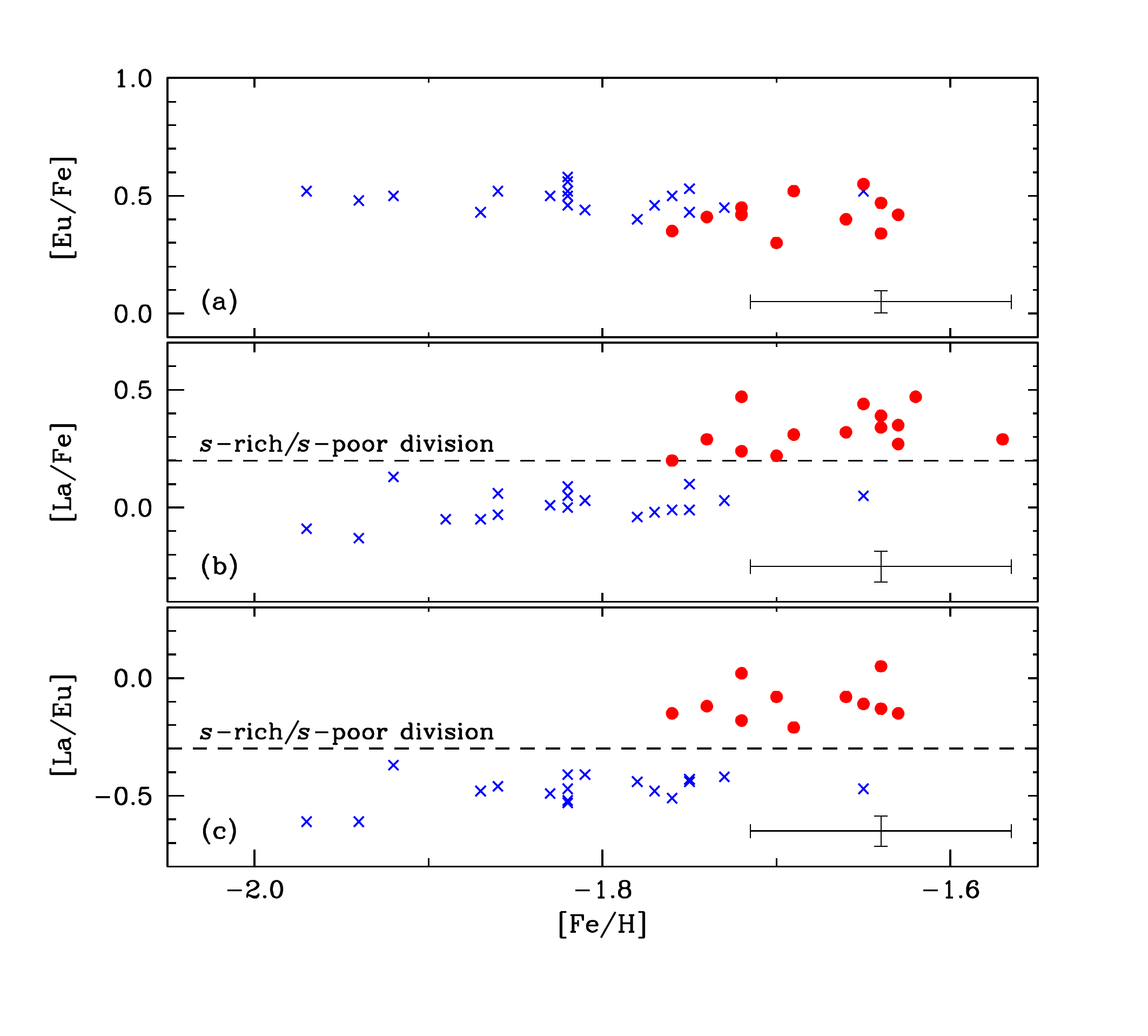}
\caption{La and Eu abundances of M22 as functions of [Fe/H].
  In panels (a) and (b) we repeat the [Eu/Fe] and [La/Fe] panels of
  Fig.~\ref{provafig1}.
  In panel (b) we have added a dashed line at [La/Fe]~=~$+$0.2 to show
  the division in this abundance ratio between \sri\ and \spo\ stars;
  see text for discussion of this choice.
  In panel (c) we plot the [La/Eu] values; the separation between
  the two groups of stars is more obvious here, and the dashed line
  represents our chosen split at [La/Eu]~=~$-$0.3.
  Symbols are as in Fig.~\ref{provafig1}.}
\label{laabund2}
\end{figure}

\begin{figure}
\centering
\includegraphics[width=9.6cm]{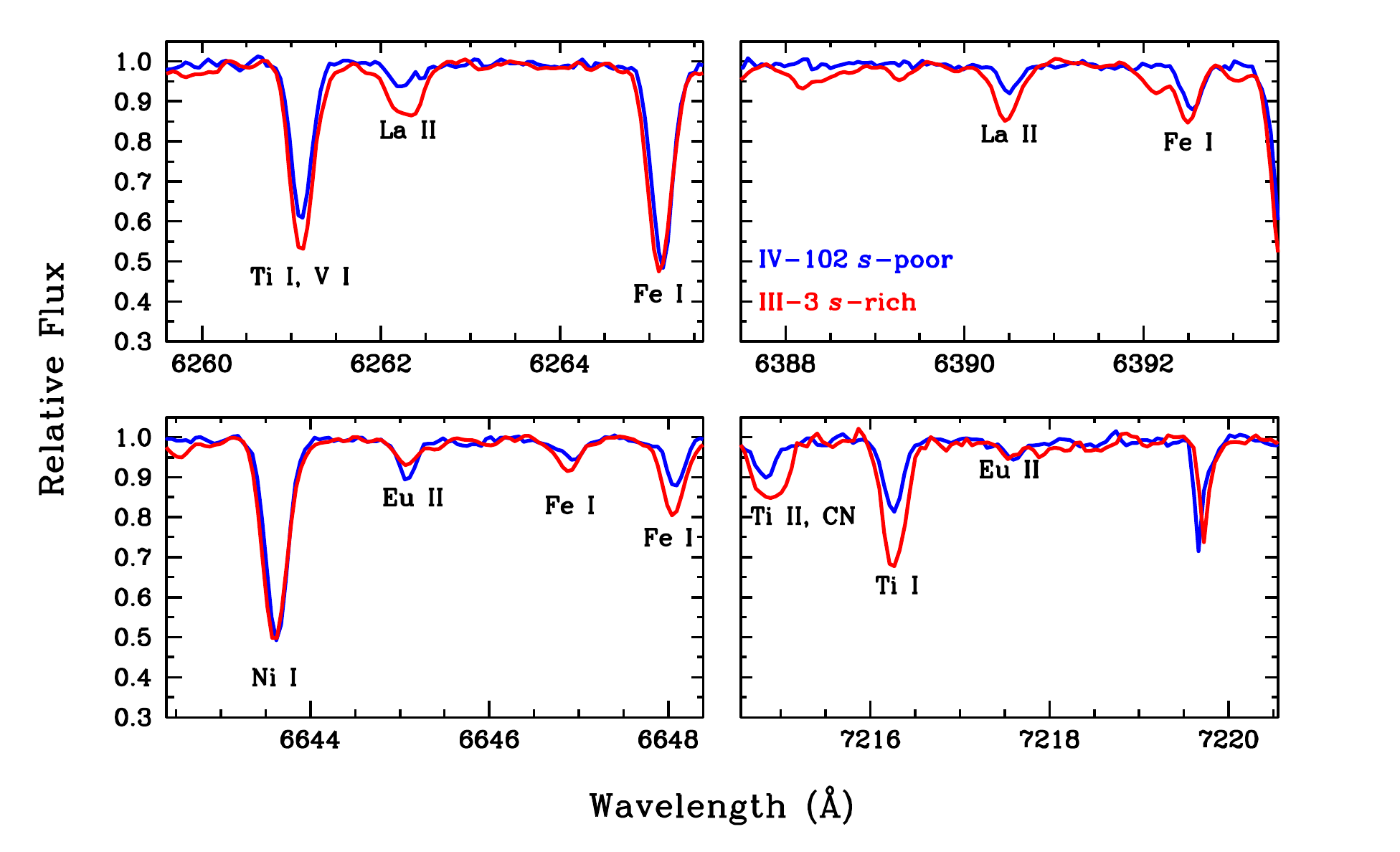}
\caption{Comparison of the spectra of La lines (top panels) and
  Eu lines (bottom panels) in two stars with similar atmospheric parameters
  but substantially different derived La abundances.
  The displayed data are taken from the MCD spectra.
  The spectrum in red is that of the \sri\ star III-3, and the one in
  blue is that of the \spo\ star IV-102.}
\label{laeuspec}
\end{figure}

\begin{figure}
\centering
\includegraphics[width=9.3cm]{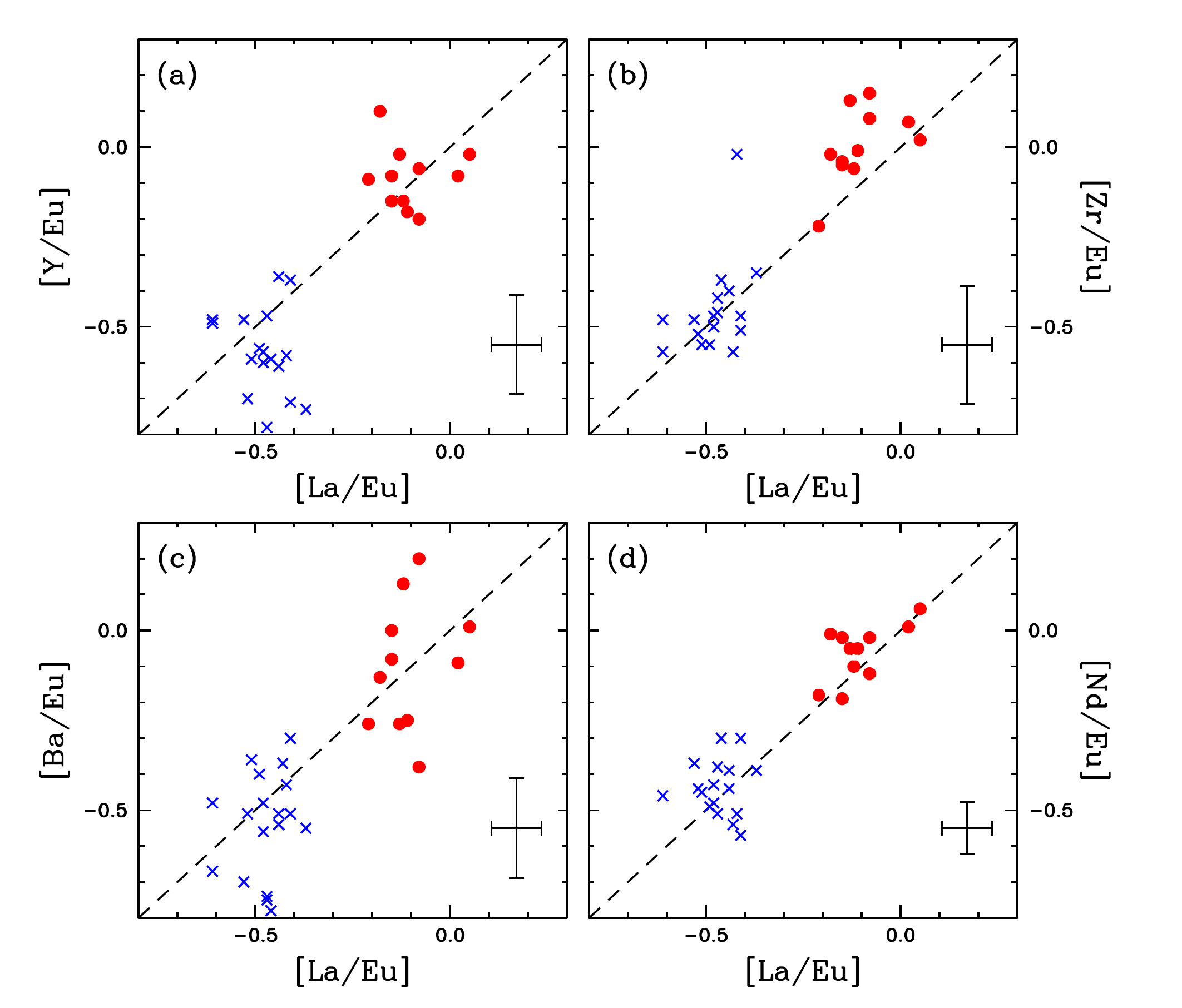}
\caption{Abundance ratios [X/Eu] for elements Y, Zr, Ba, and Nd
  as functions of [La/Eu].
  In each panel, the dashed line represents equality of the displayed
  abundance ratios.
  Symbols are as in Fig.~\ref{provafig1}.}
\label{otherncap}
\end{figure}

\begin{figure}
\centering
\includegraphics[width=9.1cm]{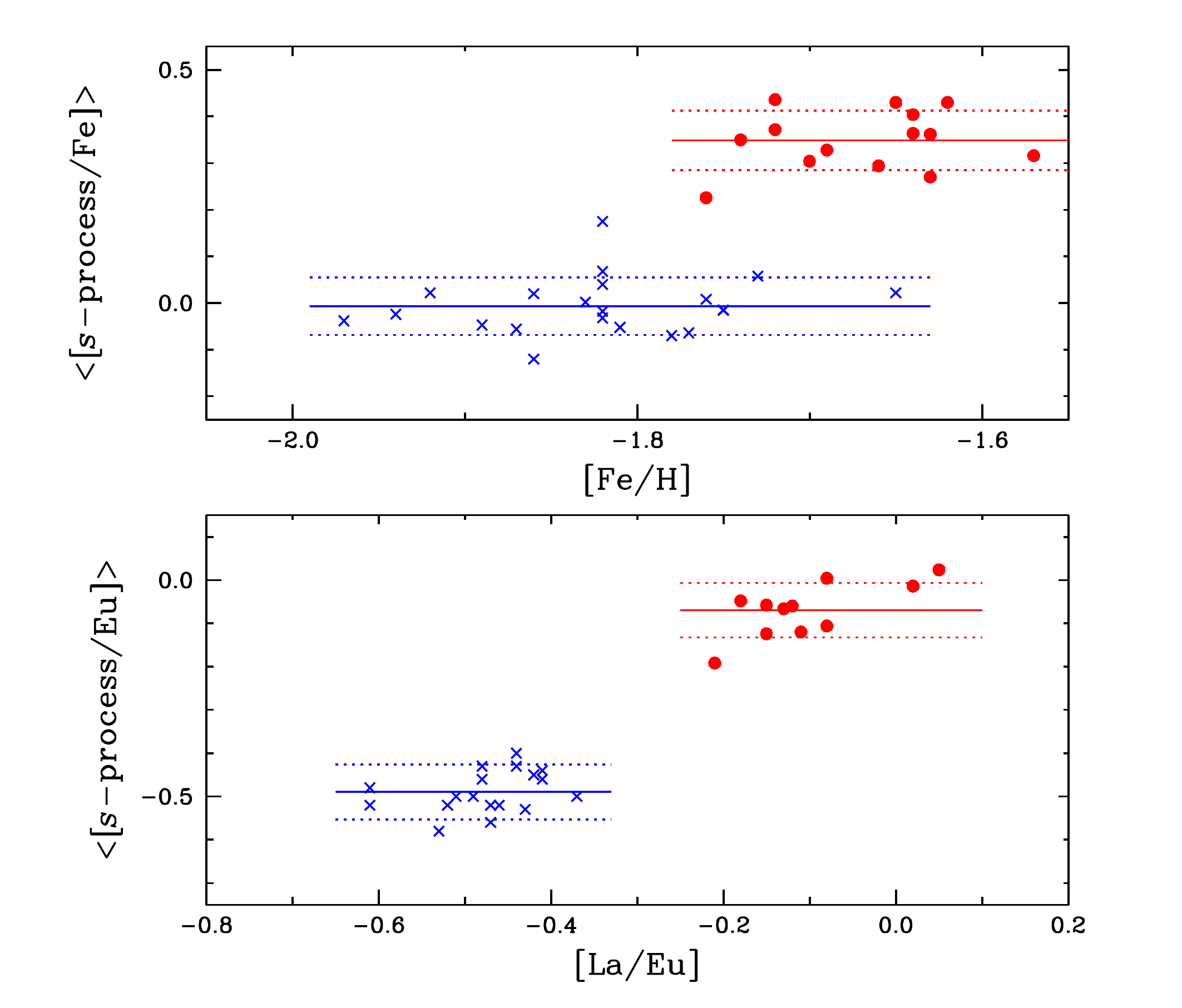}
\caption{{\it Upper panel}:
  average \spro\ abundance ratios with respect to Fe,
  plotted as functions of [Fe/H].
  {\it Bottom panel}:  average \spro\ abundance ratios with respect
  to Eu, plotted as functions of [La/Eu].
  See the text for how the averages were computed.
  In each panel, the solid lines represent the means for each group,
  and the dotted lines represent the sample deviations $\sigma$.
  Symbols are as in Fig.~\ref{provafig1}.}
\label{avgspro}
\end{figure}

%%%%%%%%%%%%%%%%%%%%%%%%%%%%%%%%%%%%%%%%%%%%%%%%%%%%%%%%%%%%%%%%%%%%%%%%%%
\subsection{The Heavy Fe-Peak Elements\label{fepeak}}
%%%%%%%%%%%%%%%%%%%%%%%%%%%%%%%%%%%%%%%%%%%%%%%%%%%%%%%%%%%%%%%%%%%%%%%%%%

Copper is very underabundant in M22, just as it is in other low metallicity
field stars (Sneden, Gratton, \& Crocker 1991, Mishenina \etal\ 2002) and
globular clusters (Simmerer \etal\ 2003).
However, [Cu/Fe] appears to vary in concert with the \spro\ elements,
being higher in the \sri\ than the \spo\ group by
$\Delta^{rich}_{poor}$[Cu/Fe]~= $+$0.15$\pm$0.04 (Table~\ref{tab-mean}).
In the top panel of Fig.~\ref{CuZn} we illustrate the Cu distributions
in the two groups, plotting them versus their \spro\ enrichment.
The much larger spread in individual [Cu/Fe] values in the \spo\ stars
compared to the \sri\ stars is worth noting.
For each $s$-group we represented the mean Cu abundance and the associated
rms. The difference among the two groups is at 3$\sigma$ level.
However, given the uncertainties associated with individual Cu
abundance measurements, interpretation of this trend should be viewed
with caution.
Certainly the \sri/\spo\ [Cu/Fe] difference is much less than that
observed for the \spro\ elements discussed in \S\ref{ncapel}.

If the [Cu/Fe] trend is real, it could put some new constraints on
scenarios for the origin of this element.
The nucleosynthetic sites of copper have been discussed by
Sneden \etal\ (1991), who suggested that much of the Cu in
metal-poor stars forms in the weak component of the \spro, at which time
neutron captures on Fe-peak elements take place during the late stages of
core He-burning (Couch, Schmiedekamp, \& Arnett 1974; Raiteri \etal 1991)
in massive stars.
Because the weak component of the \spro\ is a secondary mechanism for
nucleosynthesis, this site for the formation of copper agrees qualitatively
with a relationship of increasing [Cu/Fe] as a function of [Fe/H]
in metal-poor stars.
The Cu distribution that we observe in M22 provides mild support for
such an \spro\ origin, due to the weak component in massive stars, for
the more metal-rich stars.

As a check on the Cu results, note that our Zn abundances exhibit no
trend with \spro\ enrichment within the observational errors:
$\Delta^{rich}_{poor}$[Zn/Fe]~= $+$0.06$\pm$0.04 (Table~\ref{tab-mean}).
The individual abundances are plotted versus \spro\ enrichment
in the bottom panel of Fig.~\ref{CuZn}.
The \spo\ and \sri\ stars occupy the same [Zn/Fe] space in this
figure.
No observable difference in Zn between \sri\ and \spo\ groups would
have been expected from theory.
However the observational errors (represented by error bars in 
Fig.~\ref{CuZn}) are large enough to mask possible small abundance 
variations within the two groups.
The small positive value of $\delta$[Zn/Fe] probably illuminates the
limit of our abundance set to provide meaningful nucleosynthesis
scenarios for M22.

\begin{figure}
\centering
\includegraphics[width=9.cm]{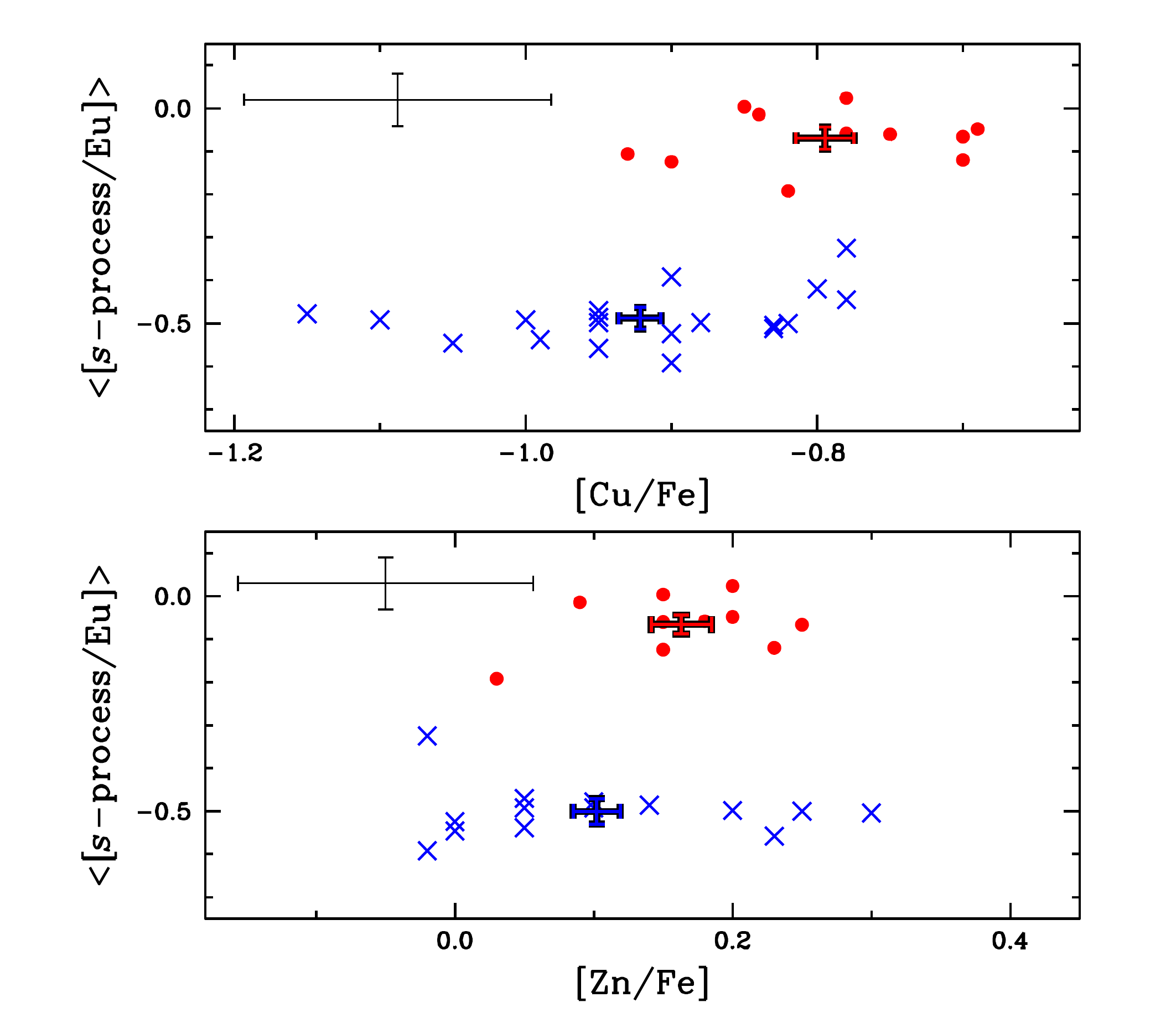}
\caption{Average \spro\ abundance ratios with respect to Eu, plotted as
  functions of the ratios of the heavy Fe-peak elements Cu and Zn to Fe.
  Error bars in blue and red represent the mean values for the \sri\
  and \spo\ stars.
  Symbols are as in Fig.~\ref{provafig1}}.
\label{CuZn}
\end{figure}

%%%%%%%%%%%%%%%%%%%%%%%%%%%%%%%%%%%%%%%%%%%%%%%%%%%%%%%%%%%%%%%%%%%%%%%%%%
\subsection{The $\alpha$ Elements\label{alphael}}
%%%%%%%%%%%%%%%%%%%%%%%%%%%%%%%%%%%%%%%%%%%%%%%%%%%%%%%%%%%%%%%%%%%%%%%%%%

The observed $\alpha$ elements in M22 are Si, Ca, and Ti.
We do not include O and Mg in this group because in globular clusters
their abundances can be affected by proton capture nucleosynthesis, 
i.e.\ different stars in GCs can have lower O, and in some cases lower Mg.
All $\alpha$ elements are overabundant in M22.
From the data in Table~\ref{tab-mean} for the whole 35-star sample,
we obtain $<$[$\alpha$/Fe]$>$~= $+$0.33~$\pm$~0.01 ($\sigma$~=~0.04).
These $\alpha$-element enhancements are in excellent agreement with
those reported by M09.
Note that the mean O and Mg abundances are nearly the same as the
other $\alpha$'s:
$<$[O/Fe]$>$~=~$+$0.34 and $<$[Mg/Fe]$>$~=~$+$0.39.
These elements will be discussed in detail in \S\ref{proton}.

No correlation with \spro\ peculiarity can be detected for
Si, and Ti; their [X/Fe] ratios are identical in \sri\ and
\spo\ stars, within the abundance measurement uncertainties.
The \ion{Ti}{I}$-$\ion{Ti}{II} abundance differences have a small 
trend with temperature, due almost entirely to a variation in\ion{Ti}{I}; 
the \ion{Ti}{II} abundance distribution is constant with \teff.
This behavior may reflect NLTE effects in the \ion{Ti}{I} abundances.
NLTE overionization of \ion{Ti}{I} is likely to be occurring, for
the average \ion{Ti}{II} abundances are higher than \ion{Ti}{I} ones. 
Probably \ion{Ti}{II} abundances are more reliable than \ion{Ti}{I} ones 
(see Bergemann 2011 for a detailed discussion).

From UVES spectra of 17 stars, M09 claimed that a small but
statistically significant positive correlation existed between
[Ca/Fe] and [Fe/H] (or [\spro/Fe]).
As shown in Fig.~\ref{Ca}, our larger sample confirms this trend at
approximately the same level:  
$\Delta^{rich}_{poor}$[Ca/Fe]~= $+$0.10$\pm$0.02 (Table~\ref{tab-mean}).

\begin{figure}
\centering
\includegraphics[width=9cm]{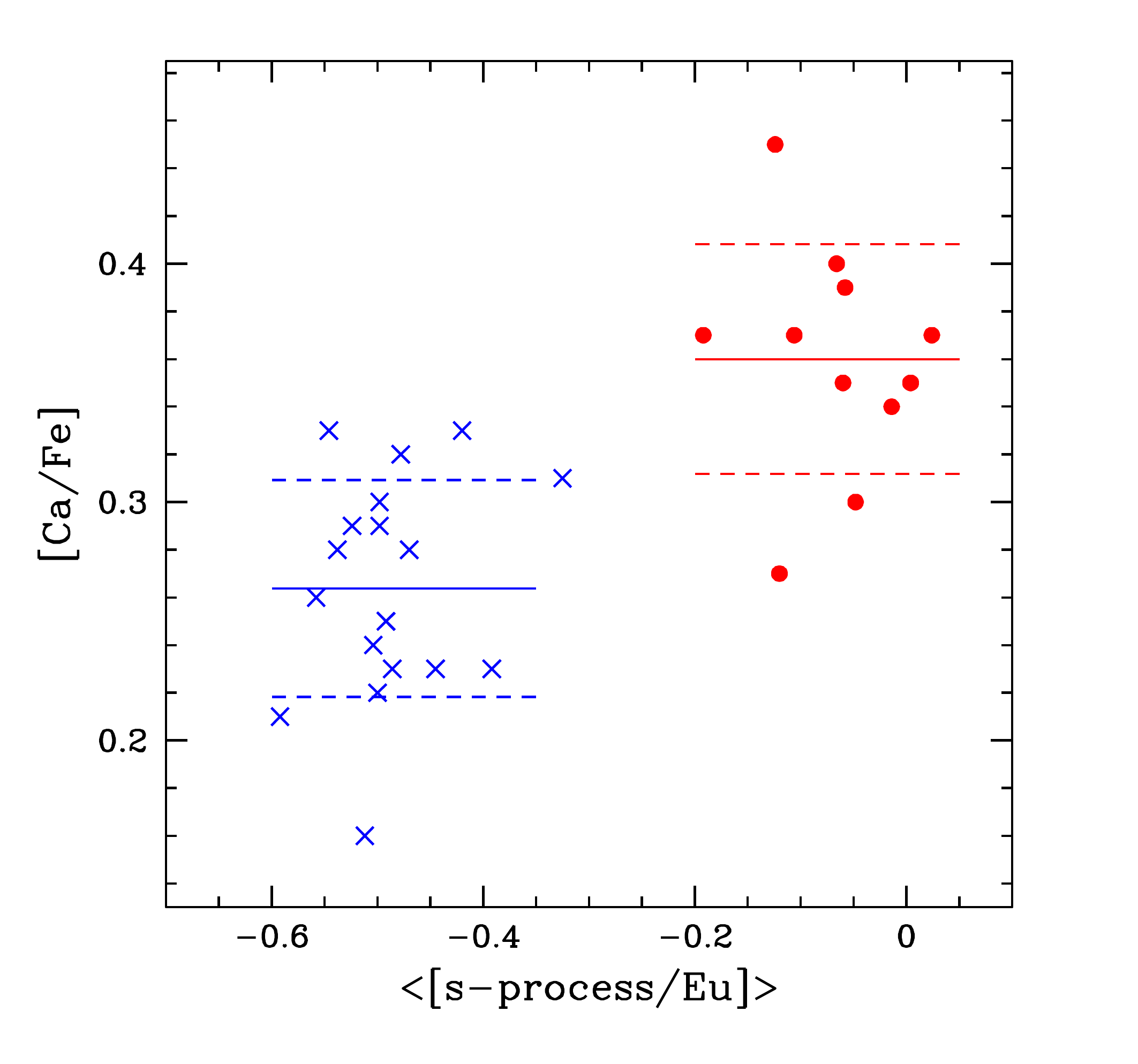}
\caption{Calcium abundance ratios plotted as a function of the average
\spro\ abundance ratios with respect to Eu.
Symbols are as in Fig.~\ref{provafig1}.}
\label{Ca}
\end{figure}

%%%%%%%%%%%%%%%%%%%%%%%%%%%%%%%%%%%%%%%%%%%%%%%%%%%%%%%%%%%%%%%%%%%%%%%%%%
\subsection{The Proton-Capture Elements\label{proton}}
%%%%%%%%%%%%%%%%%%%%%%%%%%%%%%%%%%%%%%%%%%%%%%%%%%%%%%%%%%%%%%%%%%%%%%%%%%

We now consider the abundances for C, N, O, Na, Mg, and Al, all of
which can be affected by proton capture reactions.
In Fig.~\ref{pcapture} we display the [X/Fe] values obtained for each
of these elements as a function of the mean $<$[\spro/Eu]$>$ abundance.
As in previous figures, we use different symbols to represent the
\spo\ and \sri\ groups.
The means and the $\sigma$ values obtained for each group are also shown.
Since for 5 out 8 stars in the \spo\ groups we were able to measure only
upper limits for the N abundance, the plotted mean is of course also an
upper limit.

Inspection of Fig.~\ref{pcapture} suggests that \sri\ stars have, on
average, higher C and N contents, and from Table~\ref{tab-mean} we compute
$\Delta^{rich}_{poor}$[C/Fe]~= $+$0.35$\pm$0.13 and
$\Delta^{rich}_{poor}$[N/Fe]~= $+$0.47$\pm$0.11.
As noted by M09, correlations (with large star-to-star scatter) are observed
between the \spro\ groups with Na and Al:
$\Delta^{rich}_{poor}$[Na/Fe]~= $+$0.23$\pm$0.07 and
$\Delta^{rich}_{poor}$[Al/Fe]~= $+$0.21$\pm$0.10.
In contrast, the \textit{p}-capture O and Mg elements have nearly
the same abundances in both groups: $\Delta^{rich}_{poor}$[O/Fe]~=
$-$0.04$\pm$0.07
and $\Delta^{rich}_{poor}$[Mg]~= $+$0.01$\pm$0.04.

More illuminating are the abundance comparisons among the $p$-capture
elements.
As shown by M09, Na and O abundances are anticorrelated in M22
(see their Fig.~3) just as they are in all GCs that have been
studied so far.
In  Fig.~\ref{fig4} we show the Na and O data for our larger
  sample, by using both the LTE Na values (right panel) and the
  NLTE ones (right panel).
Our sample makes it clear that the NaO 
anticorrelation exists in both M22 populations, but in somewhat
different domains of O-Na space.
The \sri\ and \spo\ stars span a similar range in oxygen,
$-$0.1~$\lesssim$ [O/Fe]~$\lesssim$ $+$0.6, but not in sodium:
$-$0.3~$\lesssim$ [Na/Fe]$_{\small {\spo}}$~$\lesssim$ $+$0.5 and
$+$0.0~$\lesssim$ [Na/Fe]$_{\small {\sri}}$~$\lesssim$ $+$0.6.
To better visualize the difference between the Na-O pattern in the
two \spro\ groups, we have drawn by hand a fiducial line tracing
the Na-O anticorrelation shape for \spo\ stars, and superimpose this
line to the \sri\ stars (upper panels of Fig.~\ref{fig4}).
With this aid, we estimate that at any [O/Fe] ratio, the
[Na/Fe] ratio in an \sri\ star is $\sim$0.2~dex larger than in
an \spo\ star.
The different behavior of \spo\ and \sri\ stars in the Na-O
anticorrelation, even if with some differences, recalls the one in
$\omega$~Cen (see Johnson \& Pilachowski 2010; Marino \etal\ 2011).

Both M22 populations exhibit positive abundance correlations of Al with
Na, as illustrated in the top panel of Fig.~\ref{almg}.
However, here also the \spo\ and \sri\ stars are distinguishable:
only the \spo\ stars have [Na or Al/Fe]~$<$~0.
No Mg variation with either Al or Na were discovered by M09,
and our expanded sample confirms this result (see the left
panel of Fig.~\ref{almg}).
M09 suggested that an intrinsic Mg variation might be too small to
be detected in the face of observational errors.
For our complete 35-star sample, as well as the \spo\ and \sri\
subsamples, $\sigma$[Mg/Fe]~$\simeq$~0.11$-$0.12 (Table~\ref{tab-mean}).
Such values are not much larger than those for [Si/Fe], [Ca/Fe],
[Ti/Fe], and [Zn/Fe], which we claim are invariant in M22 stars.
A possible Mg-Al anticorrelation may be inferred from the
half-dozen stars in Fig.~\ref{almg} with the lowest [Mg/Fe] ratios;
these all have high [Al/Fe] ratios.
However, this is not a statistically defensible conclusion, and pursuit
of this point would need careful analysis of a larger M22 sample.

The lack of a clear Mg-Al anticorrelation does not necessarily mean
that proton captures on Mg are ruled out.
If we suppose that the higher observed Mg abundances are
representative of ``primordial'' M22 material (that is, prior to any
$p$-capture synthesis events), and if primordial Al is indicated
by the lower observed Al abundances, then for this material
[Mg/Al]~$\sim$~+0.5, or log~$\epsilon$(Mg/Al)~$\sim$~+1.6.
Then if (for example) 10\% of this Mg were to be converted to
\iso{27}{Al} by $p$-capture in the primordial material, the
resulting Al would go up by a factor of four, nearly the range
covered by our data.
The 10\% decrease in Mg would be nearly impossible to be detected.
Additionally, if the ab initio abundance of Mg contains substantial
amounts of \iso{25}{Mg} and/or \iso{26}{Mg},
then the final Al abundance would be even larger after $p$-captures.

In Fig.~\ref{cn} we present correlations between C abundances and
N, Na, O abundances.
No trends among these elements are apparent if we consider our sample
of stars as a whole.
However, segregation of points into two different $s$-Fe groups
provides evidence for unique C-N (left-hand panel) and C-Na (middle
panel) anticorrelations within each group, as expected from CNO-cycle
enrichment within each M22 population separately.
However, we cannot discern any obvious nucleosynthetic signature
in the C-O plot (right-hand panel), either from the whole sample or
in the two populations individually.

In Fig.~\ref{cno} the CNO abundance sum is represented as
a function of [$s$/Fe], as usual calling the reader's attention to
the \spo\ and \sri\ groups with different symbols in the figure.
These CNO totals are shown as [(C+N+O)/Fe] in the left-hand panel
and as log$\epsilon$(C+N+O) in the right-hand panel.
The M22 population split is evident: \sri\ stars have on average a
higher [(C+N+O)/Fe] abundance, with 
$\Delta^{rich}_{poor}$[(C+N+O)/Fe]~= $+$0.13$\pm$0.03.
The mean difference in the [(C+N+O)/Fe] content is at the level of
$\gtrsim$3$\sigma$ (see Table~7),
but note that we could measure only an upper limit to
the N abundance for almost all the \spo\ stars, suggesting that the real
difference could be larger.

A final piece of evidence in the M22 abundance puzzle comes from our
analysis of the 8000~\AA\ region of six MCD stars for which we were
able to derive \iso{12}{C}/\iso{13}{C} ratios (Table~\ref{tab-ablight}).
Note that star IV-102 has very weak CN red-system bands at all wavelengths,
and our derived \iso{12}{C}/\iso{13}{C} value should be treated with caution.
The APO and LICK spectra also cover the 8000~\AA\ spectral region.
However, neither of these data sets have sufficient resolving power
and S/N to permit reliable carbon isotopic ratios.

Although the six stars with derived carbon isotopic ratios constitute
only a small subset of our M22 giants, all of them have
3.0~$<$~\iso{12}{C}/\iso{13}{C}~$<$~5.0
(Table~\ref{tab-ablight}).
Our low value for III-3 is supported by the earlier Brown \etal\ (1990)
analysis of this star, for which they obtained \iso{12}{C}/\iso{13}{C}~=~4.
Star IV-20, with no MCD data in the present study, also has a very low
isotopic ratio according to Brown \etal: \iso{12}{C}/\iso{13}{C}~=~4.
Finally, Smith \& Suntzeff (1989) used low-resolution spectra of
the CO infrared first-overtone vibration-rotation bands to derive
\iso{12}{C}/\iso{13}{C} estimates for five M22 stars.
For three stars (III-3, III-12, and IV-102) their ratios
are in excellent agreement with ours.
However, for stars IV-97 and V8 (with no available values in our study) Smith
\& Suntzeff derived higher 
\iso{12}{C}/\iso{13}{C} values: $\geq$10 and $\geq$40, respectively.
In summary, the carbon isotopic values for most M22 giants appear to be
very low, normal for the stars similar to the ones studied here, 
but a larger-sample study will be needed to determine if
stars with substantially larger ratios are few in number or common.

\begin{figure}
\centering
\includegraphics[width=9cm]{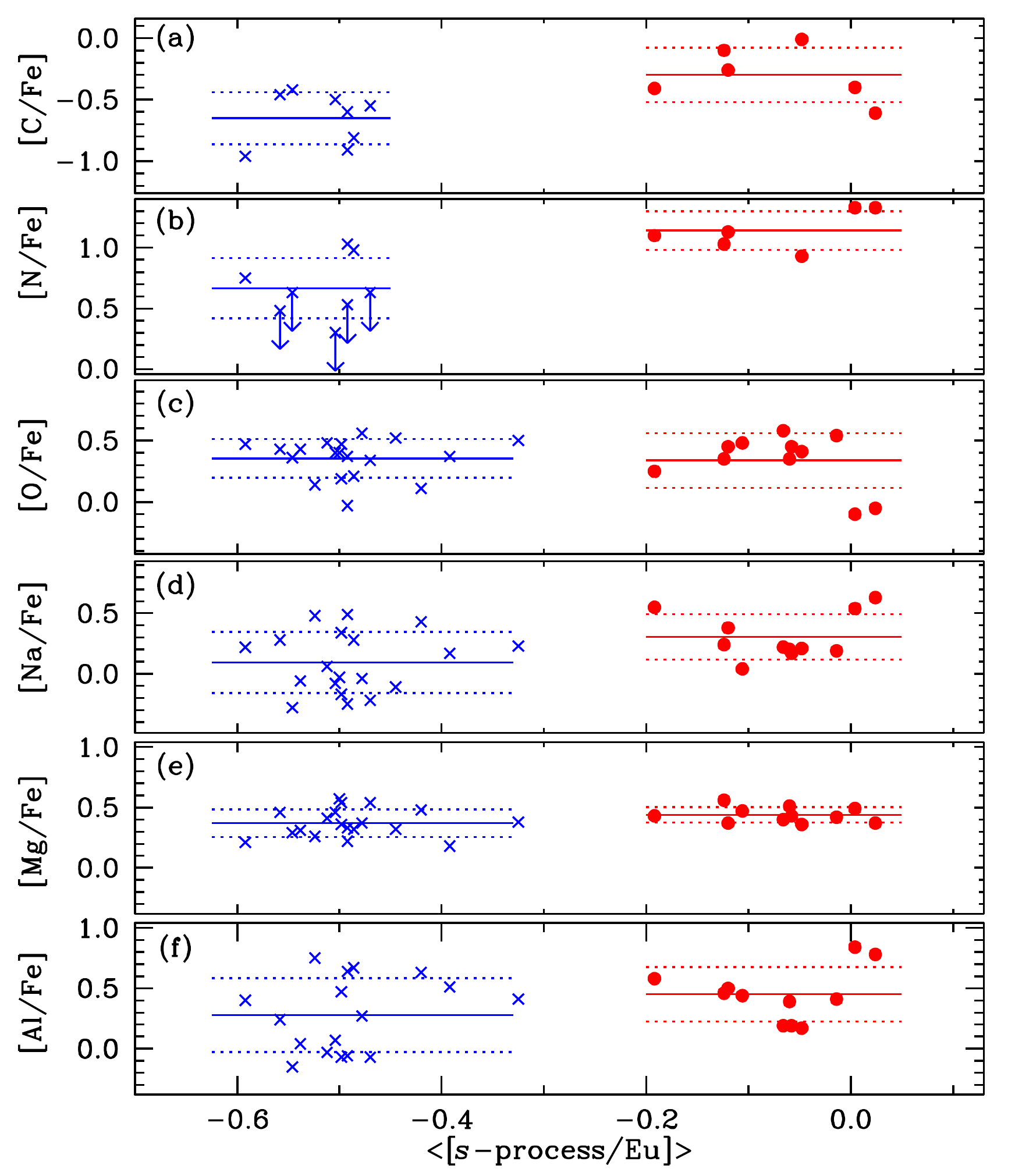}
\caption{ [C, N, O, Na, Mg, Al/Fe] abundance ratios as a function of the
 mean [\textit{s}-process/Eu] content.
 Symbols are as in Fig.~\ref{provafig1}.}
\label{pcapture}
\end{figure}

\begin{figure}
\centering
\includegraphics[width=9.5cm]{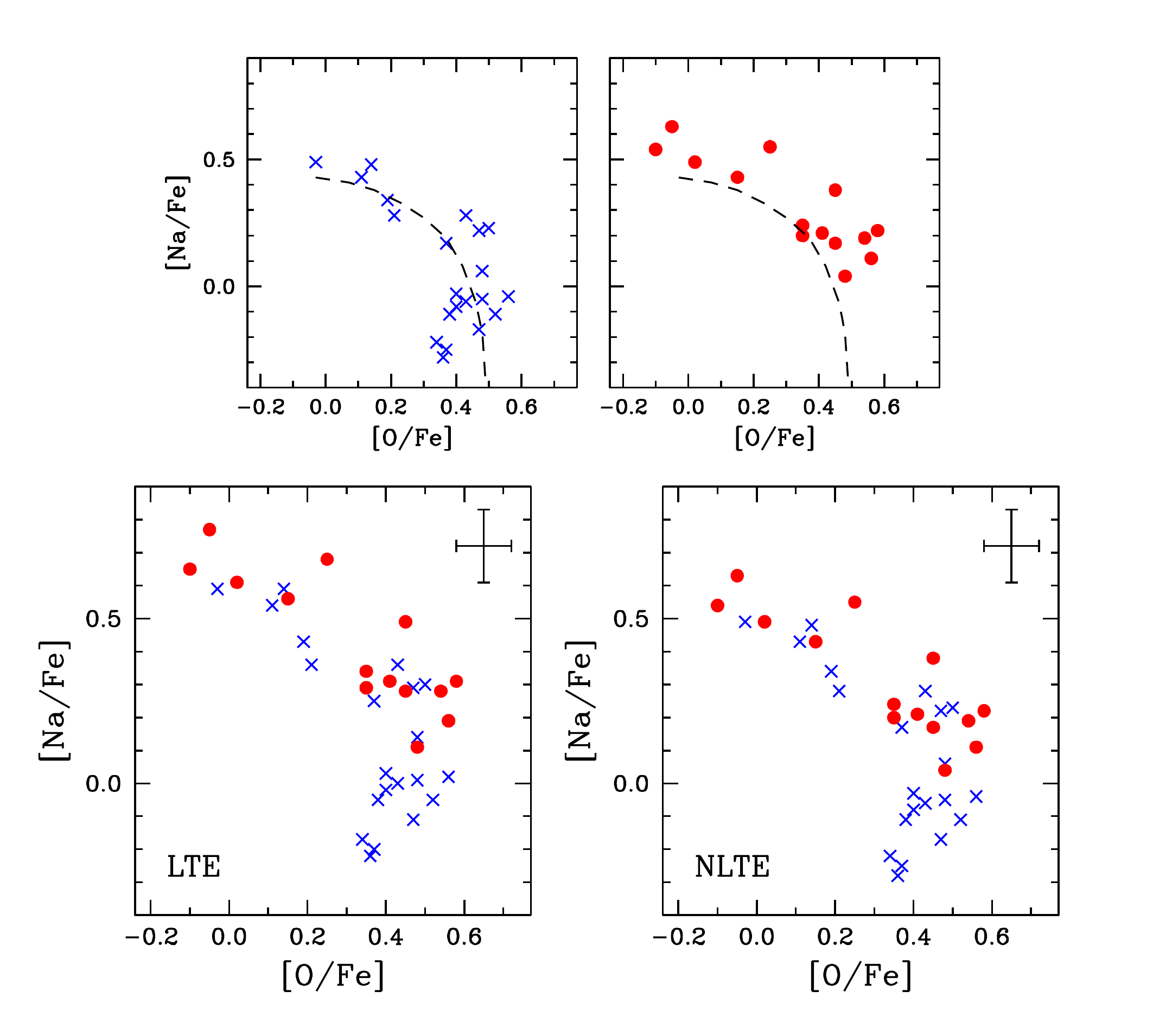}
\caption{[Na/Fe] as a function of [O/Fe].
  The bottom panels display all stars in the Na-O plane by using LTE
  (left) and NLTE Na abundances (right). The top left-hand
  panel contains only \spo\ stars and the top right-hand panel
  contains only \sri\ stars.
  The dashed lines in the top panels represent a free-hand
  representation of the mean Na-O trend of only the \spo\ stars.
  Symbols are as in Fig.~\ref{provafig1}.}
\label{fig4}
\end{figure}

\begin{figure}
\centering
\includegraphics[width=9.2cm]{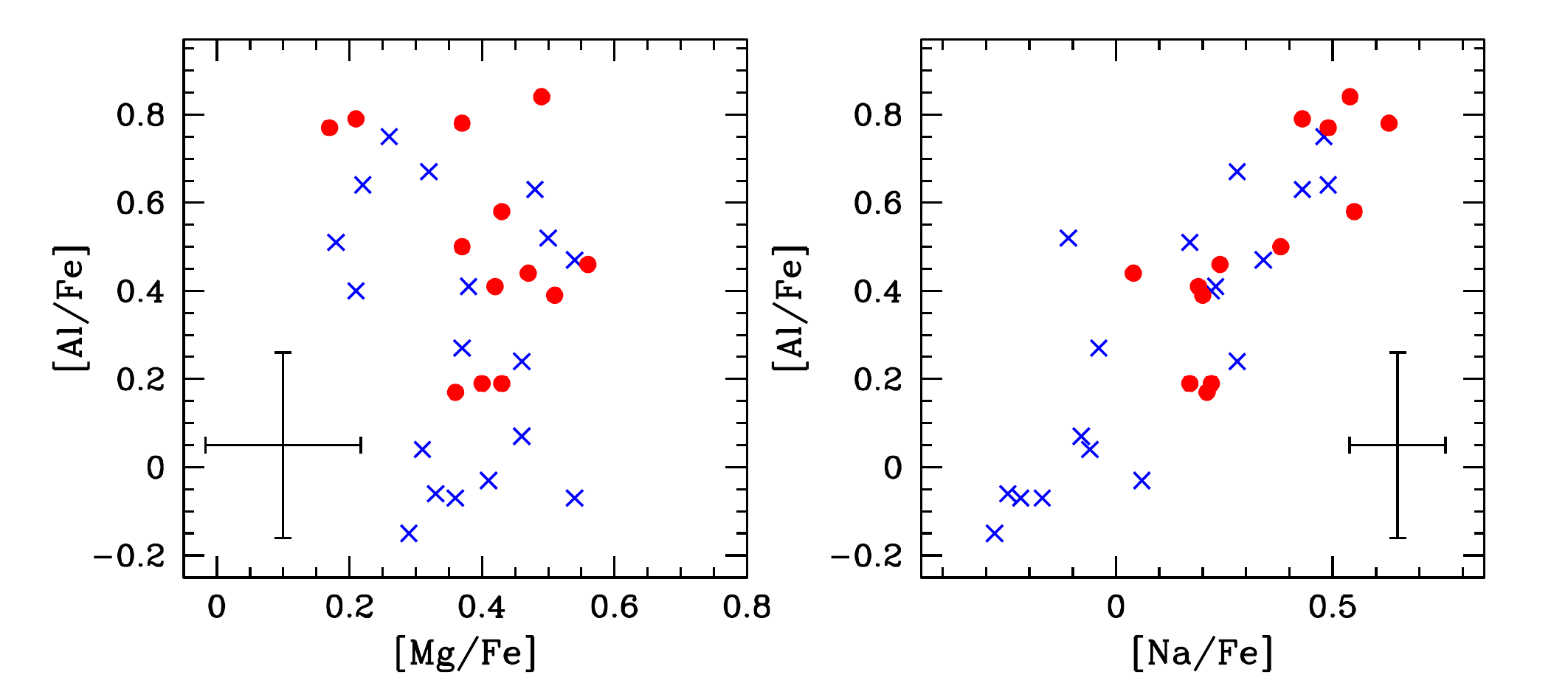}
\caption{[Al/Fe] abundance ratios as a function of [Na/Fe] (left panel),
  and [Mg/Fe] (right panel).
  Symbols are as in Fig.~\ref{provafig1}.
}
\label{almg}
\end{figure}

\begin{figure*}
\centering\includegraphics[width=14cm]{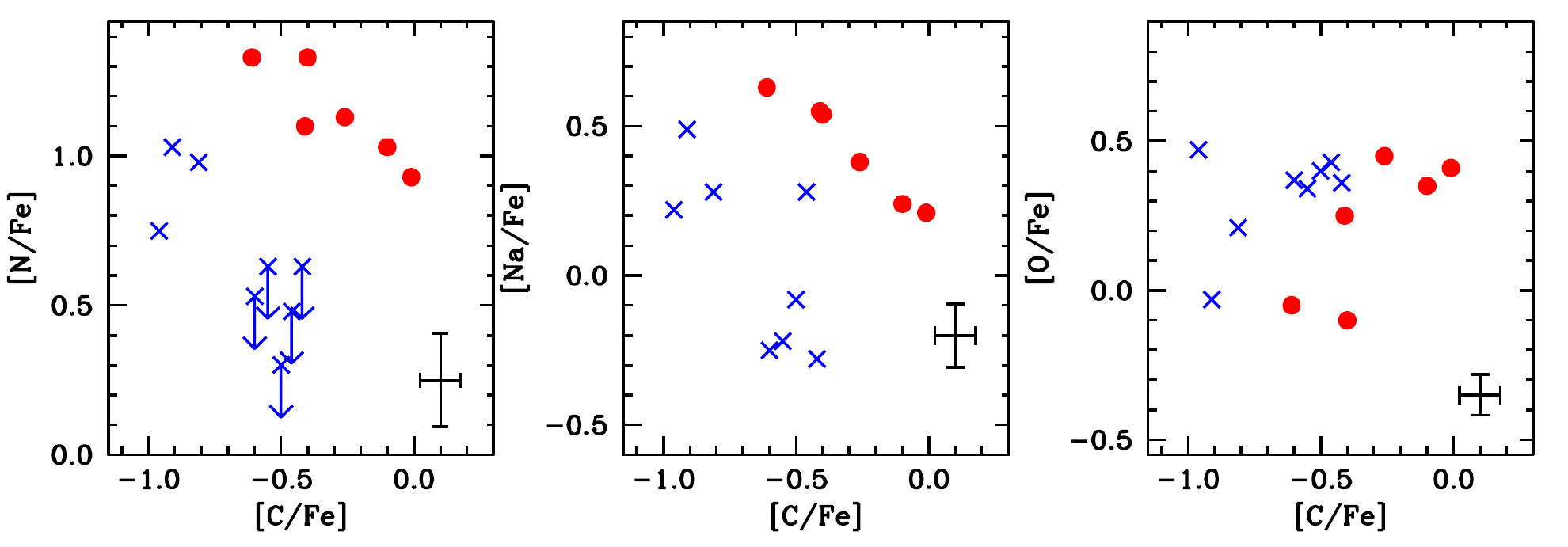}
\caption{From the left to the right: nitrogen, sodium and oxygen
  abundance ratios as a function of [C/Fe].
  Symbols are as in Fig.~\ref{provafig1}.}
\label{cn}
\end{figure*}

\begin{figure}
% CNO vs s/Eu
\centering
\includegraphics[width=9.cm]{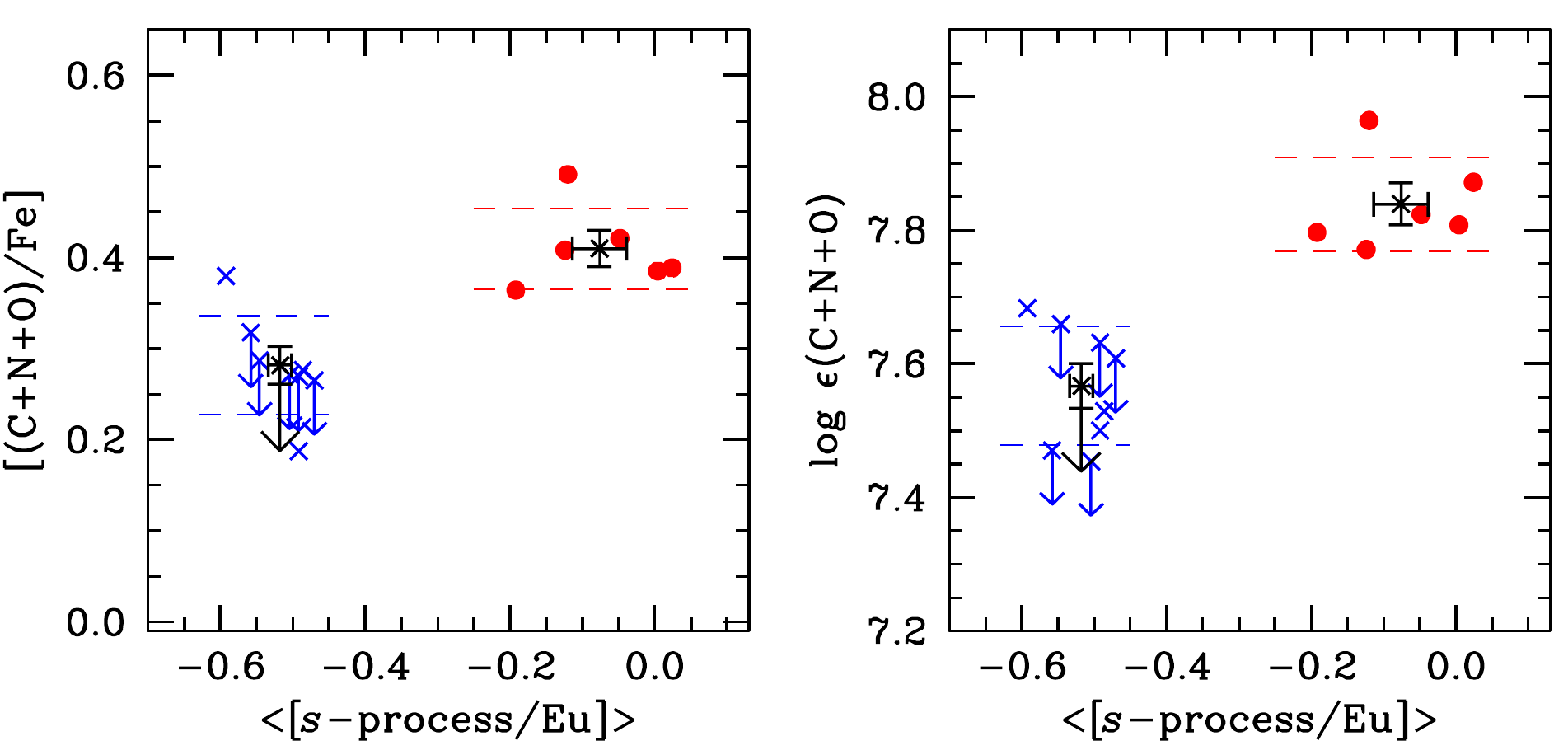}
\caption{[(C+N+O)/Fe] (left panel) and log~$\epsilon$(C+N+O) (right panel)
  as a function of $<$[$s$-process/Eu]$>$.
  Symbols are as in Fig.~\ref{provafig1}.
The dashed lines represent the error associated with single
 measures, the black error bars represent the mean CNO abundance
 contents for the two $s$-groups and the statistical error associated
with the mean.}
\label{cno}
\end{figure}

%%%%%%%%%%%%%%%%%%%%%%%%%%%%%%%%%%%%%%%%%%%%%%%%%%%%%%%%%%%%%%%%%%%%%%%%%%
\section{M22 LOW RESOLUTION SPECTROSCOPY AND PHOTOMETRY\label{rgb}}
%%%%%%%%%%%%%%%%%%%%%%%%%%%%%%%%%%%%%%%%%%%%%%%%%%%%%%%%%%%%%%%%%%%%%%%%%%

Norris \& Freeman (1983) gathered low resolution blue spectra of
130 giants in M22, and from these data determined three absorption indices:
$S$(3839), for the CN 3883 bandhead strength; $A$(Ca), for
the \ion{Ca}{II} H\&K strength; and $W$(G), for the CH G-band strength.
The CN and the Ca indices were corrected to first order for the
natural changes in absorption strengths due to \teff\ and \logg\
differences along the M22 giant branch.
This was accomplished by first drawing fiducial lines to express the
strength changes as functions of $V$ magnitude, and then
measuring offsets $\delta S(3839)$ and $\delta A$(Ca) from these lines.
Norris \& Freeman showed that a positive correlation exists between
$\delta S(3839)$ and $\delta A$(Ca), as well as between $W$(G) and
$\delta A$(Ca).
In Fig~\ref{norris1} we compare these indices with some abundances 
derived in this paper. 
The upper two panels, showing the $S$(3839) index as a function of 
[Fe/H] and [La/Eu], clearly demonstrate that our derived [Fe/H] 
metallicities and the \spro\ abundances also track the CN strength indices.
In the middle panels we show that there is a mid correlation between 
$W$(G) with [C/Fe] and $S$(3839) with [N/Fe].
Note also the positive correlation between the $A$(Ca) index and our
derived abundance ratios [Ca/H] and [Ca/Fe] (lower two panels).

In addition we have coupled our abundances to the Str\"omgren photometric
data of Richter \etal\ (1999).
Those authors demonstrated the existence of a bimodal distribution
in the $m_{1}$ index of M22 giants, which they associated with CN variations.
Here we match our spectroscopic results with photometry in two different ways.
In the top panels of Fig.~\ref{hilker} purely photometric data are
displayed, combining the Str\"omgren colours with $I$ magnitudes taken
from Stetson's database,
after correcting the data for differential reddening.
Our stars, coded as in previous figures, show little differences in
the $b-y$ versus $I$ plot (top-left panel).
However, a clear split is apparent between the $s$(Fe)-poor and $s$(Fe)-rich
stars in $m_{1}$ versus $I$ plot (top-right panel).
As the $m_{1}$ index is strongly affected by the blue CN bands and
overall metallicity, we expected this bimodal distribution as a
consequence of the higher mean abundance in both C and N of the \sri\
stars, as shown in Fig.~\ref{pcapture}.
Hence, the \sri\ stars populate the RGB sequence associated to the
stars enriched in CN, and \spo\ stars the branch associated to weaker CN
band strengths (see Richter \etal\ 1999).

Of course, our sample of stars could be contaminated by AGB stars, but
the presence of few AGB stars does not affect the results,
as discussed also in M09.  
A visual inspection of the stars on the CMD on the right-upper
panel of Fig.~19,
suggests that some AGB stars could be present among both the \spo\
and the \sri\ RGB. 
The AGB lies blueward of the RGB at a given luminosity.
But identification of blue-offset stars in the $I$ vs.\ $b-y$ CMD 
(left-upper panel of Fig.~19) as probable members of the AGB 
could be misleading, because we also expect to find the slightly more
metal-poor, \spo\ stars on the blue side of the RGB.
Unambiguous assignment of AGB stars in this manner is not easy.

On the other hand, systematically redder colours for the 
\sri\ rich stars would lead to a low probability of being shifted 
by photometric errors into the AGB region.
Indeed in M09, six out of seven probable AGB stars,
belong to the $s$-poor group. 
Note that the possible presence of AGB stars in M09 was based only 
on a visual inspection of the stars in the 
$I$ vs.\ $(B-I)$ CMD, and doesn't necessarily mean that AGB stars 
preferentially belong to the \spo\ group.

In any case, assuming that all the probable AGB stars in M09 are 
indeed true AGB members, their \spo\ sample would include half RGB 
and half AGB stars, tracing the primordial composition of the cluster.   
The chemical $s$-element abundances of probable AGB belonging to the
\spo\ group of M22 could not reflect "in situ" phenomena, as the third
dredge-up that would lead to enhanced $s$-process abundances in the
atmosphere (and additionally is expected for much more massive AGB
stars), but must reflect the primordial composition of the \spo\ group.
Moreover, we expect only a $\sim$10\% contamination of giant 
population by AGB stars 
(Lloyd Evans 1975), hence it is unlikely that the 50\% of stars 
in the \spo\ sample of M09 are real AGB.
From this discussion we conclude that the presence of few AGB stars 
does not influence our results, so for the present study we 
consider all the sample as composed by RGB, with a small contamination 
by AGB stars, without identifying individual candidate AGB stars.

To illuminate the relation between the ${\it m}_{1}$ values of \sri\ and
\spo\ stars and their average $s$-element contents, we used the following
procedure.
First,we drew a ridge line for the blue RGB, by putting a spline through
the median ${\it m}_{1}$ found in successive short intervals of $I$
magnitude, and obtained the dashed-dotted line shown in the
upper-right panel of Fig.~\ref{hilker}.
We then calculated for each star its ${\it m}_{1}$ residuals from
the ridge line (called here $\Delta {\it m}_{1}$).
We plotted the $I$ magnitude versus $\Delta {\it m}_{1}$ as illustrated
in the lower-left panel of Fig.~\ref{hilker}.
Finally, in the lower-right panel we show the average $s$-element
abundance as a function of $\Delta {\it m}_{1}$.
Clearly, the mean $<$[$s$/Fe]$>$ increases with increasing
$\Delta {\it m}_{1}$.

\begin{figure}
\centering
\includegraphics[width=9.2cm]{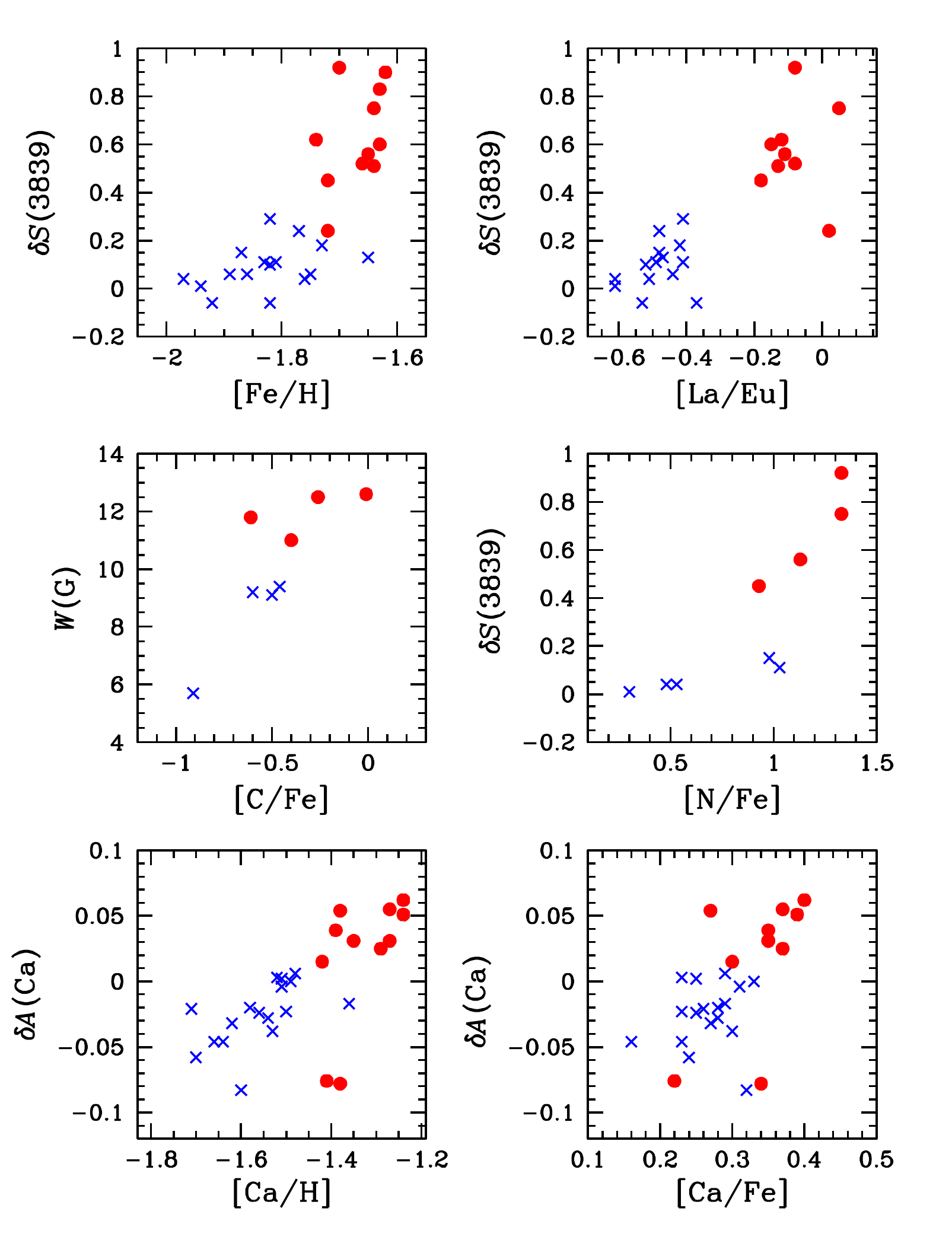}
\caption{Comparison of the chemical abundances derived in this paper with the
  indices by Norris \& Freeman (1983): CN band-strength indices $\delta
  S(3839)$ plotted as a function of the [Fe/H] (left-upper panel) and
  [La/Eu] (right-upper panel), {\it W}(G) and $\delta S(3839)$ as a function
  of [C/Fe] (left-middle panel) and [N/Fe] (right-middle panel), and $\delta
  A$(Ca) as a function of [Ca/H] (left-bottom panel) and [Ca/Fe] (right-bottom
  panel).
  Symbols are as in Fig.~\ref{provafig1}.}
\label{norris1}
\end{figure}

\begin{figure*}
\centering
\includegraphics[width=13cm]{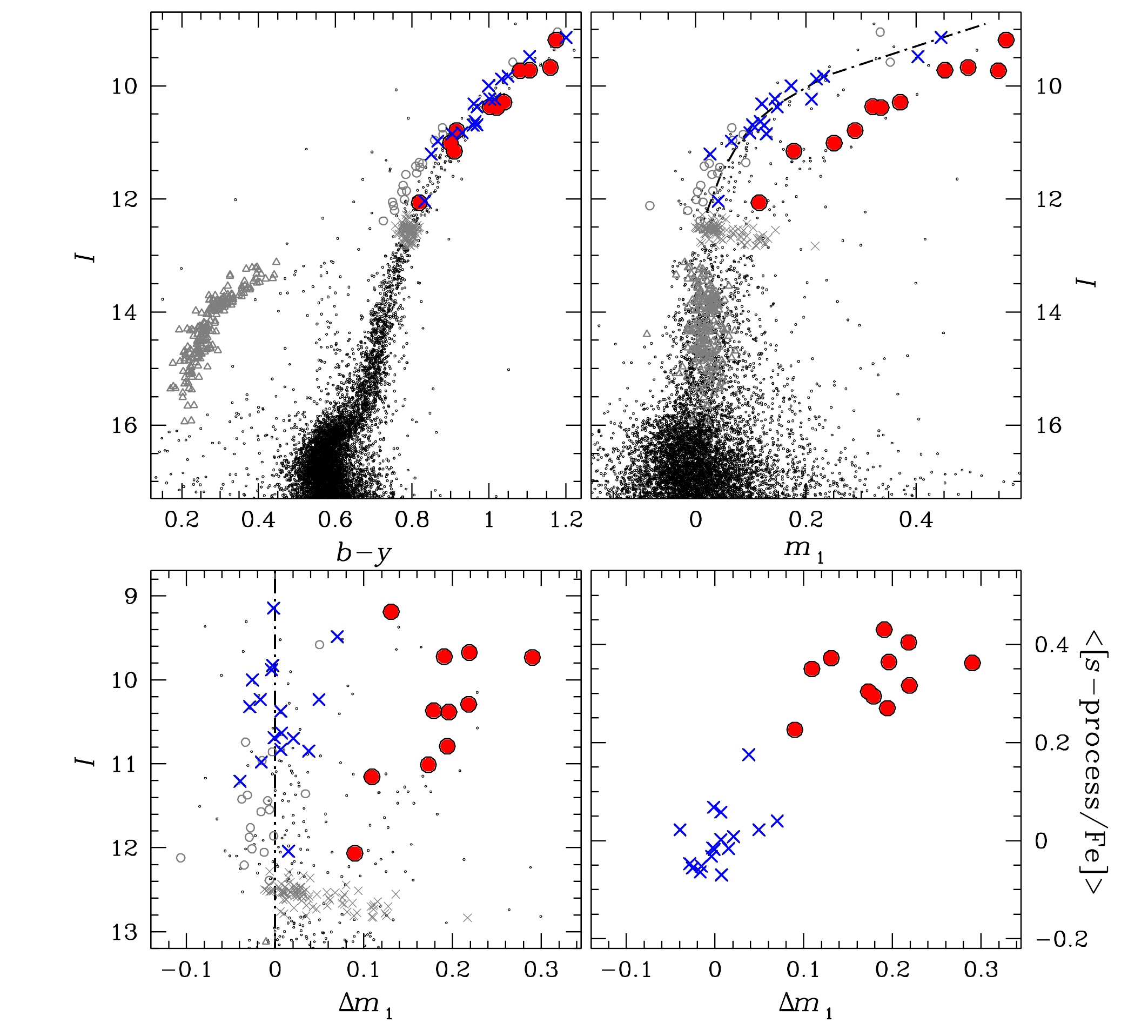}
\caption{ {\it Top panels}: $I$-$b-y$ (left) and  $I$-$m_{1}$ (right)
CMD for M22, corrected for differential reddening.
The grey symbols represent HB stars (triangles), 
probable AGB stars (circles) and RGB bump stars (crosses) selected 
in the $I$-$(b-y)$ CMD.
Spectroscopic data are superimposed with \sri\ stars represented by
red circles, and \spo\ stars by blue crosses, according with the
other figures.
{\it Bottom panels}: colour difference $\Delta m_1$ between each
analyzed star and a reference fiducial line represented as a dotted
line (left panel).
On the right panel the $<$[$s$/Fe]$>$ average abundances are shown
as a function of  $\Delta m_1$.
The sources for the photometric data are given in \S~\ref{data}.}
\label{hilker}
\end{figure*}

%%%%%%%%%%%%%%%%%%%%%%%%%%%%%%%%%%%%%%%%%%%%%%%%%%%%%%%%%%%%%%%%%%%%%%%%%%
\section{SUMMARY AND DISCUSSION\label{conclusions}}
%%%%%%%%%%%%%%%%%%%%%%%%%%%%%%%%%%%%%%%%%%%%%%%%%%%%%%%%%%%%%%%%%%%%%%%%%%

We have presented a high resolution spectroscopic analysis of 35 RGB
stars in the GC M22 from an heterogeneous sample of data, homogeneously
analyzed.
We have confirmed and extended the results of M09 that M22 hosts two
groups of stars whose mean [Fe/H] metallicities differ by 0.15~dex 
(see also in DC09).
These two groups turn out to have different chemical properties.
First, they show a different $s$/$r$ abundance ratio.
The two groups appear to be homogeneous in the \rpro\ element Eu,
that is $\Delta^{rich}_{poor}$[Eu/Fe]~$\sim$0.
However, the $s$/$r$ distribution, clearly traced by the [La/Eu]
abundance ratio, appears to be bimodal, with the \sri\ stars having
La/Eu ratios about 2.5 times larger than in the \spo\ ones,
or $\Delta^{rich}_{poor}$[La/Eu]~$\sim$0.4.
A bimodal split between the \sri\ and \spo\ of about the same amount
with respect to Eu is observed also in the other \spro\ dominated
species (Ba, Y, Nd, Zr).
This demonstrates that the stars with higher metallicity are also more
enriched in material processed through $s$ processes.

Since the most obvious M22 abundance anomaly is the spread
in \spro\ abundances, it is good to re-emphasize that [Eu/Fe] remains
constant, within observational errors, independent of the Fe abundance.
It is generally believed that massive stars are responsible for the
Fe abundance and $\alpha$ elements.
However, only a subset of these same stars appear to be responsible
also for the production of \rpro\ material.
This can be seen in large-sample abundance surveys of metal-poor
stars, which show relatively small star-to-star scatter in [$\alpha$/Fe]
ratios but an enormous range in [Eu/Fe] ratios (see the summary
of many studies in Figure~14 of Sneden, Cowan, \& Gallino 2008).
The constancy of [Eu/Fe] points M22 to a common ratio of
\rpro-donating massive stars to Fe and $\alpha$-process stars in
the ab initio IMF, irrespective of the Fe-metallicity of the two groups.
This constancy is not easily achieved, and may require fine tuning of
the evolutionary scenarios for M22.

Evidence for a small increase in [Cu/Fe] with increasing metallicity
has been detected for our sample of stars: \sri\ stars have higher [Cu/Fe],
while their [Zn/Fe] values are essentially the same at all metallicities.
This hints at an \spro\ contribution from the weak component in the 
production of Cu, but the relatively large observational errors 
associated with [Cu/Fe] suggests caution in this interpretation.

The relative abundances of the $\alpha$ elements (Si and Ti) with respect
to Fe are constant, within observational errors, with metallicity.
The curious trend of larger [Ca/Fe] ratios by 
$\Delta^{rich}_{poor}$[Ca/Fe]~$\sim$~0.1 in the higher metallicity
M22 stars found in M09, is confirmed here.
Previous work, including Norris \& Freeman (1983) and Lehnert \etal\ 
(1991), saw a range in [Ca/H] of about 0.3-0.4 dex.
Our observed range in [Ca/H] is larger than the one in [Fe/H], and is 
comparable to what was reported in Norris \& Freeman (1983) and Lehnert
\etal\ (1991).
This increase in [Ca/Fe] with [Fe/H] is comparable to what is seen in
$\omega$~Cen over a similar [Fe/H] range (see two upper left panels of 
Fig.~12 of Johnson \& Pilachowski 2010).  
The difference in the mean [Ca/Fe] in $\omega$~Cen between the metal-poor 
and the metal-intermediate groups from Johnson \& Pilachowski is 0.08 dex 
(with a small mean error due to the large sample), which is very similar 
to the ~0.1 dex difference we see in M22.

Among the light $p$-capture elements, M22 stars show the same sort
of Na-O anticorrelation and Na-Al correlations that have been
extensively cataloged in mono-metallic clusters (e.g. Carretta \etal\
2009b).
This indicates that the Ne$\rightarrow$Na and the
Mg$\rightarrow$Al conversions have been active in M22.
The Na-O anticorrelation is present in each $s$-group, with \sri\ and
\spo\ stars spanning a similar range in [O/Fe], but a different range
in [Na/Fe], i.e. the average [Na/Fe] at a given [O/Fe] is higher
in the \sri\ stars than the \spo\ by about 0.2~dex.
Carbon and nitrogen reveal the typical anticorrelation expected from
extensive CN-cycle processing, but this becomes apparent only when we
separate stars in the two $s$-groups.
On average, the \sri\ stars have higher C, N, and Na abundances, while
they have similar O and Mg.
Hence, the average overall CNO abundance between the two $s$-groups
differs by at least a factor of two.

The complexity of the chemical properties of M22 also reflects on the CMD.
A double RGB is visible when using the $m_{1}$ Str\"omgren index, sensitive
to the CN bands and to metallicity (Richter \etal\ 1999):
the lower [Fe/H] \spo\ stars populate a sequence on the blue side, while
the higher [Fe/H] \sri\ ones obviously occupy a redder branch.
A bimodal distribution is visible also on the SGB (Piotto 2009, M09),
likely associated with the double RGB.
We thus expect (but cannot prove with our data) that M22 SGB stars have
the same bimodality in \spro\ elements and Fe exhibited by the giants
studied in this work.
If true, the overall CNO differences will play the dominant role
in producing the SGB split, since M09 demonstrated that simple
Fe-peak metallicity variations are not sufficient to generate enough of
the observed photometric breadth in this region of the colour-magnitude
diagram.
Cassisi \etal\ (2008) and Ventura \etal\ (2009) have suggested that
a bulk CNO abundance difference can account for a similar SGB split
in NGC1851 (Milone \etal\ 2008).
However, spectroscopic results for this cluster are contradictory,
as evidence for CNO variations have been found by Yong \etal\ (2009),
but not by Villanova, Geisler, \& Piotto (2010).

Our work highlights the peculiarity of M22 among GCs.
We consider "{\it normal}" GCs to be those that are:
\textit{(i)} essentially mono-metallic, i.e., all their stars appear
             to have the same [Fe/H];
\textit{(ii)} chemically homogeneous in the heavy elements; but
\textit{(iii)} chemically inhomogeneous only in the light element
             abundances, as revealed by variations in the CH, CN,
             NH bands, and in the O, Na, Al, and Mg abundances.
Nearly all GCs have these characteristics. 
However, in M22, in addition to the O-Na anti-correlation there is a 
spread in the heavier elements, a characteristic that is seen only in a 
small number of other systems. 
Such systems include $\omega$~Cen, where the range in heavy elements is 
large and well established, the Galactic Bulge cluster Ter~5 
(Ferraro \etal\ 2009), M54, the central star cluster of the Sagittarius
dwarf (Sarajedini \& Layden 1995, Bellazzini \etal\ 2008, Carretta
\etal\ 2010a), the outer halo cluster NGC~2419 (Cohen \etal\ 2010), and 
perhaps NGC~1851 (Carretta \etal\ 2010b). 
The presence of heavy element abundance ranges in these systems 
necessarily means their nucleosynthetic history must be more complicated 
than for ``normal'' GCs, though whether it is an extension or a 
different process remains unclear.
Current scenarios for explaining the abundance anomalies in ``normal''
GCs argue that polluters from
a first stellar generation release into the intra-cluster medium large
amounts of material from which a second generation could form.
The candidate polluters are those expected to undergo the chemical
processes responsible for the observed enrichment in Na/N/Al and
depletions in O (and in some cases in Mg).
Candidate first-generation element donors include intermediate
mass asymptotic giant branch (AGB) stars (D'Antona \& Caloi 2004),
fast rotating massive stars (Decressin \etal\ 2007),
and/or massive binaries (de~Mink \etal\ 2009).
Note also that Marcolini \etal\ (2009) have proposed an alternative
scenario in which the first generation is the one with enhanced Na
and depleted O, at odds with what is proposed in most models.

It is difficult to fit M22's chemical properties into any of these
cluster enrichment history proposals; its history has been more
complicated than normal GCs. 
The pattern of differences in [Fe/H], in the \spro\ dominated elements, 
and in [Ca/Fe] in M22 resembles the case of $\omega$~Cen, albeit in M22 
the range in [Fe/H] is more than a factor of 20 lower. 
However, the M22 \spro\ abundances apparently differ from those
in $\omega$~Cen. 
In the latter, the \spro\ abundance ratios rise seemingly monotonically 
with [Fe/H] before reaching a plateau at constant [\spro/Fe] 
(Norris \& Da Costa 1995, Smith \etal 2000, Marino \etal 2011,
Stanford \etal\ 2010, Johnson \& Pilachowski 2010). 
However, in M22, while the range in [\spro/Fe] is comparable to that 
in $\omega$~Cen, our results do
not suggest a monotonic increase of [La/Fe] with [Fe/H], rather there
appears to be an overlap, possibly due to observational errors, in [Fe/H]
values between the \spo\ and \sri\
groups, with the location of the metal-rich, \spo\ star II-31 could being
particularly striking.
These differences may suggest different nucleosynthetic histories for the two
clusters, and could contrast with that proposed by 
Da Costa \& Marino (2010)
who concluded  that the $s$ enrichment processes in $\omega$~Cen and M22 were
similar, at least in the [Fe/H] range common to both clusters.
Indeed, aside from the spread in [Fe/H] present in the two \spro\ groups, 
and their difference in [Ca/Fe] values, it appears, considering the 
[\spro/Fe] abundance ratios and those for other heavy elements in each 
group as constant, that two groups each behave separately in a similar 
way to normal mono-metallic GCs.

Interpretation of the M22 abundance pattern would be much easier without
the need to account for the Ca, Fe and \spro\ variations.
The presence of a group of stars with higher [Fe/H] that also have little
change in Eu/Fe and most $\alpha$/Fe ratios could simply argue that multiple
episodes of core-collapse supernova (SNII) played a role in the evolution of
M22.
The larger [\spro/Fe] abundances in the higher metallicity M22 stars
is a significant complication.
In current self-enrichment models, the only ways to account for
these observations is to consider an unique source for Fe and \spro\
elements, or a fine tuning in the times of accumulation of the material
from which successive generations form or not evolving as an isolated system
so that external gas flows can contribute to the enrichment processes.

In the Sun, the \spro\ contribution is mainly due to two components:
\textit{(i)} the ``main'' component attributed to low mass AGB stars
($\sim$2-4~M$_{\odot}$; Busso, Gallino, \& Wasserburg 1999),
and \textit{(ii)} the ``weak'' component attributed to massive stars
(Raiteri \etal\ 1993, and references therein).
If the \spro\ enrichment in the \sri\ stars in M22 is due to the
main component, this would imply a relatively large difference in
age among the \sri\ and \spo\ stars, since the low mass AGB stars
evolve in times of the order of some Gyrs.
Low mass AGB stars are also expected to increase the total CNO
abundances, which would be consistent with our results.
A major \spro\ contribution from the weak component would be consistent
with a much faster evolution of the cluster, and with a smaller age
difference between the $s$-groups.
Note that the weak component mainly produces the lighter nuclei in
the $s$-chain, like \iso{58}{Fe}, \iso{63}{Cu}, and \iso{65}{Cu}.
Since, as suggested by Sneden \etal\ (1991), much of the
Cu in metal-poor stars can be produced in the weak component of the
$s$-processes, a possible Cu increase with Fe qualitatively
supports the idea that massive stars contribute to the pollution of
the intra-cluster medium in M22.
The advantage of this scenario is that the same stars can be the sources
for increases in the metallicity and \spro\ elements.
However, if this scenario is correct it would imply that massive stars
also produce enough
of the heavier \spro\ nuclei to quantitatively account for the
observed Ba, La, and Nd abundances in the \sri\ group of  M22 stars.
Simply following this scenario, after the evolution of massive stars
belonging to the metal/\spro-poor group (possibly from both
Na-poor/O-rich and Na-rich/O-poor populations), a second generation
slightly enhanced in Fe, and enriched in \spro\ elements formed
from the material created by these massive stars that end as SNII.
Then, the Na-poor/O-rich stars from the \sri\ stars form their own Na-O
anticorrelation (and Al-Na correlations), similarly to {\it normal} GCs.

Instructive at this point, is again the comparison with $\omega$~Cen, and
its constancy in [Cu/Fe].
In this extreme cluster the enormous elemental variations in $n$-capture have
been interpreted as due to the contributions from low-mass AGB stars via
\spro\ nucleosynthesis (e.g., Norris \& Da Costa 1995), thus the observed
constant values of [Cu/Fe] found by Cunha et al.\ (2002) do not fit a picture
in which Cu is produced in AGB stars.
This could mean that we shouldn't be looking to the mechanism that produces
the \spro\ abundance difference between the two M22 groups to explain
the [Cu/Fe]. In turn, this could suggest that in M22 we may well have evidence
that weak-$s$ in massive stars is contributing to the nucleosynthesis,
and possibly is not in $\omega$~Cen.
As previously said, we express caution with our results on [Cu/Fe]
abundances as the associated observational errors are relatively
large. However, the contribution from the weak-$s$ component in M22 could 
be supported also by the lower [heavy-$s$/light-$s$] ratio, traced by 
the [La/Y] abundances, in the \sri\ stars that have also higher [Cu/Fe]. 
Indeed, the weak-$s$ component produces more light-$s$, than heavier 
\spro\ elements.

As an alternative scenario,
we could suppose the following sequence of events:
1) SNII in the cluster explode and expel material far from the cluster
center at high velocity;
2) intermediate AGB stars pollute the medium of material enriched
in Na/N/Al and depleted in O, this material goes into the cluster
central region via cooling flow (as predicted by D'Ercole \etal\ 2008),
hence a second generation of stars (Na-rich, O-poor) formed;
3) at the end, low mass AGB stars evolve and expelled material
enriched in \spro-elements.
A second cooling flow involves the material ejected from low mass
AGB and the one expelled by the first SNII.
The latter should need  a longer time to be re-attracted
towards the center of the cluster because it was ejected at
higher velocity.
The material from low mass AGB and SNII mix together in the center
of the cluster, and another star formation event occurs forming
stars enriched in $s$ elements and Fe at the same time.

These evolutionary scenarios require special circumstances to
occur for M22 but not for the vast majority of GCs.
Another qualitative attempt to explain M22's abundance set is to
suppose that the present M22 is composed of the merger of two originally
clusters.
Of course, in this hypothesis we are assuming that any spread is present
in the [Fe/H] abundances in each \spro\ group, and hence the 
two $s$-groups being mono-metallic.
This is attractive because it does not require now-departed members of
the more metal-poor group to have been responsible for the creation
of the more metal-rich group.
This idea would probably fail immediately if no normal GCs could be found
with the chemical mixes of M22's two stellar groups.
However the well-studied M5 (Ivans \etal\ 2001,
Ram\'irez \& Cohen 2003) and M4 (Ivans \etal\ 1999, Marino \etal\ 2008)
have relative abundance mixes that resemble those of the lower and
higher metallicity M22 groups, respectively.
The mean literature values for Ba and La in M5 range from $-$0.08 to 0.18, 
and 0.02 to 0.18, respectively. 
On the other hand, M4 has substantially higher values of these abundance
ratios: [La/Fe]~=~0.45, [Ba/Fe]~=~0.60 (Ivans \etal\ 1999); [Ba/Fe]~=~0.41 
(Marino \etal\ 2008); these are unusually high for GCs.
Additionally, the [Fe/H] metallicity difference between M5 and M4 is
about 0.1~dex, similar to the mean difference in the two M22 groups.
However, that spread is not a requirement of the cluster merger scenario.
Finally, the mean Na and Al abundances are slightly higher in M4 than in M5.
A significant uncertainty for the cluster merging idea might be the
unknown probability of such a merger to have happened early in our Galaxy's
history.
The merger probability of two GCs in the field halo is likely to be very small
given the volume, but
that may not true in a dwarf galaxy/merger object - there the clusters exist
in a much smaller volume, the relative velocities are lower and maybe
dynamical friction can bring two clusters together in the center, with the
dwarf galaxy subsequently disrupted (see Bekki 2010, and
references therein).

Another GC has been recently interpreted in cluster merging
hypothesis: NGC~1851 (Carretta et al. 2010b). Like M22,
there is a bimodality in $s$-process elements (Yong \& Grundahl 2008),
a possible spread in [Fe/H] metallicity and [Ca/H], a segregation
of stars in color-magnitude diagram quantities that correlates
with abundance variations (Milone \etal\ 2008, Han \etal\ 2009).

Our findings show that M22 represents an important piece in the
understanding of GC evolution.
It shares similarities both with {\it normal} mono-metallic GCs and
with the most extreme case of $\omega$~Cen, and thus may be a
bridge to better understanding what has made $\omega$~Cen so unique.
A full understanding of the M22 chemical evolution should be important
in shedding light on the multiple stellar population phenomenon in GCs.

\begin{acknowledgements}
We are grateful for the observing time allocated to this program by
the directors and telescope allocation committees of several observatories.
GW and CS thank the Padua Astronomical Observatory for its hospitality
while parts of this paper were being completed.
We thank Melike Af\c{s}ar, Evan Kirby, and Ian Roederer for comments
and suggestions on the manuscript,
B. Plez \& V. Smith for providing their linelists for molecular bands,
K. Lind and M. Bergemann for useful discussions on NLTE
  abundances, and the referee for his/her suggestions that have improved the manuscript.
This publication makes use of data products from: (a) the Two Micron All
Sky Survey, which is a joint project of the University of Massachusetts
and the Infrared Processing and Analysis Center/California Institute
of Technology, funded by the National Aeronautics and Space
Administration and the National Science Foundation; and
(b) the Canadian Astronomy Data Centre, operated by the National
Research Council of Canada with the support of the Canadian Space Agency.
This research has made use
of the SIMBAD database, operated at CDS, Strasbourg, France.
We gratefully acknowledge the sources of financial support for this
study:
to AFM, APM, and GP for PRIN2007 (grant prot. n. 20075TP5K9);
to MZ for Fondecyt Regular 1085278 and the Mideplan Milky Way
Millennium Nucleus P07-021-F; to CS for several grants from the US
National Science Foundation, currently AST~09-08978; and to JEN for
Australian Research Council grant DP0984924 for studies of the
populations of Galactic stellar systems.
\end{acknowledgements}
%
%--------------------------------------------------------------------
%
\bibliographystyle{aa}

\begin{table*}
\begin{center}
\caption{M22 Targets: Instruments, Positions, and Photometric Data.\label{tab-datastars}}
\begin{tabular}{crcccrrrrrrr}
\hline\hline
ID\tablefootmark{a}&ID(M09)&SOURCE\tablefootmark{b}&$\alpha$(2000)&$\delta$(2000)&$B$\tablefootmark{c}&$V$\tablefootmark{c}&$I$\tablefootmark{c}&$K$\tablefootmark{d}&$b-y$\tablefootmark{e}&$m_1$\tablefootmark{e}&$A_V$\\
\hline
 I-12  & 200031 &AAT, MCD,UVES      & 18:36:27.36 & $-$23:51:26.1 & 13.177 & 11.672 & 10.000  &  7.893 & 0.999 &    0.173 &  0.998 \\
 I-27  & 200083 &APO, UVES     & 18:36:30.58 & $-$23:52:48.7 & 13.662 & 12.393 & 10.903  &  8.952 & 0.480 & $-$0.001 &  1.017 \\
 I-36  &        &   LICK       & 18:36:30.07 & $-$23:53:37.5 & 13.342 & 11.931 & 10.320  &  8.337 & 0.961 &    0.120 &  1.024 \\
 I-37  &        &   LICK       & 18:36:30.02 & $-$23:53:50.1 & 13.450 & 12.008 & 10.372  &  8.279 & 0.969 &    0.148 &  1.025 \\
 I-53  &        &   UVES       & 18:36:38.36 & $-$23:54:01.3 & 14.051 & 12.690 & 11.155  &  9.207 & 0.910 &    0.178 &  1.068 \\
 I-57  &        &AAT, APO      & 18:36:38.27 & $-$23:53:13.0 & 13.578 & 11.981 & 10.290  &  8.089 & 1.039 &    0.371 &  1.034 \\
 I-80  &     88 &   UVES       & 18:36:36.06 & $-$23:50:16.1 & 13.910 & 12.527 & 11.011  &  9.064 & 0.900 &    0.251 &  1.066 \\
 I-85  & 200080 &   UVES       & 18:36:32.82 & $-$23:51:10.9 & 13.747 & 12.473 & 10.978  &  9.117 & 0.867 &    0.065 &  1.083 \\
 I-86  & 200068 &AAT, UVES     & 18:36:32.13 & $-$23:51:31.4 & 13.749 & 12.328 & 10.689  &  8.660 & 0.969 &    0.103 &  1.064 \\ 
 I-92  &        &   LICK       & 18:36:33.28 & $-$23:52:32.5 & 13.141 & 11.561 &  9.828  &  7.654 & 1.050 &    0.232 &  1.019 \\
 II-1  &        &   LICK       & 18:36:23.89 & $-$23:52:43.5 & 13.585 & 12.043 & 10.367  &  8.204 & 1.003 &    0.321 &  1.089 \\
II-31  &        &   LICK       & 18:36:09.93 & $-$23:52:52.6 & 13.436 & 11.927 & 10.234  &  8.124 & 1.015 &    0.210 &  1.026 \\
II-96  &        &    MCD       & 18:36:17.24 & $-$23:54:11.3 & 13.134 & 11.604 &  9.879  &  7.786 & 1.033 &    0.219 &  1.042 \\
II-104 &     71 &   UVES       & 18:36:07.72 & $-$23:50:55.4 & 13.764 & 12.336 & 10.696  &  8.658 & 0.959 &    0.124 &  1.083 \\
III-3  & 200006 &APO, MCD, UVES& 18:36:17.51 & $-$23:54:26.2 & 12.912 & 11.107 &  9.188  &  6.783 & 1.175 &    0.563 &  1.039 \\
III-12 &        &AAT, MCD      & 18:36:14.26 & $-$23:54:31.1 & 13.236 & 11.540 &  9.723  &  7.401 & 1.106 &    0.452 &  1.024 \\
III-14 &        &AAT, LICK     & 18:36:15.10 & $-$23:54:54.6 & 12.964 & 11.134 &  9.145  &  6.743 & 1.201 &    0.445 &  1.034 \\
III-15 &        &   LICK       & 18:36:15.61 & $-$23:55:01.2 & 13.057 & 11.362 &  9.484  &  7.138 & 1.106 &    0.403 &  1.007 \\
III-25 & 200104 &   UVES       & 18:36:20.16 & $-$23:55:53.9 & 13.873 & 12.688 & 11.207  &  9.397 & 0.850 &    0.026 &  1.014 \\
III-33 &     61 &   UVES       & 18:36:20.13 & $-$23:56:45.4 & 13.640 & 12.249 & 10.632  &   8.618  & 0.964 &    0.117 &  1.018 \\
III-35 & 200076 &   UVES       & 18:36:20.51 & $-$23:56:24.4 & 13.738 & 12.404 & 10.828  &  8.854 & 0.930 &    0.098 &  1.042 \\
III-47 &        &    APO       & 18:36:14.80 & $-$23:55:15.7 & 13.699 & 12.385 & 10.848  &  8.979 & 0.904 &    0.128 &  0.992 \\
III-50 &    224 &   UVES       & 18:36:13.54 & $-$23:54:54.5 & 14.699 & 13.493 & 12.066  & 10.319 & 0.819 &    0.115 &  0.981 \\
III-52 & 200025 &LICK, UVES    & 18:36:10.18 & $-$23:54:21.8 & 13.238 & 11.526 &  9.732  &  7.459 & 1.081 &    0.549 &  1.016 \\
III-96 &        &    AAT       & 18:36:02.20 & $-$23:56:50.1 & 13.511 & 12.138 & 10.575  &  8.583 &  ...  &     ...  &  0.996 \\
IV-20  &     51 &AAT, UVES     & 18:36:25.99 & $-$23:55:58.3 & 13.622 & 12.071 & 10.383  &  8.164 & 1.021 &    0.336 &  1.036 \\
IV-59  & 200043 &   UVES       & 18:36:32.16 & $-$23:56:03.9 & 13.408 & 11.927 & 10.233  &  8.152 & 1.002 &    0.144 &  1.139 \\
IV-68  &    221 &   UVES       & 18:36:32.97 & $-$23:54:59.0 & 14.682 & 13.490 &  ...    & 10.222 & 0.833 &    0.041 &  1.065 \\
IV-76  &        &    AAT       & 18:36:35.41 & $-$23:54:39.9 & 13.709 & 12.299 & 10.790  &  8.852 & 0.916 &    0.289 &  1.066 \\
IV-88  &        &    AAT       & 18:36:44.97 & $-$23:54:57.3 & 13.718 & 12.228 & 10.608  &  8.500 &  ...  &     ...  &  1.065 \\
IV-97  & 200083 &   UVES       & 18:36:41.06 & $-$23:58:18.9 & 12.799 & 11.043 &  9.065  &  6.759 &  ...  &     ...  &  1.102 \\
IV-102 &        &AAT, LICK, MCD& 18:36:36.19 & $-$23:59:38.9 & 12.881 & 11.051 &  9.111  &  6.769 &  ...  &     ...  &  0.986 \\
  V-2  &        &   LICK       & 18:36:28.02 & $-$23:55:01.6 & 13.260 & 11.498 &  9.674  &  7.276 & 1.160 &    0.494 &  1.036 \\
    C%\tablefootmark{f}
       &        &    MCD       & 18:36:10.21 & $-$23:48:44.0 & 13.309 & 11.309 &  9.253  &  6.737 &  ...  &     ...  &  0.989 \\
 C513%\tablefootmark{f}
       &        &    APO       & 18:35:50.02 & $-$23:57:40.6 & 13.052 & 11.356 &  9.567  &  7.359 &  ...  &     ...  &  1.072 \\
\hline
\end{tabular}
\end{center}

\tablefoottext{a}{When available, we used the identification scheme of Arp \& Melbourne (1959) or its extension by Lloyd-Evans (1975), with Roman numerals I-IV for the quadrants and V for the center. If no other name is available, we used identifications from Cudworth (1986), preceded by C. These names are entered into the SIMBAD database ({\sf http://simbad.u-strasbg.fr/simbad/}) as `NGC 6656 abbb'', where ``a'' is the roman numeral and ``bbb'' is the number; for example, star IV-97 can be found as NGC 6656 4097 in SIMBAD.}\\
\tablefoottext{b}{See text for definitions.}\\
\tablefoottext{c}{Stetson photmetric database.}\\
\tablefoottext{d}{2MASS database.}\\
\tablefoottext{e}{Richter \etal\ (1999).}\\

\end{table*}

\begin{table*}
\begin{center}
\caption{Atmospheric Paramenters.\label{tab-model}}
\begin{tabular}{ccrr cr}
\hline\hline
Star     &Source  &\teff   &\logg &\vmicro&[Fe/H]\\      
          &        &[K]     &      &[km s$^{-1}$] & \\
\hline
  I-12 &     MCD &     4260&     0.45&     1.55&        $-$1.90\\
        &     AAT &     4300&     0.70&     1.70&        $-$1.86\\
        &    UVES &     4300&     0.75&     1.55&        $-$1.85\\
        &     AVG.&     4285&     0.65&     1.60&        $-$1.87\\
   I-27 &     APO &     4420&     1.40&     1.60&        $-$1.82\\
        &    UVES &     4490&     1.46&     1.65&        $-$1.63\\
        &     AVG.&     4455&     1.45&     1.60&        $-$1.73\\
   I-36 &    LICK &     4400&     0.80&     1.70&        $-$1.89\\
   I-37 &    LICK &     4370&     0.90&     1.55&        $-$1.73\\
   I-53 &    UVES &     4500&     1.35&     1.55&        $-$1.74\\
   I-57 &     APO &     4300&     1.20&     1.70&        $-$1.65\\
        &     AAT &     4250&     0.90&     1.65&        $-$1.62\\
        &     AVG.&     4275&     1.05&     1.65&        $-$1.64\\
   I-80 &    UVES &     4460&     1.15&     1.55&        $-$1.70\\
   I-85 &    UVES &     4600&     1.00&     1.45&        $-$1.81\\
   I-86 &     AAT &     4420&     1.10&     1.20&        $-$1.80\\
        &    UVES &     4500&     1.30&     1.50&        $-$1.84\\
        &     AVG.&     4460&     1.20&     1.35&        $-$1.82\\
   I-92 &    LICK &     4240&     0.75&     1.55&        $-$1.75\\
   II-1 &    LICK &     4300&     0.75&     1.50&        $-$1.66\\
  II-31 &    LICK &     4380&     1.20&     1.65&        $-$1.65\\
  II-96 &     MCD &     4400&     1.00&     2.10&        $-$1.82\\
 II-104 &    UVES &     4460&     1.15&     1.45&        $-$1.76\\
  III-3 &     MCD &     4000&     0.30&     2.25&        $-$1.72\\
        &     APO &     4010&     0.40&     2.25&        $-$1.78\\
        &    UVES &     3990&     0.20&     2.10&        $-$1.66\\
        &     AVG.&     4000&     0.30&     2.20&        $-$1.72\\
 III-12 &     MCD &     4150&     0.70&     1.95&        $-$1.69\\
        &     AAT &     4220&     1.25&     2.00&        $-$1.61\\
        &     AVG.&     4185&     1.00&     1.95&        $-$1.65\\
 III-14 &    LICK &     4010&     0.40&     2.15&        $-$1.84\\
        &     AAT &     4050&     0.30&     2.15&        $-$1.80\\
        &     AVG.&     4030&     0.35&     2.15&        $-$1.82\\
 III-15 &    LICK &     4070&     0.40&     1.85&        $-$1.82\\
 III-25 &    UVES &     4700&     1.35&     1.75&        $-$1.92\\
 III-33 &    UVES &     4430&     1.05&     1.70&        $-$1.78\\
 III-35 &    UVES &     4500&     1.25&     1.35&        $-$1.83\\
 III-47 &     APO &     4600&     1.20&     2.00&        $-$1.82\\
 III-50 &    UVES &     4700&     1.70&     1.45&        $-$1.76\\
 III-52 &    LICK &     4050&     0.50&     1.70&        $-$1.63\\
        &    UVES &     4100&     0.65&     1.80&        $-$1.62\\
        &     AVG.&     4075&     0.60&     1.75&        $-$1.63\\
 III-96 &     AAT &     4480&     1.30&     1.65&        $-$1.86\\
  IV-20 &     AAT &     4320&     1.05&     1.70&        $-$1.65\\
        &    UVES &     4260&     0.90&     1.60&        $-$1.63\\
        &     AVG.&     4290&     1.00&     1.65&        $-$1.64\\
  IV-59 &    UVES &     4400&     1.00&     1.70&        $-$1.77\\
  IV-68 &    UVES &     4750&     1.65&     1.20&        $-$1.75\\
  IV-76 &     AAT &     4730&     1.50&     2.30&        $-$1.63\\
  IV-88 &     AAT &     4400&     1.20&     1.70&        $-$1.62\\
  IV-97 &    UVES &     4000&     0.05&     2.00&        $-$1.94\\
 IV-102 &    LICK &     4020&     0.25&     2.15&        $-$1.96\\
        &     MCD &     4050&     0.10&     2.35&        $-$1.95\\
        &     AAT &     3990&     0.20&     2.15&        $-$2.01\\
        &     AVG.&     4020&     0.20&     2.20&        $-$1.97\\
      C &     MCD &     3960&     0.30&     2.25&        $-$1.69\\
   C513 &     APO &     4100&     0.40&     1.65&        $-$1.86\\
    V-2 &    LICK &     4130&     0.65&     1.75&        $-$1.57\\
\hline
\end{tabular}
\end{center}

\end{table*}

\clearpage

\longtab{4}{
\begin{longtable}{lccccc}
\caption{Sensitivity of abundance ratios on atmopheric parameters changes. \label{tab-errors}}\\
\hline\hline
[X/Fe]\tablefootmark{a}&$\Delta$(\teff) &$\Delta$(\logg) &$\Delta$([A/H]) &$\Delta$(\vmicro) &total\tablefootmark{b}\\
\hline
\endfirsthead
\caption{continued.}\\
\hline\hline
[X/Fe]\tablefootmark{a}&$\Delta$(\teff) &$\Delta$(\logg) &$\Delta$([A/H]) &$\Delta$(\vmicro) &total\tablefootmark{b}\\\\
\hline
\endhead

\endfoot
\multicolumn{6}{c}{AAT: IV-20} \\
             & $\pm$60 K& $\pm$0.15  & $\pm$0.10 & $\pm$0.12 km/s &   \\
O    & $\pm$0.00 & $\pm$0.05 & $\mp$0.06 & $\pm$0.00 & 0.08 \\
Na   & $\pm$0.05 & $\mp$0.02 & $\mp$0.10 &$\mp$0.03 & 0.12 \\
Mg   & $\mp$0.06 & $\mp$0.01 & $\pm$0.00 & $\pm$0.03 & 0.07 \\
Al   & $\mp$0.05 & $\mp$0.01 & $\pm$0.00 & $\pm$0.04 & 0.06 \\
Si   & $\mp$0.09 & $\pm$0.02 & $\pm$0.02 & $\pm$0.05 & 0.11 \\
Ca   & $\mp$0.03 & $\mp$0.02 & $\pm$0.00 &$\pm$0.00 & 0.04 \\
$\rm {Ti_{I}}$& $\pm$0.00 & $\mp$0.02 & $\pm$0.00 & $\pm$0.03 & 0.04 \\
%$\rm {Ti_{II}}$& \nodata & \nodata & \nodata & \nodata  & \nodata & \nodata \\
$\rm {Fe_{I}}$&$\pm$0.11  & $\pm$0.00 & $\mp$0.01 & $\mp$0.05 & 0.12 \\
$\rm {Fe_{II}}$&$\mp$0.05 & $\pm$0.07 & $\pm$0.02 & $\mp$0.02 & 0.09 \\
Cu   &$\pm$0.10  &$\pm$0.00  &$\mp$0.10  &$\pm$0.00  & 0.14  \\
% Zn           & \nodata &\nodata  & \nodata & \nodata & \nodata \\
Ba   & $\mp$0.07 & $\pm$0.04 & $\pm$0.04 &$\mp$0.05 & 0.10 \\
La   & $\pm$0.02 & $\pm$0.07 & $\mp$0.06 &$\pm$0.00 & 0.09 \\
\hline
\\
\multicolumn{6}{c}{APO: I-57} \\
             & $\pm$100 K& $\pm$0.17  & $\pm$0.10 & $\pm$0.13 km/s &   \\
O    & $\pm$0.03 & $\pm$0.07 & $\mp$0.05 & $\pm$0.00 & 0.09 \\
Na   & $\pm$0.12 & $\mp$0.00 & $\mp$0.10 &$\pm$0.00 & 0.16 \\
Mg   & $\mp$0.05 & $\mp$0.03 & $\mp$0.03 & $\pm$0.02 & 0.07 \\
Al   & $\mp$0.05 & $\mp$0.01 & $\mp$0.01 & $\pm$0.04 & 0.07 \\
Si   & $\mp$0.13 & $\pm$0.00 & $\mp$0.01 & $\pm$0.03 & 0.13 \\
Ca   & $\pm$0.00 & $\mp$0.04 & $\mp$0.03 &$\mp$0.02 & 0.05 \\
$\rm {Ti_{I}}$&$\pm$0.09 & $\mp$0.03 &$\mp$0.03& $\pm$0.00 & 0.10 \\
$\rm {Ti_{II}}$&$\mp$0.12& $\pm$0.02 &$\pm$0.00& $\mp$0.04 & 0.13 \\
$\rm {Fe_{I}}$&$\pm$0.13  & $\pm$0.01 & $\pm$0.01 & $\mp$0.04 & 0.14 \\
$\rm {Fe_{II}}$&$\mp$0.08 & $\pm$0.07 & $\pm$0.09 & $\mp$0.01 & 0.14 \\
Cu   & $\pm$0.06  &$\pm$0.00  &$\mp$0.10 &$\pm$0.00  & 0.12 \\
Zn   & $\pm$0.05  &$\pm$0.02  &$\mp$0.10 &$\mp$0.06  & 0.13 \\
Y    & $\mp$0.11 & $\pm$0.03 & $\pm$0.00 &$\pm$0.01 & 0.11 \\
Zr   & $\pm$0.00 & $\pm$0.05 & $\mp$0.04 &$\pm$0.00 & 0.06 \\
Ba   & $\mp$0.10 & $\pm$0.04 & $\pm$0.03 &$\mp$0.06 & 0.13 \\
La   & $\pm$0.05 & $\pm$0.08 & $\mp$0.06 &$\pm$0.00 & 0.11 \\
Nd   & $\mp$0.10 & $\pm$0.04 & $\pm$0.01 &$\pm$0.02 & 0.11 \\
Eu   & $\pm$0.00 & $\pm$0.08 & $\mp$0.05 &$\pm$0.00 & 0.09 \\
\hline
\\
\multicolumn{6}{c}{LICK: II-31} \\
               & $\pm$70 K & $\pm$0.18 & $\pm$0.10 & $\pm$0.12 km/s &  \\
O              &$\pm$0.00  & $\pm$0.08 & $\mp$0.05 & $\pm$0.00 & 0.09 \\
Na           & $\pm$0.05 & $\mp$0.02 & $\mp$0.10 &$\pm$0.02 & 0.12 \\
Mg             &$\mp$0.03  & $\mp$0.01 & $\pm$0.02 & $\pm$0.02 & 0.04 \\
Al             &$\mp$0.01  &$\mp$0.01  & $\pm$0.01 & $\pm$0.03 & 0.03 \\
Si             &$\mp$0.08  & $\pm$0.02 & $\pm$0.01 & $\pm$0.03 & 0.09 \\
 Ca           & $\pm$0.02 & $\mp$0.01 & $\pm$0.03 &$\mp$0.02 & 0.04  \\
$\rm {Ti_{I}}$ &$\pm$0.03  & $\mp$0.01 & $\pm$0.01 & $\mp$0.01 & 0.03 \\
$\rm {Ti_{II}}$&$\mp$0.10  & $\pm$0.06 & $\mp$0.01 & $\mp$0.01 & 0.12 \\
$\rm {Fe_{I}}$ &$\pm$0.09  & $\pm$0.00 & $\mp$0.01 & $\mp$0.03 & 0.10  \\
$\rm {Fe_{II}}$&$\mp$0.06  & $\pm$0.08 & $\pm$0.02 & $\mp$0.01 & 0.10 \\
 Cu           & $\pm$0.04 &  $\pm$0.00 & $\mp$0.05 & $\pm$0.00 & 0.06 \\
 Zn           & $\pm$0.00 &  $\pm$0.02 & $\mp$0.05 & $\mp$0.02 & 0.06  \\
 Y            & $\mp$0.09 & $\pm$0.06 & $\mp$0.03 &$\mp$0.03 & 0.12 \\
 Zr           & $\pm$0.00 & $\pm$0.10 & $\mp$0.05 &$\pm$0.00 & 0.11 \\
 Ba           & $\mp$0.07 & $\pm$0.05 & $\pm$0.05 &$\mp$0.06 & 0.12 \\
 La           & $\pm$0.00 & $\pm$0.07 & $\mp$0.05 &$\pm$0.00 & 0.09 \\
 Nd           & $\mp$0.08 & $\pm$0.06 & $\pm$0.04 &$\pm$0.02 & 0.11 \\
 Eu           & $\pm$0.00 & $\pm$0.06 & $\mp$0.08 &$\pm$0.00 & 0.10 \\
\hline 
\\
\multicolumn{6}{c}{MCD: II-96} \\
              & $\pm$50 K & $\pm$0.10  & $\pm$0.10 & $\pm$0.12 km/s &  \\
 O    & $\pm$0.00 & $\pm$0.05 & $\mp$0.04 & $\pm$0.00 & 0.06 \\
Na   & $\pm$0.05 & $\mp$0.00 & $\mp$0.10 &$\pm$0.00 & 0.11 \\
 Mg   & $\pm$0.02 & $\mp$0.02 & $\pm$0.00 & $\pm$0.02 & 0.03 \\
 Al   & $\mp$0.03 & $\mp$0.01 & $\pm$0.00 & $\pm$0.02 & 0.04 \\
 Si   & $\pm$0.05 & $\pm$0.00 & $\pm$0.02 & $\pm$0.02 & 0.06 \\
 Ca   & $\pm$0.01 & $\mp$0.01 & $\pm$0.00 &$\pm$0.02 & 0.02 \\
$\rm {Ti_{I}}$&$\mp$0.02 & $\mp$0.02 & $\pm$0.01 & $\pm$0.01 & 0.03 \\
$\rm {Ti_{II}}$&$\pm$0.09& $\pm$0.04 & $\pm$0.04 & $\pm$0.00 & 0.11 \\
$\rm {Fe_{I}}$&$\pm$0.07  & $\pm$0.00 & $\mp$0.02 & $\mp$0.03 & 0.08 \\
$\rm {Fe_{II}}$&$\mp$0.04 & $\pm$0.08 & $\pm$0.02 & $\mp$0.01 & 0.09 \\
 Cu   & $\pm$0.05 & $\pm$0.00 &$\mp$0.05  &$\mp$0.06 & 0.09 \\
 Zn   & $\pm$0.00 & $\pm$0.00 & $\mp$0.07 &$\mp$0.09 & 0.11  \\
 Y    & $\pm$0.07 & $\pm$0.04 & $\pm$0.04 &$\pm$0.01 & 0.09 \\
 Zr   & $\pm$0.00 & $\pm$0.05 & $\mp$0.05 &$\pm$0.00 & 0.07 \\
 Ba   & $\pm$0.04 & $\pm$0.04 & $\pm$0.04 &$\mp$0.06 & 0.09 \\
 La   & $\pm$0.00 & $\pm$0.05 & $\mp$0.05 &$\pm$0.00 & 0.07 \\
 Nd   & $\mp$0.06 & $\pm$0.04 & $\pm$0.04 &$\pm$0.02 & 0.08 \\
Eu   & $\pm$0.00 & $\pm$0.07 & $\mp$0.05 &$\pm$0.00 & 0.09 \\
\hline
\\
\multicolumn{6}{c}{UVES: I-53} \\
              & $\pm$50 K& $\pm$0.14  & $\pm$0.10 & $\pm$0.13 km/s& \\
 O        & $\pm$0.00       & $\pm$0.06 & $\mp$0.05 & $\pm$0.00 & 0.08\\
Na       &  $\mp$0.02      &  $\mp$0.01  &  $\mp$0.01  & $\pm$0.01& 0.03\\
 Mg      &  $\mp$0.03      &  $\pm$0.00 &  $\mp$0.01 & $\pm$0.01  &0.03\\
 Al       &  $\mp$0.03      &  $\mp$0.01  &  $\mp$0.01 & $\pm$0.02  &0.04\\
 Si       &   $\mp$0.05      &  $\pm$0.01  &  $\pm$0.01 & $\pm$0.01 & 0.05\\
 Ca      &   $\mp$0.01      &  $\mp$0.01  &  $\mp$0.01 & $\mp$0.01&0.02\\
$\rm {Ti_{I}}$  &  $\pm$0.04     &  $\mp$0.01   &  $\mp$0.02 & $\pm$0.00 &0.05\\
$\rm {Ti_{II}}$ &   $\pm$0.02     &  $\mp$0.01  &  $\mp$0.01 & $\mp$0.02  &0.03\\
$\rm {Fe_{I}}$   &  $\pm$0.07      & $\pm$0.00  &   $\mp$0.01 & $\mp$0.02 &0.07 \\
$\rm {Fe_{II}}$  &  $\mp$0.02      &  $\pm$0.05 &  $\pm$0.02 & $\mp$0.02 &0.06\\
Cu        & $\pm$0.05       & $\pm$0.00 &  $\mp$0.04  & $\mp$0.05&0.08\\
Zn        & $\pm$0.00      & $\pm$0.00   & $\mp$0.08  &$\mp$0.08  &0.11 \\
Y          &  $\pm$0.03      & $\mp$0.01  &   $\pm$0.00 & $\mp$0.02 &0.04\\
Zr         &  $\pm$0.03      & $\pm$0.00 &   $\pm$0.00 & $\pm$0.02  &0.04\\
Ba        &  $\pm$0.05      &  $\mp$0.01 &   $\pm$0.00 & $\mp$0.09 &0.10\\
La        &  $\pm$0.05      &  $\pm$0.07  & $\mp$0.03 & $\pm$0.02  &0.09\\
Nd       &  $\mp$0.08      &  $\pm$0.00  & $\pm$0.00 &  $\pm$0.04 &0.09\\
Eu        &  $\pm$0.02      &  $\pm$0.05  & $\mp$0.08 & $\pm$0.00  &0.10\\ 
\hline
\end{longtable}
\tablefoottext{a}{For $\rm {Fe_{I}}$ and $\rm {Fe_{II}}$ the [X/H]
  sensitivities are given; for and all remaining elements  we list the
  sensitivities for the abundance ratios [el/Fe].} 
\tablefoottext{b}{Quadratic sum of the contributions given by each atmospheric parameter.}
}

\longtab{5}{
\begin{longtable}{ccrrrrrrrrrrrrc}
\caption{Light Element Abundances\label{tab-ablight}}\\
\hline\hline
Star &Source &Fe\tablefootmark{a} &C\tablefootmark{b}&N\tablefootmark{c}&O&Na$_{\rm {\small LTE}}$&Mg&Al&Si&Ca&TiI&TiII&$\phantom{}^{12}$C/$\phantom{}^{13}$C&$s$-rich?\\
\hline
\endfirsthead
\caption{continued.}\\
\hline\hline
Star&Source&Fe\tablefootmark{a}&C\tablefootmark{b}&N\tablefootmark{c}&O&Na$_{\rm {\small LTE}}$&Mg&Al&Si&Ca&TiI&TiII&$\phantom{}^{12}$C/$\phantom{}^{13}$C&$s$-rich?\\
\hline
\endhead

\endfoot

  I-12 &     MCD &    $-$1.90&  $-$0.81&     0.98&     ... &     0.38&     0.27&     0.76&     0.46&     0.29&     0.20&     0.33&  4.00&     no \\
           &     AAT &    $-$1.86&     ... &     ... &     0.24&     0.39&     0.35&     0.66&     0.33&     0.21&     0.20&     ... &  ... &     no \\
           &    UVES &    $-$1.85&     ... &     ... &     0.18&     0.31&     0.34&     0.60&     0.44&     0.20&     0.22&     0.35&  ... &     no \\
           &     AVG. &   $-$1.87&  $-$0.81&     0.98&     0.21&     0.36&     0.32&     0.67&     0.41&     0.23&     0.21&     0.34&  ...&     no \\
   I-27 &     APO &    $-$1.81&     ... &     ... &     0.55&     0.32&     0.50&     0.49&     0.58&     0.32&     0.27&     0.41&  ... &    yes \\
           &    UVES &    $-$1.63&     ... &     ... &     0.52&     0.24&     0.34&     0.33&     0.44&     0.35&     0.22&     0.41&  ... &    yes \\
           &     AVG. &   $-$1.72&     ... &     ... &     0.54&     0.28&     0.42&     0.41&     0.51&     0.34&     0.25&     0.41&  ... &    yes \\
   I-36 &    LICK &    $-$1.89&     ... &     ... &     0.38&  $-$0.05&     0.50&     0.52&     0.42&     0.27&     0.10&     0.32&  ... &     no \\
   I-37 &    LICK &    $-$1.73&     ... &     ... &     0.37&     0.25&     0.18&     0.51&     0.54&     0.23&     0.11&     0.26&  ... &     no \\
   I-53 &    UVES &    $-$1.74&     ... &     ... &     0.35&     0.29&     0.51&     0.39&     0.45&     0.35&     0.25&     0.36&  ... &    yes \\
   I-57 &     APO &    $-$1.65&     ... &     ... &     0.59&     0.33&     0.34&     0.10&     0.53&     0.42&     0.20&     0.30&  ... &    yes \\
           &     AAT &    $-$1.62&     ... &     ... &     0.57&     0.29&     0.46&     0.27&     0.51&     0.38&     0.23&     ... &  ... &    yes \\
           &     AVG. &   $-$1.64&     ... &     ... &     0.58&     0.31&     0.40&     0.19&     0.52&     0.40&     0.22&     0.30&  ... &    yes \\
   I-80 &    UVES &    $-$1.70&  $-$0.40&     1.33&  $-$0.10&     0.65&     0.49&     0.84&     0.41&     0.35&     0.19&     0.28&  ... &    yes \\
   I-85 &    UVES &    $-$1.81&  $-$0.91&     1.03&  $-$0.03&     0.59&     0.22&     0.64&     0.46&     0.25&     0.18&     0.28&  ... &     no \\
   I-86 &     AAT &    $-$1.80&     ... &     ... &     0.40&     0.00&     0.38&     ... &     0.47&     0.34&     ... &     ... &  ... &     no \\
           &    UVES &    $-$1.84&     ... &     ... &     0.45&  $-$0.01&     0.23&     0.04&     0.35&     0.22&     0.23&     0.35&  ... &     no \\
           &     AVG. &   $-$1.82&     ... &     ... &     0.43&     0.00&     0.31&     0.04&     0.41&     0.28&     0.23&     0.35&  ... &     no \\
   I-92 &    LICK &    $-$1.75&     ... &     ... &     0.52&  $-$0.05&     0.32&     ... &     0.47&     0.23&     0.16&     0.35&  ... &     no \\
   II-1  &    LICK &    $-$1.66&     ... &     ... &     0.48&     0.11&     0.47&     0.44&     0.54&     0.37&     0.21&     0.43&  ... &    yes \\
  II-31 &    LICK &    $-$1.65&     ... &     ... &     0.47&  $-$0.11&     0.36&  $-$0.07&     0.33&     0.29&     0.13&     0.29&  ... &     no \\
  II-96 &     MCD &    $-$1.82&  $-$0.96&     0.75&     0.47&     0.29&     0.21&     0.40&     0.43&     0.21&     0.07&     0.26&  4.00&     no \\
 II-104&    UVES &    $-$1.76&  $-$0.60&     0.53&     0.37&  $-$0.20&     0.33&  $-$0.06&     0.43&     0.25&     0.20&     0.35&  ... &     no \\
  III-3  &     MCD &    $-$1.72&  $-$0.01&     0.93&     0.49&     0.36&     0.39&     0.14&     0.58&     0.26&     0.27&     0.15&  3.50&    yes \\
           &     APO &    $-$1.78&     ... &     ... &     0.50&     0.25&     0.34&     0.18&     0.62&     0.30&     0.44&     0.35&  ... &    yes \\
           &    UVES &    $-$1.66&     ... &     ... &     0.25&     0.31&     0.34&     0.20&     0.42&     0.34&     0.33&     0.38&  ... &    yes \\
           &     AVG.&    $-$1.72&  $-$0.01&     0.93&     0.41&     0.31&     0.36&     0.17&     0.54&     0.30&     0.35&     0.29&  ...&    yes \\
 III-12 &     MCD &    $-$1.69&  $-$0.26&     1.13&     0.40&     0.52&     0.43&     0.55&     0.33&     0.33&     0.17&     0.28&  4.00&    yes \\
           &     AAT &    $-$1.61&     ... &     ... &     0.50&     0.46&     0.30&     0.45&     0.43&     0.21&     0.25&     ... &  ... &    yes \\
           &     AVG. &   $-$1.65&  $-$0.26&     1.13&     0.45&     0.49&     0.37&     0.50&     0.38&     0.27&     0.21&     0.28&  ...&    yes \\
 III-14 &    LICK &    $-$1.84&     ... &     ... &     0.50&     0.14&     0.42&  $-$0.10&     0.53&     0.17&     0.26&     0.39&  ... &     no \\
           &     AAT &    $-$1.80&     ... &     ... &     0.46&     0.13&     0.40&     0.05&     0.42&     0.15&     0.24&     ... &  ... &     no \\
           &     AVG.&    $-$1.82&     ... &     ... &     0.48&     0.14&     0.41&  $-$0.03&     0.48&     0.16&     0.25&     0.39&  ... &     no \\
 III-15 &    LICK &    $-$1.82&     ... &     ... &     0.11&     0.54&     0.48&     0.63&     0.62&     0.33&     0.24&     0.40&  ... &     no \\
 III-25 &    UVES &    $-$1.92&     ... &     ... &     0.56&     0.02&     0.37&     0.27&     0.44&     0.32&     0.17&     0.31&  ... &     no \\
 III-33 &    UVES &    $-$1.78&  $-$0.55&     0.63&     0.34&  $-$0.17&     0.54&  $-$0.07&     0.40&     0.28&     0.19&     0.28&  ... &     no \\
 III-35 &    UVES &    $-$1.83&     ... &     ... &     0.19&     0.43&     0.54&     0.47&     0.37&     0.30&     0.20&     0.36&  ... &     no \\
 III-47 &     APO &    $-$1.82&     ... &     ... &     0.50&     0.30&     0.38&     0.41&     0.39&     0.31&     0.11&     0.18&  ... &     no \\
 III-50 &    UVES &    $-$1.76&  $-$0.10&     1.03&     0.35&     0.34&     0.56&     0.46&     0.44&     0.45&     0.27&     0.27&  ... &    yes \\
 III-52 &    LICK &    $-$1.63&     ... &     ... &     0.40&     0.30&     0.42&     0.11&     0.37&     0.35&     0.26&     0.23&  ... &    yes \\
        &    UVES &    $-$1.62&     ... &     ... &     0.49&     0.26&     0.44&     0.27&     0.48&     0.42&     0.38&     0.44&  ... &    yes \\
        &     AVG.&    $-$1.63&     ... &     ... &     0.45&     0.28&     0.43&     0.19&     0.43&     0.39&     0.32&     0.34&  ... &    yes \\
 III-96 &     AAT &    $-$1.86&     ... &     ... &     0.48&     0.01&     0.51&     ... &     0.30&     0.28&     0.25&     ... &  ... &     no \\
  IV-20 &     AAT &    $-$1.65&     ... &     ... &     ... &     0.80&     0.29&     0.82&     0.32&     0.33&     0.32&     ... &  ... &    yes \\
        &    UVES &    $-$1.63&  $-$0.61&     1.33&  $-$0.05&     0.73&     0.44&     0.74&     0.47&     0.40&     0.28&     0.36&  ... &    yes \\
        &     AVG.&    $-$1.64&  $-$0.61&     1.33&  $-$0.05&     0.77&     0.37&     0.78&     0.40&     0.37&     0.30&     0.36&  ... &    yes \\
  IV-59 &    UVES &    $-$1.77&     ... &     ... &     0.14&     0.59&     0.26&     0.75&     0.42&     0.29&     0.25&     0.34&  ... &     no \\
  IV-68 &    UVES &    $-$1.75&  $-$0.42&     0.63&     0.36&  $-$0.22&     0.29&  $-$0.15&     0.41&     0.33&     0.29&     0.25&  ... &     no \\
  IV-76 &     AAT &    $-$1.63&     ... &     ... &     0.56&     0.19&     0.31&     ... &     0.44&     0.22&     0.14&     ... &  ... &    yes \\
  IV-88 &     AAT &    $-$1.62&     ... &     ... &     0.02&     0.61&     0.17&     0.77&     0.48&     0.35&     0.26&     ... &  ... &    yes \\
  IV-97 &    UVES &    $-$1.94&  $-$0.50&     0.30&     0.40&  $-$0.02&     0.46&     0.07&     0.44&     0.24&     0.26&     0.42&  ... &     no \\
 IV-102 &    LICK &    $-$1.96&     ... &     ... &     0.44&     0.41&     0.45&     0.19&     ... &     0.33&     0.30&     0.27&  ... &     no \\
        &     MCD &    $-$1.95&  $-$0.46&     0.48&     0.41&     0.28&     0.42&     0.29&     0.42&     0.25&     0.21&     0.26&  3.00&     no \\
        &     AAT &    $-$2.01&     ... &     ... &     0.44&     0.40&     0.52&     ... &     0.49&     0.21&     0.33&     ... &  ... &     no \\
        &     AVG.&    $-$1.97&  $-$0.46&     0.48&     0.43&     0.36&     0.46&     0.24&     0.46&     0.26&     0.28&     0.27&  ...&     no \\
      C &     MCD &    $-$1.69&  $-$0.41&     1.10&     0.25&     0.68&     0.43&     0.58&     0.38&     0.37&     0.38&     0.20&  5.00&    yes \\
   C513 &     APO &    $-$1.86&     ... &     ... &     0.40&     0.03&     0.57&     ... &     0.40&     0.22&     0.14&     0.34&  ... &     no \\
    V-2 &    LICK &    $-$1.57&     ... &     ... &     0.15&     0.56&     0.21&     0.79&     0.40&     0.45&     0.30&     0.30&  ... &    yes\\

\hline
\end{longtable}
\tablefoottext{a}{For this column only, metallicities [Fe/H] are given.}

\tablefoottext{b}{For this and all remaining abundances, [el/Fe] values are given.}

\tablefoottext{c}{For the stars II-104, III-33, IV-68, IV-97, and IV-102, we were able to measure only upper limits for the nitrogen abundance.}
}

\longtab{6}{
\begin{longtable}{ccrrrrrrrrrc}
\caption{Heavy Element Abundances\label{tab-abheavy}}\\
\hline\hline
Star &Source &Fe\tablefootmark{a} &Cu\tablefootmark{b}  &Zn   &Y  &Zr  &Ba  &La  &Nd &Eu &$s$-rich?\\
\hline
\endfirsthead
\caption{continued.}\\
\hline\hline
Star &Source &Fe\tablefootmark{a} &Cu\tablefootmark{b}  &Zn   &Y  &Zr  &Ba  &La  &Nd &Eu &$s$-rich?\\
\hline
\endhead

\endfoot

  I-12 &     MCD &    $-$1.90&    $-$0.93&     0.13&  $-$0.20&  $-$0.05&     0.03&  $-$0.12&  $-$0.03&     ... &      no \\
        &     AAT &    $-$1.86&    $-$0.93&     ... &     ... &     ... &  $-$0.20&     0.03&     ... &     ... &      no \\
        &    UVES &    $-$1.85&    $-$1.00&     0.15&  $-$0.07&  $-$0.03&     0.01&  $-$0.07&     0.03&     0.43&      no \\
        &     AVG.&    $-$1.87&    $-$0.95&     0.14&  $-$0.14&  $-$0.04&  $-$0.05&  $-$0.05&     0.00&     0.43&      no \\
   I-27 &     APO &    $-$1.81&    $-$0.90&     0.08&     0.34&     0.58&     0.20&     ... &     0.48&     ... &     yes \\
        &    UVES &    $-$1.63&    $-$0.78&     0.10&     0.40&     0.46&     0.52&     0.47&     0.44&     0.45&     yes \\
        &     AVG.&    $-$1.72&    $-$0.84&     0.09&     0.37&     0.52&     0.36&     0.47&     0.46&     0.45&     yes \\
   I-36 &    LICK &    $-$1.89&    $-$1.38&     ... &  $-$0.15&     ... &  $-$0.03&  $-$0.05&     0.04&     ... &      no \\
   I-37 &    LICK &    $-$1.73&    $-$0.90&     ... &  $-$0.13&     0.43&     0.02&     0.03&  $-$0.06&     0.45&      no \\
   I-53 &    UVES &    $-$1.74&    $-$0.75&     0.15&     0.26&     0.35&     0.54&     0.29&     0.31&     0.41&     yes \\
   I-57 &     APO &    $-$1.65&    $-$0.75&     0.25&     0.45&     0.60&     0.22&     0.41&     0.42&     0.47&     yes \\
        &     AAT &    $-$1.62&    $-$0.65&     ... &     ... &     ... &     0.20&     0.27&     ... &     ... &     yes \\
        &     AVG.&    $-$1.64&    $-$0.70&     0.25&     0.45&     0.60&     0.21&     0.34&     0.42&     0.47&     yes \\
   I-80 &    UVES &    $-$1.70&    $-$0.85&     0.15&     0.24&     0.38&     0.50&     0.22&     0.18&     0.30&     yes \\
   I-85 &    UVES &    $-$1.81&    $-$1.10&     0.10&  $-$0.27&  $-$0.03&     0.14&     0.03&  $-$0.13&     0.44&      no \\
   I-86 &     AAT &    $-$1.80&    $-$0.98&     0.05&     ... &     ... &     0.12&  $-$0.05&     ... &     ... &      no \\
        &    UVES &    $-$1.84&    $-$1.00&     ... &  $-$0.18&     0.00&  $-$0.10&     0.05&     0.08&     0.52&      no \\
        &     AVG.&    $-$1.82&    $-$0.99&     0.05&  $-$0.18&     0.00&     0.01&     0.00&     0.08&     0.52&      no \\
   I-92 &    LICK &    $-$1.75&    $-$0.78&     ... &     0.07&     ... &  $-$0.11&  $-$0.01&  $-$0.01&     0.43&      no \\
   II-1 &    LICK &    $-$1.66&    $-$0.93&     ... &     0.20&     0.55&     0.02&     0.32&     0.38&     0.40&     yes \\
  II-31 &    LICK &    $-$1.65&    $-$0.88&     ... &     0.05&     0.10&  $-$0.23&     0.05&     0.14&     0.52&      no \\
  II-96 &     MCD &    $-$1.82&    $-$0.90&  $-$0.02&  $-$0.22&     0.10&  $-$0.18&     0.09&     0.05&     0.56&      no \\
 II-104 &    UVES &    $-$1.76&    $-$1.00&     0.05&  $-$0.09&  $-$0.05&     0.14&  $-$0.01&     0.05&     0.50&      no \\
  III-3 &     MCD &    $-$1.72&    $-$0.78&     0.13&     0.56&     0.40&     0.24&     0.23&     0.40&     0.45&     yes \\
        &     APO &    $-$1.78&    $-$0.67&     0.33&     0.42&     0.44&     0.31&     0.33&     0.49&     0.35&     yes \\
        &    UVES &    $-$1.66&    $-$0.63&     0.15&     0.58&     0.36&     0.33&     0.16&     0.35&     0.45&     yes \\
        &     AVG.&    $-$1.72&    $-$0.69&     0.20&     0.52&     0.40&     0.29&     0.24&     0.41&     0.42&     yes \\
 III-12 &     MCD &    $-$1.69&    $-$0.72&     0.23&     0.37&     0.54&     0.29&     0.39&     0.50&     0.55&     yes \\
        &     AAT &    $-$1.61&    $-$0.68&     ... &     ... &     ... &     0.31&     0.48&     ... &     ... &     yes \\
        &     AVG.&    $-$1.65&    $-$0.70&     0.23&     0.37&     0.54&     0.30&     0.44&     0.50&     0.55&     yes \\
 III-14 &    LICK &    $-$1.84&    $-$0.70&     ... &     0.10&     0.10&     0.04&     0.12&     0.21&     0.58&      no \\
        &     AAT &    $-$1.80&    $-$0.95&     ... &     ... &     ... &  $-$0.28&  $-$0.03&     ... &     ... &      no \\
        &     AVG.&    $-$1.82&    $-$0.83&     ... &     0.10&     0.10&  $-$0.12&     0.05&     0.21&     0.58&      no \\
 III-15 &    LICK &    $-$1.82&    $-$0.80&     ... &     0.09&  $-$0.05&  $-$0.05&     0.05&     0.16&     0.46&      no \\
 III-25 &    UVES &    $-$1.92&    $-$1.15&     0.10&  $-$0.23&     0.15&  $-$0.05&     0.13&     0.11&     0.50&      no \\
 III-33 &    UVES &    $-$1.78&    $-$0.95&     0.05&  $-$0.21&     0.00&  $-$0.11&  $-$0.04&     0.01&     0.40&      no \\
 III-35 &    UVES &    $-$1.83&    $-$0.95&     0.20&  $-$0.06&  $-$0.05&     0.10&     0.01&     0.01&     0.50&      no \\
 III-47 &     APO &    $-$1.82&    $-$0.78&  $-$0.02&     0.16&     0.50&     0.01&     ... &     0.03&     0.50&      no \\
 III-50 &    UVES &    $-$1.76&    $-$0.90&     0.15&     0.20&     0.30&     0.27&     0.20&     0.16&     0.35&     yes \\
 III-52 &    LICK &    $-$1.63&    $-$0.84&     ... &     0.32&     0.36&     0.38&     0.24&     0.30&     0.40&     yes \\
        &    UVES &    $-$1.62&    $-$0.73&     0.18&     0.36&     0.40&     0.46&     0.29&     0.49&     0.43&     yes \\
        &     AVG.&    $-$1.63&    $-$0.78&     0.18&     0.34&     0.38&     0.42&     0.27&     0.40&     0.42&     yes \\
 III-96 &     AAT &    $-$1.86&    $-$0.90&     ... &     ... &     ... &  $-$0.21&  $-$0.03&     ... &     ... &      no \\
  IV-20 &     AAT &    $-$1.65&    $-$0.78&     ... &     ... &     ... &     0.30&     0.49&     ... &     ... &     yes \\
        &    UVES &    $-$1.63&    $-$0.78&     0.20&     0.32&     0.36&     0.40&     0.29&     0.40&     0.34&     yes \\
        &     AVG.&    $-$1.64&    $-$0.78&     0.20&     0.32&     0.36&     0.35&     0.39&     0.40&     0.34&     yes \\
  IV-59 &    UVES &    $-$1.77&    $-$0.90&     0.00&  $-$0.14&  $-$0.04&  $-$0.10&  $-$0.02&  $-$0.02&     0.46&      no \\
  IV-68 &    UVES &    $-$1.75&    $-$1.05&     0.00&  $-$0.29&  $-$0.04&     0.16&     0.10&  $-$0.01&     0.53&      no \\
  IV-76 &     AAT &    $-$1.63&    $-$0.88&     ... &     ... &     ... &     0.19&     0.35&     ... &     ... &     yes \\
  IV-88 &     AAT &    $-$1.62&    $-$0.75&     ... &     ... &     ... &     0.39&     0.47&     ... &     ... &     yes \\
  IV-97 &    UVES &    $-$1.94&    $-$0.83&     0.30&  $-$0.01&     0.00&     0.00&  $-$0.13&     0.02&     0.48&      no \\
 IV-102 &    LICK &    $-$1.96&    $-$1.18&     ... &     0.01&  $-$0.10&  $-$0.06&  $-$0.13&     0.07&     ... &      no \\
        &     MCD &    $-$1.95&    $-$0.95&     0.23&     0.06&     0.00&  $-$0.26&  $-$0.07&     0.05&     0.52&      no \\
        &     AAT &    $-$2.01&    $-$0.73&     ... &     ... &     ... &  $-$0.14&  $-$0.07&     ... &     ... &      no \\
        &     AVG.&    $-$1.97&    $-$0.95&     0.23&     0.04&  $-$0.05&  $-$0.15&  $-$0.09&     0.06&     0.52&      no \\
      C &     MCD &    $-$1.69&    $-$0.82&     0.03&     0.43&     0.30&     0.26&     0.31&     0.34&     0.52&     yes \\
   C513 &     APO &    $-$1.86&    $-$0.82&     0.25&  $-$0.07&     0.15&  $-$0.26&     0.06&     0.22&     0.52&      no \\
    V-2 &    LICK &    $-$1.57&    $-$0.63&     ... &     0.26&     0.30&     0.22&     0.29&     0.51&     ... &     yes\\
\hline
\end{longtable}

\tablefoottext{a}{For this column only, metallicities [Fe/H] are given.}

\tablefoottext{b}{For this and all remaining abundances, [el/Fe] values are given.}
}

\begin{table*}
\begin{center}
\caption{Mean Abundances\label{tab-mean}}

\begin{tabular}{lrrrrrrrrrrrrrr}

\hline\hline
Abundance &mean   &$\pm$  &$\sigma$ &\# & & mean &$\pm$ &$\sigma$ &\# & &mean&$\pm$&$\sigma$ &\#  \\
          &\multicolumn{4}{c}{total}&&\multicolumn{4}{c}{\spo}&&\multicolumn{4}{c}{\sri}\\
\hline
$\rm [Fe/H]$    &  $-$1.76&     0.02&     0.10&       35& & $-$1.82&     0.02&     0.07&       21& &  $-$1.67&     0.01&     0.05&       14\\
$\rm [C/Fe]$    &  $-$0.50&     0.08&     0.28&       14& & $-$0.65&     0.08&     0.21&        8& &  $-$0.30&     0.10&     0.22&        6\\
$\rm [N/Fe]$    &     0.87&     0.09&     0.32&       14& &    0.67&     0.09&     0.25&        8& &     1.14&     0.07&     0.16&        6\\
$\rm [O/Fe]$    &     0.34&     0.03&     0.18&       35& &    0.36&     0.03&     0.15&       21& &     0.32&     0.06&     0.23&       14\\
$\rm [Na/Fe]_{NLTE}$   &     0.17&     0.04&     0.25&       35& &    0.08&     0.05&     0.24&       21& &     0.31&     0.05&     0.19&       14\\
$\rm [Na/Fe]_{LTE}$    &     0.26&     0.05&     0.27&       35& &    0.15&     0.06&     0.26&       21& &     0.42&     0.06&     0.20&       14\\
$\rm [Mg/Fe]$   &     0.39&     0.02&     0.11&       35& &    0.38&     0.03&     0.12&       21& &     0.39&     0.03&     0.11&       14\\
$\rm [Al/Fe]$   &     0.38&     0.05&     0.29&       31& &    0.29&     0.07&     0.30&       18& &     0.50&     0.07&     0.24&       13\\
$\rm [Si/Fe]$   &     0.44&     0.01&     0.06&       35& &    0.43&     0.02&     0.07&       21& &     0.45&     0.02&     0.06&       14\\
$\rm [Ca/Fe]$   &     0.30&     0.01&     0.07&       35& &    0.26&     0.01&     0.04&       21& &     0.36&     0.02&     0.06&       14\\
$\rm [Ti I/Fe]$ &     0.22&     0.01&     0.07&       35& &    0.19&     0.02&     0.06&       21& &     0.26&     0.02&     0.07&       14\\
$\rm [Ti II/Fe]$&     0.32&     0.01&     0.06&       32& &    0.32&     0.01&     0.06&       20& &     0.32&     0.02&     0.06&       12\\
$\rm [Cu/Fe]$   &  $-$0.88&     0.02&     0.14&       35& & $-$0.94&     0.03&     0.14&       21& &  $-$0.79&     0.02&     0.09&       14\\
$\rm [Zn/Fe]$   &     0.13&     0.02&     0.09&       24& &    0.10&     0.03&     0.11&       14& &     0.16&     0.02&     0.07&       10\\
$\rm [Y/Fe]$    &     0.07&     0.04&     0.24&       32& & $-$0.08&     0.03&     0.13&       20& &     0.33&     0.03&     0.10&       12\\
$\rm [Zr/Fe]$   &     0.21&     0.04&     0.22&       30& &    0.07&     0.04&     0.16&       18& &     0.41&     0.03&     0.11&       12\\
$\rm [Ba/Fe]$   &     0.09&     0.04&     0.22&       35& & $-$0.05&     0.03&     0.12&       21& &     0.31&     0.04&     0.13&       14\\
$\rm [La/Fe]$   &     0.14&     0.03&     0.18&       34& &    0.01&     0.01&     0.06&       20& &     0.33&     0.02&     0.09&       14\\
$\rm [Nd/Fe]$   &     0.17&     0.03&     0.19&       32& &    0.05&     0.02&     0.09&       20& &     0.37&     0.03&     0.11&       12\\
$\rm [Eu/Fe]$   &     0.46&     0.01&     0.07&       30& &    0.49&     0.01&     0.05&       19& &     0.42&     0.02&     0.08&       11\\
$\rm [La/Y]$    &     0.06&     0.03&     0.15&       31& &    0.11&     0.04&     0.16&       19& &  $-$0.01&     0.03&     0.11&       12\\
$\rm [(C+N+O)/Fe]$ &     0.33&     0.02&     0.08&       14& &    0.28&     0.02&     0.05&       8 & &   0.41&     0.02&     0.04&     6  \\
$\rm {log\epsilon(C+N+O)}$ &     7.68&     0.04&     0.16&       14& &    7.57&     0.03&     0.09&       8 & &   7.84&     0.03& 0.07    &     6  \\

\hline

\end{tabular}

\end{center}
\end{table*}

%\clearpage
%\input{tab-datastars}

%\clearpage
%\input{tab-model}

%\clearpage
%\input{tab-errors}

%\clearpage
%\input{tab-ablight}

%\clearpage
%\input{tab-abheavy}

%\clearpage
%\input{tab-mean}

\end{document}